\documentclass[%
 reprint,
 amsmath,amssymb,
 aps,
]{revtex4-2}

\newcommand{\xmax}[1]{#1_{\text{max}}}

\usepackage{bm}
\usepackage{graphicx}
\usepackage[]{subfigure}
\usepackage{color}

\begin{document}

\title{Direct statistical simulation of Lorenz96 system in model reduction approaches}%

\author{Kuan Li}%
\email[Email:]{kuan.li.81@gmail.com}
\affiliation{Department of Applied Mathematics, University of Leeds, Leeds, LS2 9JT, UK}

\author{J.B. Marston}%
\affiliation{ 
Department of Physics, Box 1843, Brown University, Providence, Rhode Island 02912-1843, USA, \& Brown Theoretical Physics Center, Brown University, Providence, Rhode Island 02912-S USA}%

\author{Steven M. Tobias}
\affiliation{Department of Applied Mathematics, University of Leeds, Leeds, LS2 9JT, UK}

\date{}%

\begin{abstract}
Direct statistical simulation (DSS) of nonlinear dynamical systems bypasses the traditional route of accumulating statistics by lengthy direct numerical simulations (DNS) by solving the equations that govern the statistics themselves.  DSS suffers, however, from the curse of dimensionality as the statistics (such as correlations) generally have higher dimension than the underlying dynamical variables.  Here we investigate two approaches to reduce the dimensionality of DSS, illustrating each method with numerical experiments with the Lorenz-96 dynamical system. The forms of DSS chosen here involve approximate closures at second and third order in the equal time cumulants.  We demonstrate significant reduction in computational effort that can be achieved without sacrificing the accuracy of DSS.  The methods developed here can be applied to turbulent fluid and magnetohydrodynamical systems.

\end{abstract}

\maketitle

\section{Introduction}

Astrophysical and geophysical fluids are usually highly anisotropic and inhomogeneous.  When the fluids are turbulent, direct numerical simulation (DNS) is of limited use due to the enormous range of spatio-temporal scales that are active \cite{Glatzmaier_2013}.  Statistical approaches are often more fruitful \cite{frisch_1995} in these cases.  Direct statistical simulation (DSS) eschews the numerical simulation of the underlying dynamics in favour of solving directly for the statistics without imposing assumptions of homogeneity or isotropy. DSS comes in many forms but the one we focus on here is based upon expansions in low-order equal-time cumulants \cite{farrell_ioannou_2003, marston65conover}.  The flows may often be decomposed into slowly varying coherent structures at large scales and rapidly varying non-coherent turbulence at small scales \cite{tobias_2021}.  Low-order statistics may therefore be smoother in space than the dynamical variables themselves and thus it is natural to seek approximations to DSS of reduced dimensionality.  We note that this approach has been applied with some success to idealized barotropic flows in Ref. \cite{allawala_2016}.

The form of DSS we investigate here is based upon ensemble averaging and expansions in the low-order equal-time cumulants.  Closures of the hierarchy of cumulants at second (CE2) and third order (CE3 and CE2.5) are considered.  Cumulant expansions have been deployed with success to study a variety of problems from the chaotic low dimensional dynamical problems governed by the ordinary differential equations (ODEs), e.g., see \cite{allawala_2016} to the weakly turbulent but spatially more complicated fluid dynamical and magnetohydrodynamical (MHD) problems governed by the partial differential equations (PDEs), e.g., see \cite{Tobias_2011, sb2015}. The effectiveness of DSS has been demonstrated for highly chaotic systems \cite{kuan_2021} that are nevertheless accurately approximated by a low-order closure (CE2.5). We note that fluid dynamical system are typically less chaotic or turbulent than many nonlinear dynamical systems, and are often well approximated by CE2. However, in certain cases CE2 fails qualitatively \cite{tm2013,marston2023}. CE2.5, CE3, or other forms of DSS are then required.


DSS is plagued, however, by the curse of dimensionality \cite{bellman_1957, bishop_2016} as the statistics (such as correlations) generally have higher dimension than the underlying dynamical variables. For a three dimensional fluid dynamical problem with a spatial resolution of $10^3\times 10^3\times 10^3$, the second and third cumulants would require $10$ terabytes and $10^4$ petabytes of storage respectively.  Here we develop two approaches to model reduction to tame the size of the cumulant expansions:  A dynamical strategy via the eigen-decomposition, and a static strategy via unitary transformations similar to that employed in Ref. \cite{atm2020}.

We explore the numerical efficiency and convergence of the two strategies for a representative low-order dynamical model, the Lorenz96 system \cite{Lz_96}, in a variety of its dynamical regimes. The nonlinear behaviour of the Lorenz96 model is controlled by a parameter that allows us to explore how the reliability of the dimensionally reduced DSS degrades as the nonlinearity increases.   The Lorenz96 system possesses a discrete translational symmetry that mimics the continuous translational symmetry often found in geophysical and astrophysical fluid dynamical systems. It also breaks reflection symmetry and has a chiral nature analogous to properties of rotating fluids.  Thus it serves as a toy model for more complicated fluid models encountered in nature. We also investigate the  breaking of translation symmetry with the addition of inhomogeneous forcing.

The paper is organised as follows. Section \ref{L96} we introduce the Lorenz96 system and the formulation of direct statistical simulation via cumulant expansions. Model reduction strategies are discussed in Section \ref{diag}. The results of our numerical experiments are presented and discussed in Section \ref{exp}. We conclude with some suggestions for further implementations of the model reduction methods for PDE systems in Section \ref{con}.

\section{The Lorenz96 model and cumulant expansions}
\label{L96}

The Lorenz96 system is a simplified fluid dynamical system in one-dimension, which was originally developed by \cite{Lz_96} to study the predictability of the weather forecasting systems. The system is quadratically nonlinear, in which the nonlinear(inertial) term is designed to conserve energy. The model consists of $\xmax{n}$ ($\xmax{n}\geq 4$) independent variables and satisfies the periodic condition, i.e., $x_i = x_{i+\xmax{n}}$. In the vector form, the governing equation may be written as
\begin{equation}
	\frac{d}{dt} {\bf x} = {\bf x}^T {\cal Q} \cdot {\bf x} + {\cal L} \cdot {\bf x} + {\bf f},
	\label{govEq}
\end{equation}
where the nonlinear operator, ${\cal Q}$, is a sparse tensor of rank-three with the nonzero entries, ${\cal Q}_{(i,i+1,i-1)} = 1$ and ${\cal Q}_{(i,i-2,i-1)} = -1$ for ${\xmax n} \geq i\geq 1$ and the linear operator, ${\cal L}$, is an identity matrix with negative sign, i.e., ${\cal L} = \text{diag}[-1, -1, \cdots]$. Specifically, the $i$-th component of the equation reads
\begin{eqnarray}
	    d_t  x_i &=& (x_{i+1}-x_{i-2}) x_{i-1}  - x_i + f_i.
\label{Lz96eq}
\end{eqnarray}
The external force, $f_i$, is an external force and is assumed to be either a constant in time or the independent random vector with each component satisfying the Gaussian distribution, $f_i \sim {\cal N}(\mu_i, \sigma^2_i)$, where $\mu_i$ and $\sigma_i^2$ are the statistical mean and variance of $f_i$. Two dynamical regimes of the Lorenz96 system are of interests in this study, 1) the periodic state, if the external force is approximately $f_i \sim {\cal O}(1)$ and 2) the chaotic state if $f_i$ is large, i.e., $f_i\geq 5$.

The Lorenz96 system resembles a set of typical dynamical systems governed by the ordinary/partial differential equations defined on a latitude circle. The Lorenz96 system satisfies the translational symmetry, if the external force, $f_i$ is an equal constant or satisfying the same probability distribution function (PDF) over all nodes. In this study, we consider the case that all unknown variables, $x_i$, operate in the same time scale without sub-grid couplings \cite{Fatkullin_2004} and we set the spatial resolution of the system to be ${\xmax n} = 8$ for comparison purposes.

\subsection{The cumulant representation of the Lorenz96 system in statistical space}

The dynamics modelled by the governing dynamical equations, such as Eq.~(\ref{govEq}) for the Lorenz96 system, can be equivalently described by the probability distribution function, PDF, governed by the statistical equations. Direct statistical simulation is a general mathematical framework, which converts the spatial-temporal dependent differential/partial differential equations in physical space to the evolution of PDFs in statistical phase space. The statistical equilibrium of the dynamical states given by the solution of the cumulants equations, are thereafter invariant or evolving slowly in time.

\subsubsection{The cumulant hierarchy}
\label{cum}

The statistics, namely cumulants, which is a measure of the shape of a probability distribution function in statistical hierarchy \cite{Kendall_1987}, is the fundamental building block in DSS approximation of the dynamical systems. Applying the Reynolds decomposition by assuming the state vector, ${\bf x}$, of the dynamical system to be a random variable, one can write ${\bf x}$ as the sum of the coherent component, $C_{\bf x} = \langle {\bf x} \rangle $, and a non-coherent counterpart, $\delta {\bf x}$, i.e., 
	\begin{equation}
			{\bf x} = C_{\bf x} + \delta {\bf x},
		\label{rede}
	\end{equation}
where the coherent component, $C_{\bf x}$, is also known as the first cumulant of the state vector, ${\bf x}$ and $\delta {\bf x}$ is treated as the random fluctuation, which vanishes in statistical average, $\langle \delta {\bf x} \rangle ={\bf 0}$. We also assume that the statistical average further satisfies of the Reynolds averaging rules, i.e.,
\begin{equation}
\langle \delta {\bf x} \rangle = {\bf 0}, \hspace{0.05\hsize} \langle \delta {\bf x} \otimes C_{\bf x} \rangle = {\bf 0}, \hspace{0.02\hsize} \text{and} \hspace{0.02\hsize} \langle {\bf x} \otimes C_{\bf x} \rangle = C_{\bf x} \otimes C_{\bf x},
\end{equation}
where the symbol, $\otimes$, stands for the outer product of two tensors. In this paper, the ensemble average is employed to derive the cumulant equations of the low-order dynamical systems and is noted as $\langle \bullet \rangle$. By definition \cite{Kendall_1987}, the second and third cumulants of ${\bf x}$, which read  
\begin{equation}
C_{{\bf x} {\bf x}} = \langle \delta {\bf x} \otimes \delta {\bf x} \rangle \hspace{0.02\hsize} \text{and} \hspace{0.02\hsize} C_{{\bf x}{\bf x}{\bf x}} = \langle \delta {\bf x} \otimes \delta {\bf x} \otimes \delta {\bf x} \rangle
\end{equation}
in the vector form or
\begin{equation}
C_{x_i x_j} = \langle \delta x_i \delta x_j \rangle \hspace{0.02\hsize} \text{and} \hspace{0.02\hsize} C_{x_i x_j x_k} = \langle \delta x_i \delta x_j \delta x_k \rangle
\end{equation}
in the scalar form for ${\xmax n} \geq i,j,k \geq 1$ are identical to the second and third order statistical central moments. We note that the cumulant expansion of a PDF is equivalent to the expansions of statistical moments, i.e., any two PDFs with identical cumulant expansion also have identical cumulants, and vice versa. The fourth and higher order cumulants, which differ from the central moments, are chosen to represent the state vector, ${\bf x}$, in DSS framework than using the central moments, i.e., if the PDF of ${\bf x}$, is or close to a Gaussian distribution, the cumulant expansion equal to or greater than the third order vanishes or remains very small; whilst the central moments diverges as the even order of the expansion approaches to infinity. In this paper, we explicitly use cumulant expansions up to fourth order, which reads
\begin{equation}
C_{{\bf x}{\bf x}{\bf x}{\bf x}} = \langle \delta {\bf x} \otimes \delta {\bf x} \otimes\delta {\bf x} \otimes\delta {\bf x} \rangle - \left\lbrace C_{{\bf x}{\bf x}} \otimes C_{{\bf x}{\bf x}} \right\rbrace_3,
\label{defc4}
\end{equation}
where the term $C_{{\bf x}{\bf x}{\bf x}{\bf x}}$ and $\langle \delta {\bf x} \otimes \delta {\bf x} \otimes\delta {\bf x} \otimes\delta {\bf x} \rangle$ are the fourth cumulant and central moments respectively; whilst the higher order terms, such as the fifth order cumulant, which is defined as
\begin{equation}
C_{\bf xxxxx} = \langle \delta {\bf x} \otimes \delta {\bf x} \otimes\delta {\bf x} \otimes \delta{\bf x} \otimes\delta {\bf x} \rangle - \left\lbrace C_{\bf xxx} \otimes C_{\bf xx} \right\rbrace_{10}
\label{defc5}
\end{equation}
are not considered. In Eqs.~(\ref{defc4} \& \ref{defc5}), the symbol, $\lbrace \ \rbrace_n$, stands for the symmetrisation operation with $n$ permutations, e.g,
\begin{eqnarray}
\lbrace C_{x_1x_2} C_{x_3x_4} \rbrace_3 &=& C_{x_1x_2} C_{x_3x_4} + C_{x_1x_3} C_{x_2x_4} \nonumber \\ &+& C_{x_1x_4} C_{x_2x_3} \nonumber \\
\lbrace C_{x_1x_2} C_{x_3x_4x_5} \rbrace_{10} &=& C_{x_1x_2} C_{x_3x_4x_5} + C_{x_1x_3} C_{x_2x_4x_5} \nonumber \\&+& C_{x_1x_4} C_{x_2x_3x_5} +  C_{x_1x_5} C_{x_2x_3x_4} \nonumber \\ 
&+& C_{x_2x_3} C_{x_1x_4x_5} + C_{x_2x_4} C_{x_1x_3x_5} \nonumber \\
&+& C_{x_2x_5} C_{x_1x_3x_4} + C_{x_3x_4} C_{x_1x_2x_5} \nonumber \\
&+& C_{x_3x_5} C_{x_1x_2x_4} + C_{x_4x_5} C_{x_1x_2x_3} \nonumber \\
\label{symn}
\end{eqnarray}

\subsubsection{The Hopf approach of the cumulant expansion of the Lorenz96 system}
\label{hopf}

Owing to the computational expense of solving for the full distribution via the Fokker-Planck equations, it is 
natural to adopt a strategy that
 only solves the governing equations of the low-order statistics. The cumulant expansion of the dynamical system given by (\ref{govEq}) can be directly derived from the governing dynamical equations \cite{allawala_2016, kuan_2021} or via the Hopf functional approach \cite{frisch_1995}. In this paper, we adapt the latter approach to obtain the cumulant expansions of the Lorenz96 system. To begin with, we first define the generating function, ${\bf \Psi}$,
\begin{equation}
{\bf \Psi}[{\bf p}, {\bf x}] = e^{\text{i} {\bf p} \cdot {\bf x}},
\label{wavefun}
\end{equation}
which is analogous to the wave function in the quantum mechanics, where ${\bf p}$ is an auxiliary variable and $\text{i}$ is the imaginary unit. In quantum physics, the quantities, ${\bf p}$ and ${\bf x}$ are the uncertainty-pair and satisfy the relation, $p_i = - { i} \partial/ \partial x_i$. We note that the function, ${\bf \Psi}$, is a generating function for the central moments. To obtain the higher order cumulant equations with the order greater than three, the linear transform must be applied to convert the governing equation of the central moments into the cumulant, e.g., see \$\ref{cum}. The first three cumulant of ${\bf x}$ represented by the generating function, ${\bf \Psi}$, are given by 
\begin{eqnarray}
C_{\bf x} &=& \left. \nabla \left\langle e^{i {\bf p} \cdot {\bf x}} \right\rangle \right\rvert_{{\bf p}={\boldsymbol 0}}, \nonumber\\
C_{{\bf x} {\bf x}} &=& \left. \nabla_p \otimes \nabla_p \left\langle  e^{i {\bf p} \cdot ({\bf x} - C_{\bf x})} \right\rangle \right\rvert_{{\bf p}={\boldsymbol 0}} \nonumber \\  
C_{{\bf x} {\bf x} {\bf x}} &=& \left. \nabla_p \otimes \nabla_p \otimes \nabla_p \left\langle e^{i {\bf p} \cdot ({\bf x} - C_{\bf x})} \right\rangle \right\rvert_{{\bf p}={\boldsymbol 0}},
\end{eqnarray}
where the symbol, $\nabla_p$, stands for the gradient operator, i.e., $\nabla_p = -\text{i} \left[\partial_{p_{1}}, \partial_{p_{2},}\cdots \right]$. The generating function, ${\bf \Psi}$, satisfies a Schr\"odinger-like equation
\begin{equation}
	\text{i} \frac{d}{d t} {\bf \Psi} = \hat{\cal H} {\bf \Psi},
	\label{schro1}
\end{equation}
where the operator, ${\hat{\cal H}}$, which is linear and analogy to the total Hamiltonian in quantum physics, is given by
\begin{equation}
	{\hat{\cal H}} = {\bf p} \cdot \text{i} \left[{\cal L} \nabla_p + {\cal Q} \nabla_p \otimes \nabla_p \right].
	\label{schro2}
\end{equation}
Moreover, the generating equation, (\ref{schro1}) is invariant under the unitary transform and can be further written as
\begin{equation}
	\text{i} \frac{d}{d t} \left[ U{\bf \Psi} U^{\dagger} \right]= U{\hat{\cal H}} {\bf \Psi} U^{\dagger} = \left[ U{\hat{\cal H}} U^{\dagger} \right] \left[U {\bf \Psi} U^{\dagger}\right],
	\label{schro3}
\end{equation}
where the $U$ is an arbitrary unitary transform. By taking the Taylor's expansion of the generating equation (\ref{schro1}) and equating the coefficients of the power series of ${\bf p}$ up to the third order, we obtain the first three cumulant equations in the vector form as follows
\begin{eqnarray}
\frac{d}{dt}C_{\bf x} &=& {\cal L} \cdot C_{\bf x} + {\cal Q} \cdot C_{\bf x} \cdot C_{\bf x} + {\cal Q} \cdot C_{{\bf x}{\bf x}}, \label{c1} \\
\frac{d}{dt}C_{{\bf x}{\bf x}} &=& \left\lbrace {\cal L} \cdot C_{{\bf x}{\bf x}} + 2{\cal Q} \cdot C_{\bf x} \cdot C_{{\bf x}{\bf x}} + {\cal Q} \cdot C_{{\bf x}{\bf x}{\bf x}}\right\rbrace_2, \label{c2} \\
\frac{d}{dt}C_{{\bf x}{\bf x}{\bf x}} &=& \left\lbrace {\cal L} \cdot C_{{\bf x}{\bf x}{\bf x}} + 2{\cal Q} \cdot C_{{\bf x}{\bf x}{\bf x}} \cdot C_{{\bf x}} + 2{\cal Q} \cdot C_{{\bf x}{\bf x}} \cdot C_{{\bf x}{\bf x}} \right\rbrace_3 \nonumber \\
&+& {\cal O}(C_{{\bf x}{\bf x}{\bf x}{\bf x}}),
\label{c3}
\end{eqnarray} 
where the symbol, $\lbrace \bullet \rbrace_n$ for $n=2$ and $3$, which is defined in Eq. (\ref{symn}), stands for the symmetrisation procedure with $2$ and $3$ permutations. Equivalently, in scalar form, the cumulant equation can be written as
\begin{eqnarray}
d_t C_i &=& -C_{i-2}C_{i-1} - C_{i-2,i-1} + C_{i-1}C_{i+1}\nonumber \\ && + C_{i-1,i+1}  - C_i + \mu_i,\\
d_t C_{i,j} &=& -C_{i-2}C_{i-1,j} - C_{i-2,i-1,j} - C_{i-2,j}C_{i-1} \nonumber \\
&& + C_{i-1}C_{i+1,j} + C_{i-1,i+1,j} + C_{i-1,j}C_{i+1} \nonumber \\
&& - C_{i,j-2}C_{j-1} - C_{i,j-2,j-1} - C_{i,j-1}C_{j-2} \nonumber \\
&&+ C_{i,j-1}C_{j+1} + C_{i,j-1,j+1} - 2C_{i,j} \nonumber \\ &&+ C_{i,j+1}C_{j-1}  + 2\sigma^2_i \delta_{i,j}  \label{ceI2} \\
d_t C_{i,j,k} &=& -C_{i-2}C_{i-1,j,k} - C_{i-2,j}C_{i-1,k} - C_{i-2,j,k}C_{i-1} \nonumber \\ 
&&- C_{i-2,k}C_{i-1,j} + C_{i-1}C_{i+1,j,k}  + C_{i-1,j}C_{i+1,k} \nonumber \\ 
&&+ C_{i-1,j,k}C_{i+1} + C_{i-1,k}C_{i+1,j} - C_{i,j-2} C_{j-1,k}\nonumber \\ &&
- C_{i,j-2,k}C_{j-1} - C_{i,j-1}C_{j-2,k} + C_{i,j-1}C_{j+1,k} \nonumber \\ 
&& - C_{i,j-1,k}C_{j-2} + C_{i,j-1,k}C_{j+1} + C_{i,j,k+1}C_{k-1} \nonumber \\
 && -C_{i,j,k-2}C_{k-1} + C_{i,j,k-1}C_{k+1} - C_{i,j,k-1}C_{k-2}\nonumber \\
 && - 3C_{i,j,k} + C_{i,j+1}C_{j-1,k} + C_{i,j+1,k}C_{j-1}  \nonumber \\
 &&+ C_{i,k+1}C_{j,k-1} - C_{i,k-2}C_{j,k-1} + C_{i,k-1}C_{j,k+1} \nonumber \\
 &&- C_{i,k-1}C_{j,k-2} + {\cal O}\left(\langle {\delta x_i,\delta x_j, \delta x_k, \delta x_l}\rangle\right),
 \label{lz96cum}
\end{eqnarray}
where $C_i$, $C_{i,j}$ and $C_{i,j,k}$ are the short notations of the cumulants, $C_{x_i}$, $C_{x_ix_j}$ and $C_{x_ix_jx_k}$, respectively for ${\xmax n}\geq i,j,k \geq 1$. The second and third cumulant equations are sparse equations, e.g., the sub- and super-diagonal elements, $C_{i,i+1}$, which never appear on the right hand side of the second cumulant equations, are only determined via the definition of cumulant hierarchy, $d_t \langle\delta x_i \delta x_{i+1} \rangle$. The second and third cumulant equations are symmetric, which comprise of ${\xmax n}({\xmax n}+1)/2$ and ${\xmax n}({\xmax n}+1)({\xmax n}+2)/6$ independent entries.

\subsubsection{The statistical closure of the cumulant equations}

For a quadratically nonlinear dynamical system such as (\ref{govEq}), the $n{+1}$st cumulant always appears in the $n$th order cumulant equation. However, the analytical expansions of the high order cumulant equations, e.g., see Eqs. (\ref{c1}--\ref{c3}), become more complex, of higher dimension, and more computationally intensive, leading to the curse of dimensionality. Hence, the cumulant hierarchy of the DSS equations must be truncated at the lowest possible order.  

\vspace{0.02\hsize}
\noindent{\bf {\small The CE2 closure:}} \\ 
It is possible that the PDF of the state vector, ${\bf x}$,  of the dynamical system can be effectively approximated by the Gaussian distribution, for which the cumulant hierarchy naturally truncates at second order and all higher order terms are zero. For this case the cumulant equations describing the evolution of the first and second cumulant in Eq. (\ref{c1}) and (\ref{c2}) are called the CE2 system, so called because only the equations for the first and second cumulants are solved and  where all higher order terms greater than two are neglected, see e.g., \cite{mqt2019}. The CE2 approximation is the simplest DSS system and is particularly suited for solving the fluid dynamical problems, for which the primary interaction is that between the coherent components and non-coherent fluctuations.

\vspace{0.02\hsize}
\noindent{\bf {\small The CE3 closure:}}\\ 
In many realistic problems, as one moves away from statistical equilibrium, the statistics of a dynamical systems is poorly represented by a Gaussian PDF. Many distributions exhibit strong asymmetry (skewness) or long tails (flatness) as we will see in \$\ref{exp}. This indicates the significance of physical process within the dynamical system for the interactions between the non-coherent components. For these problems, one may have to take the third order cumulants into consideration, for example setting the fourth order cumulant to zero $C_{\bf xxxx}$ =0, \cite{orszag_1970} i.e.,
\begin{equation}
0 = C_{\bf xxxx} = \langle \delta {\bf x} \otimes \delta {\bf x} \otimes \delta {\bf x} \otimes \delta {\bf x} \rangle -  \left\lbrace C_{\bf xx} \otimes C_{\bf xx}\right\rbrace_3.
\end{equation}
This may lead to an unrealisable system. To remedy this, some effects of the fourth order cumulants that are proportional to the rate of change (gradient) of $x_i$ can be included through modelling  via a diffusion process, $-C_{x_ix_mx_n}/\tau_d$ \cite{monin_1975}. The parameter, $\tau_d>0$, is known as the eddy damping parameter \cite{mqt2019}. The third order cumulant equation (\ref{c3}) may now be rewritten as
\begin{eqnarray}
\frac{d}{dt}C_{{\bf x}{\bf x}{\bf x}} &=& \left\lbrace {\cal L} \cdot C_{{\bf x}{\bf x}{\bf x}} + 2{\cal Q} \cdot C_{{\bf x}{\bf x}{\bf x}} \cdot C_{{\bf x}} + 2{\cal Q} \cdot C_{{\bf x}{\bf x}} \cdot C_{{\bf x}{\bf x}} \right\rbrace_3 \nonumber \\
 &-& \frac{1}{\tau_d} C_{\bf xxx}.
 \label{c3II}
\end{eqnarray}

\vspace{0.02\hsize}
\noindent{\bf {\small The CE2.5 closure as a simplified CE3 approximation:}} \\
The CE3 equations are complicated and involve many interactions. They may be simplified slightly by assuming that the third cumulant evolves rapidly in comparison with the first and second cumulant. This
means that Eq. (\ref{c3II}) is further simplified to a diagnostic system by setting all time derivatives for the third cumulants to be zero, i.e. $d_t C_{\bf xxx}=0$. A further simplification that leads to faster computation involves the neglect of  all terms involving the first order cumulants, $C_{\bf x}$ in the equations for the third cumulant. The third order cumulants then are the solution of the diagnostic equation,
\begin{equation}
 \frac{1}{\tau_d} C_{\bf xxx} = \left\lbrace 2{\cal Q} \cdot C_{{\bf x}{\bf x}} \cdot C_{{\bf x}{\bf x}} \right\rbrace_3.
 \label{c25}
\end{equation}
This truncation that couples Eqs. (\ref{c1}), (\ref{c2}) and (\ref{c25}) is named CE2.5 approximation \cite{mqt2019} and \cite{atm2020}.

\subsubsection{The eigenbasis of the second cumulant,  $C_{\bf xx}$}

Recalling the definition of the second cumulant (covariance) of the state vector, i.e., $C_{\bf xx} = \langle \delta {\bf x} \otimes \delta {\bf x} \rangle$, it is straightforward to show that the matrix, $C_{\bf xx}$, is a non-negative definite matrix with the dimension, ${\xmax n} \times {\xmax n}$, which can always be diagonalised by the orthogonal transform, i.e.,
\begin{equation}
	C_{\bf xx} =  {\bf V}^T \cdot {\bf D} \cdot {\bf V}, 
	\label{pod}
\end{equation}
where the symbol, ${\bf V}^T$, is the transpose of the matrix, ${\bf V}$. The diagonal matrix, ${\bf D}$, is comprised of the eigenvalues, $\lambda_i$, of the covariance matrix, ${\bf D} = \text{diag}[\lambda_1, \lambda_2, \cdots]$ and $\bf V$, which is the matrix representation of the unitary transform, ${U}$, defined in Eq.~(\ref{schro3}), is the orthogonal matrix with the columns given by the eigenvectors, ${\bf V} = [{\bf v}_1, {\bf v}_2, \cdots]$, where  ${\bf v}_i$ is a column vector. The eigenvectors of the covariance matrix are also the eigenbasis of the dynamical system. In data science, the eigen-decomposition in Eq.~(\ref{pod}) is  alternatively named as the proper orthogonal decomposition \cite{Berkooz_1993}.

We now consider the case where the dynamical system has translational symmetry. This is achieved if the forcing does not depend on node-site, i.e. if the external force, $f_i$, is an equal constant or satisfies the same statistical distribution over all nodes, we will see that in \$\ref{exp} this symmetry appears in the solutions of the Lorenz96 system in both of the periodic and chaotic states. For these systems, the Fourier series,
\begin{equation}
{\cal F}_k(\theta_j) = \sqrt{\frac{2}{\xmax{n}}} \left\lbrace
\begin{array}{rl}
\frac{1}{\sqrt{2}}  & \text{for} \ \ \ k = m = 0, \\
\cos m \theta_j    & \text{for} \ \ \ k = 2m - 1,\\
\sin m \theta_j     & \text{for} \ \ \ k = 2m, 
\end{array}
\right.
\end{equation}
for $m=1,2,3,\cdots {\xmax n}/2$, is the eigen-basis of the covariance matrix, where $m$ is the wave number and $\theta_j$, is the angle of the state variable, $x_i$, in the latitude circle, i.e., $\theta_j = 2 \pi \frac{i-1}{\xmax{n}}$ for ${\xmax n} \geq i \geq 1$. Very importantly, the eigen-decomposition of the covariance matrix is non-unique, due to the reason that the eigen-pairs of the covariance matrix, $C_{\bf xx}$, are doubly-folded and the linear combination of two eigenvectors with the same eigenvalue is also the eigenvector of the covariance matrix. The degeneracy of the eigen-pairs can be removed if the translational symmetry of the dynamical system is broken. For the case of broken translational symmetry, the Fourier series is no longer the eigenbasis of the covariance matrix.

Perhaps, it is of interests to define a dynamical basis functions for representing the dynamics of the Lorenz system in translational symmetry. At each time step, $t_i$, we project the solution of the dynamical model, ${\bf x}(t_k)$, onto the Fourier basis, ${\cal F}_k$, and obtain the following relation
\begin{eqnarray}
x_j (t_k) &=& \sum_m c_m(t_i) \cos m\ \theta_j + d_m(t_i) \sin m\  \theta_j \nonumber \\
&=& \sum_m a_m(t_i) \sin m \left[\theta_j + \Delta \theta(t_i)\right],
\end{eqnarray}
where the spectral coefficient, $a_m$, is the defined as $a_m = \sqrt{c_m^2 + d_m^2}$ and $\Delta \theta$ is the phase shift and satisfies $\sin \Delta \theta = c_m / a_m$ and $\cos \Delta \theta = d_m / a_m$. The adaptive basis function, 
\begin{equation}
{\cal D}_m = \sin m (\theta_j + \Delta \theta),
\label{dynbs}
\end{equation}
which varies in time, is also an eigenfunction of the covariance matrix, $C_{{\bf x}{\bf x}}$.

\subsubsection{The model reduction strategies for the cumulant equations}
\label{diag}

The covariance matrix, $C_{\bf xx}$, is a measure of the complexity of the spatial fluctuation of the non-coherent components of the dynamical system and the eigensystem of the matrix can be further utilised to simplify the complexity of the cumulant equations in the CE2/2.5/3 approximations. In this study, we consider two strategies for reducing the cumulant equations of the Lorenz96 systems:

\vspace{0.01\vsize}
\noindent{\bf Strategy {\it i})}\\

The state vector, ${\bf x}$, of the dynamical system can be uniquely represented by the eigenbasis of the covariance matrix; each eigenvector's  significance is quantified  by the corresponding eigenvalue. When we solve the matrix-based cumulant systems in time, e.g., Eqs. (\ref{c1}, \ref{c2} \& \ref{c25}), we filter out the eigenvectors with very small eigenvalues at each time step to reduce the complexity of the covariance.  This strategy is specially suited for computing cumulant equations with very small number of significant eigen-pairs, $n_e$, but with very large spatial resolutions, ${\xmax n}$. For these problems, we may consider to represent the covariance matrix using the Schmidt decomposition, i.e.,
\begin{equation}
	C_{\bf xx} = \sum_{k=1}^{n_e} \lambda_k {\bf v}_k \otimes {\bf v}_k,
	\label{par_sum}
\end{equation}
where $\lambda_k$ and ${\bf v}_k$ are the $k$th eigen-pair of the covariance matrix. At every time step, we store and compute $n_e$ significant eigen-pairs of the cumulant equations so that equation~\ref{par_sum} is an accurate represnetation for $n_e < n$ \cite[see e.g.][]{atm2020}. 

\vspace{0.01\vsize}
\noindent{\bf Strategy {\it ii})}\\
Using the orthogonal transform, defined in Eq.~(\ref{pod}), we first transform the governing equation~(\ref{govEq}) of the Lorenz96 system into the following form  
\begin{equation}
	\frac{d}{dt} {\bf y} = {\bf y}^T {\cal Q}' \cdot {\bf y} + {\cal L}' \cdot {\bf y} + {\bf V} \cdot {\bf f},
	\label{govEq2}
\end{equation}
where the state vectors of the original and new governing equations satisfy
\begin{equation}
	 {\bf x} = {\bf V}^{T} \cdot {\bf y} \hspace{0.05\hsize} \text{and} \hspace{0.05\hsize} {\bf y} = {\bf V} \cdot {\bf x}.
\end{equation}
The quadratically nonlinear operator, ${\cal Q}'$ and the linear operator, ${\cal L}'$ of the transformed dynamical equation are given by $ {\cal Q}' = {\bf V} \cdot {\cal Q}$ and ${\cal L}' = {\bf V} \cdot {\cal L}$. By using the technique introduced in \$\ref{hopf}, we obtain the cumulant expansion for the transform equations~(\ref{govEq2}). Then the cumulant expansion of the second order is simultaneously diagonalised. We note that the unitary transform does not create or annihilate information of the dynamical equation, i.e., the transformed cumulant equations in CE2/2.5/3 are equivalent to the original cumulants.

The transformed cumulant equation is numerically attractive as we only need to solve the diagonal elements of the second cumulant and significantly reduce the computational complexity for the third order equations, i.e., many of the nonlinear interactions involving the off-diagonal elements of the covariance are ultimately zero in all time. For the Lorenz96 system with translational symmetry, the Fourier series is the natural eigenbasis of the covariance matrix and the orthogonal transform of the dynamical system in Eq. (\ref{govEq2}) can be directly obtained. But if the eigenbasis of the covariance matrix cannot be obtained by other means than the data obtained via solving the DNS or original DSS equations, the application of this strategy may be limited.

\section{Numerical experiments}
\label{exp}

In this section, we study the effectiveness of direct statistical simulation for describing the statistics of the Lorenz96 systems in the periodic and chaotic state and compare the low-order cumulants obtained in DSS with those obtained in DNS. We consider 2 cases. For the first translational symmetry is respected, whilst for the second we use the inhomogeneous force, $f_i$, to drive the Lorenz96 system in order to break the translational symmetry and also the effectiveness of the cumulant equations for approximating the dynamical system with and without the translational symmetry. We also compare the numerical performance of the model reduction strategies, {\it i}) and {\it ii}), introduced in \$\ref{diag}, for solving the DSS equations. For all test cases in this section, the spatial resolution of the Lorenz96 system is kept to be ${\xmax n}=8$ for comparison purposes.

\subsection{The Lorenz96 system in the periodic state}

\begin{figure}[htp]
\centering
\subfigure[$f_i=1.2$]
{
	\includegraphics[width=0.45\hsize]{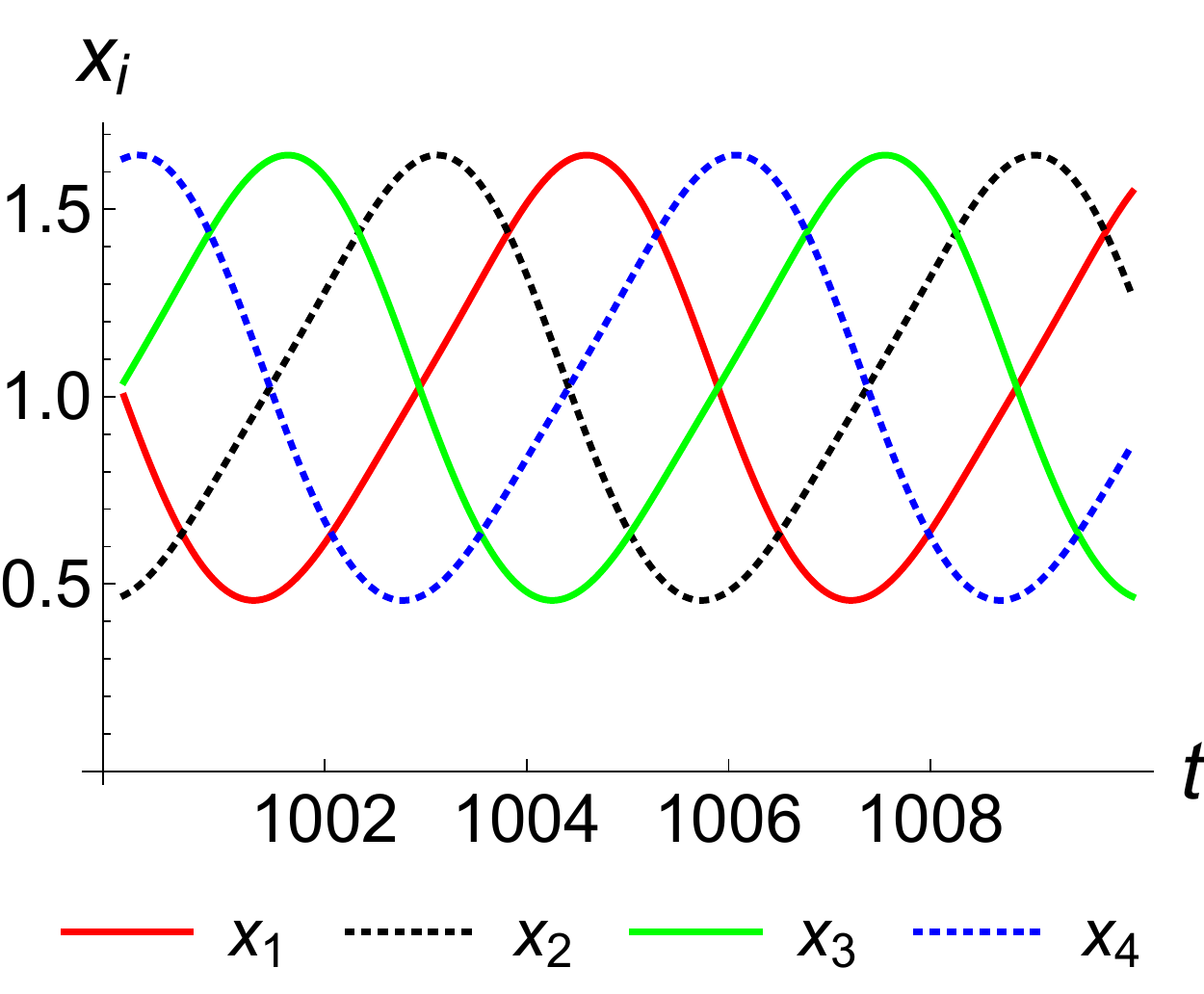}
}
\subfigure[$f_i=1.2$]
{
	\includegraphics[width=0.45\hsize]{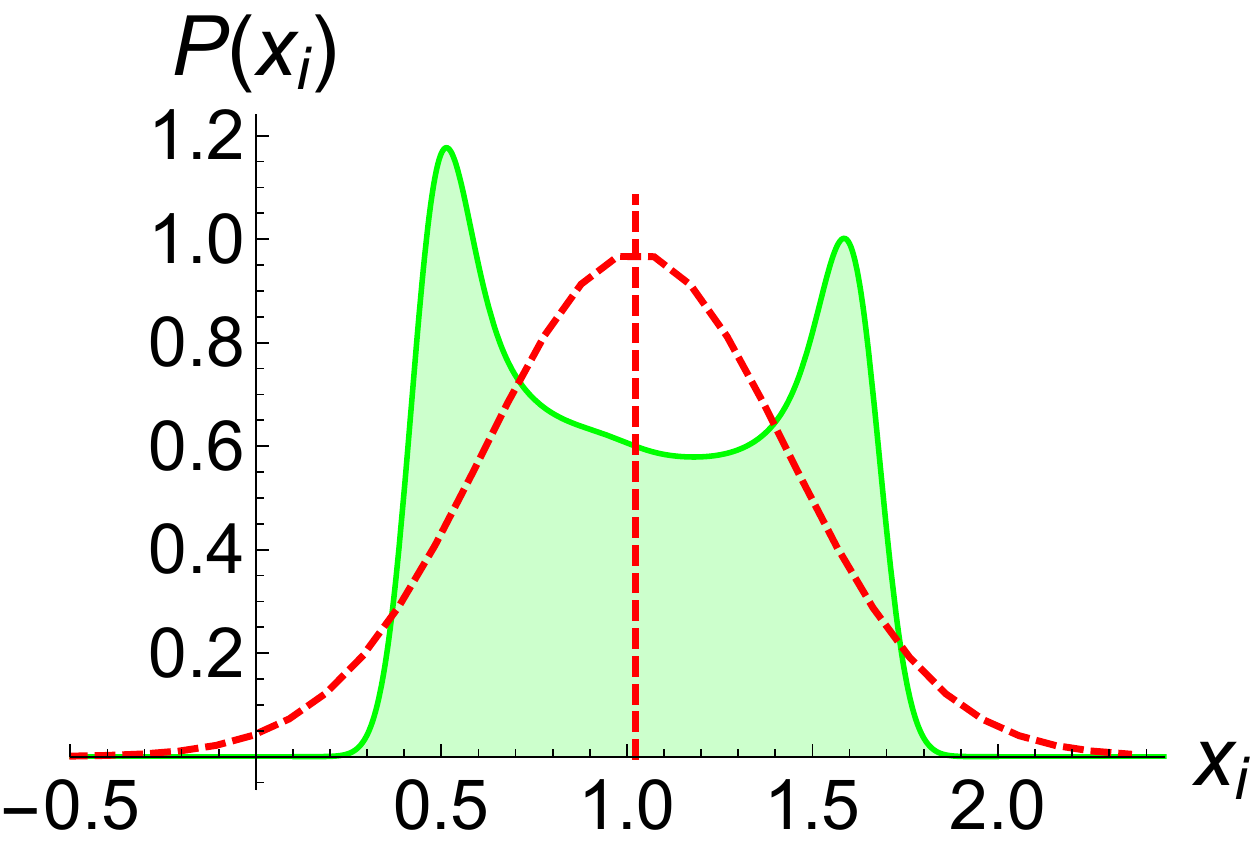}
}\\
\subfigure[$f_i=2$]
{
	\includegraphics[width=0.45\hsize]{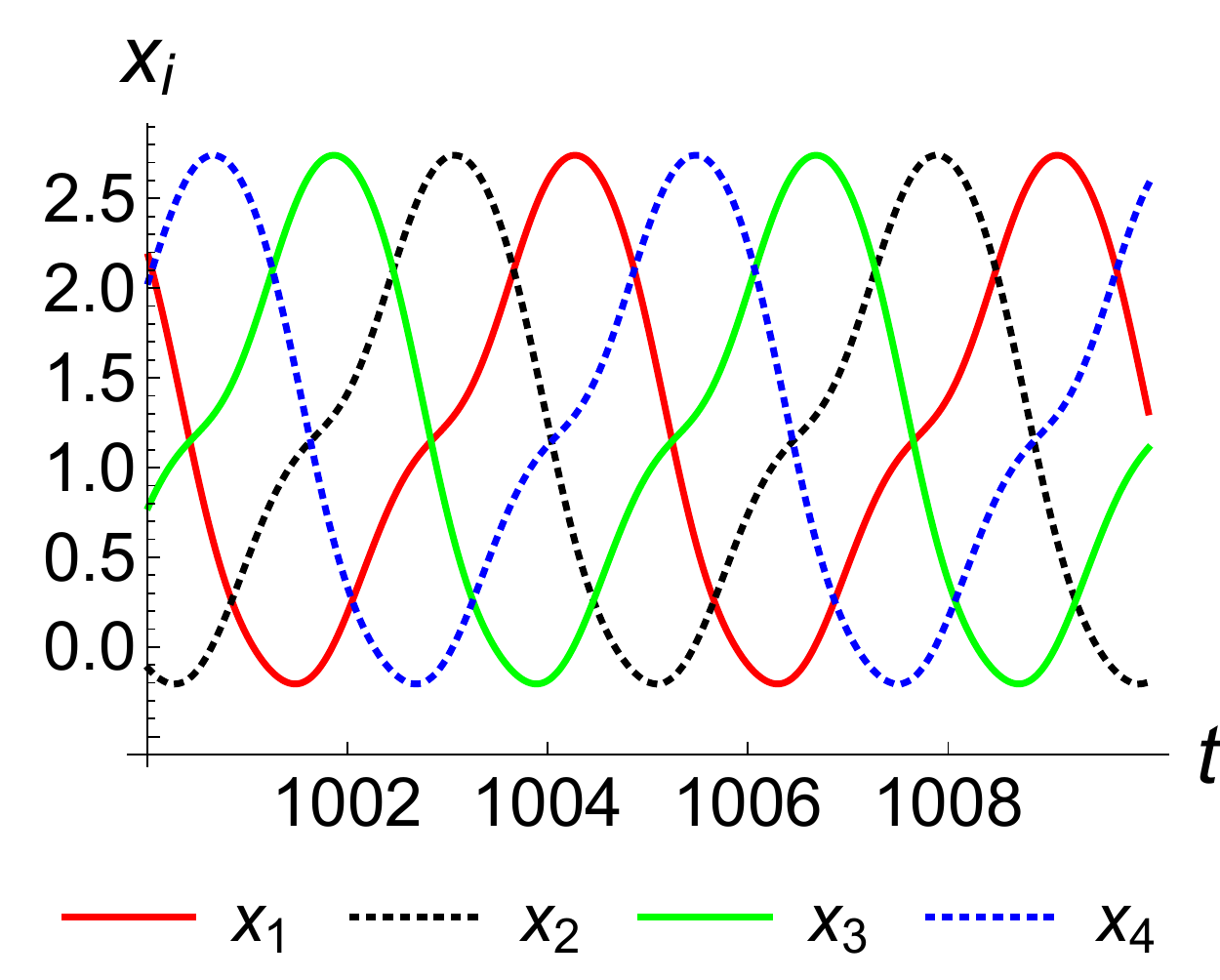}
}
\subfigure[$f_i=2$]
{
	\includegraphics[width=0.45\hsize]{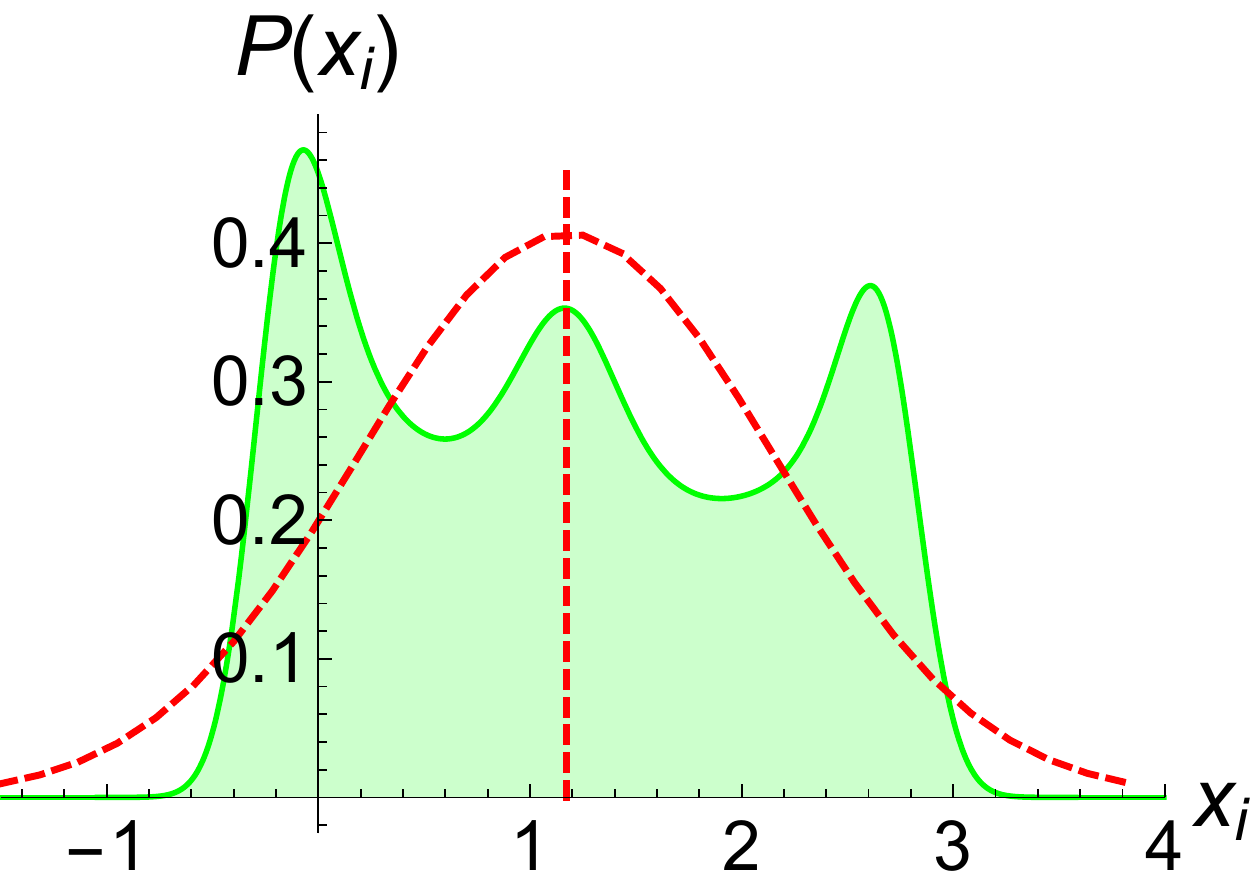}
}\\
\subfigure[$f_i=1.2$]
{
	\includegraphics[width=0.9\hsize]{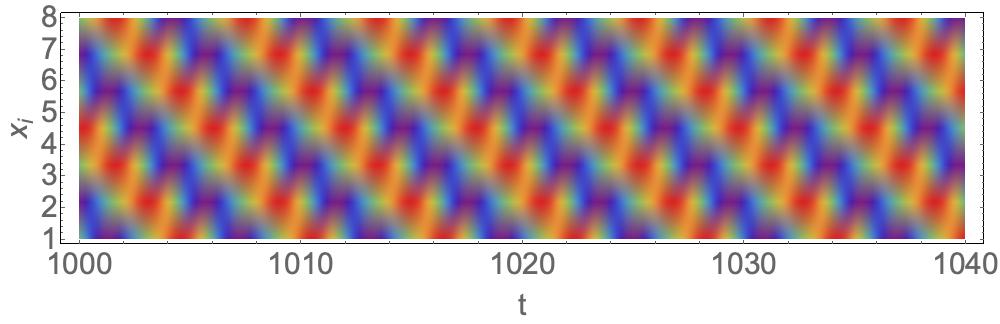}
}
\subfigure[$f_i=2$]
{
	\includegraphics[width=0.9\hsize]{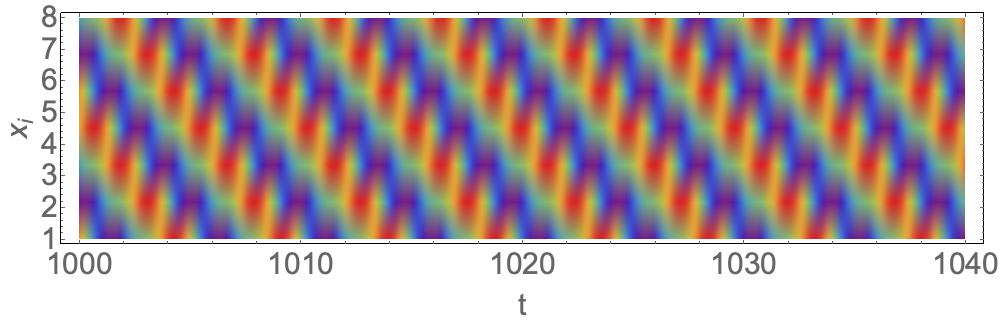}
}
\caption{The plots of the trajectory, PDFs and Hovm{\"o}ller diagram of the Lorenz96 in the periodic states, where the unknowns, $x_i$, oscillate with equal amplitude for $f_i=1.2$ and $2$ and split into two groups with the same frequency but in two different amplitudes for the even and odd indices of $i$.}
\label{plt_periodic1}
\end{figure}

We begin our numerical experiments by studying the statistical signature of the Lorenz96 system in the periodic state in DNS. If $f_i \geq 1$, the state vector, ${\bf x}$, of the Lorenz96 system becomes unstable and the wave solution is generated \cite{Kekem_2018}.  We forward integrate the dynamical equation~(\ref{govEq}) for a long time to obtain the statistical equilibrium, where the typical solutions for $f_i = 1.2$ and $f_i = 2$ are shown in Fig. (\ref{plt_periodic1}).
As $f_i$ is equally applied  over all nodes translational symmetry is preserved for all wave solutions of Lorenz96 system in this dynamical regime. As the external force, $f_i$, becomes large, we obtain the wave solutions with larger amplitude and more complicated spatio-temporal patterns.  The spatial complexity of the wave solution of the Lorenz96 system will be discussed in details in the next paragraph. In terms of statistics, all of the state variables, $x_i$, are identical to each other. In this dynamical regime, the PDFs, ${\cal P}(x_i)$, of the Lorenz96 system are distinctively different from the Gaussian distribution, i.e., ${\cal P}(x_i)$ comprises a few local maxima and is distributed in a finite domain. For the case of $f_i=1.2$, the PDF of $x_i$ has two peaks, which are located at the left and right ends of ${\cal P}(x_i)$; whilst for $f_i=2$, the third local maxima at $x=0$ is merged due to the contribution of the Fourier models for $m=0$ and $m=4$.

\begin{figure}
\centering
\subfigure[$C_{{\bf x}{\bf x}}$ for $f_i=1.2$]
{
	\includegraphics[width=0.45\hsize]{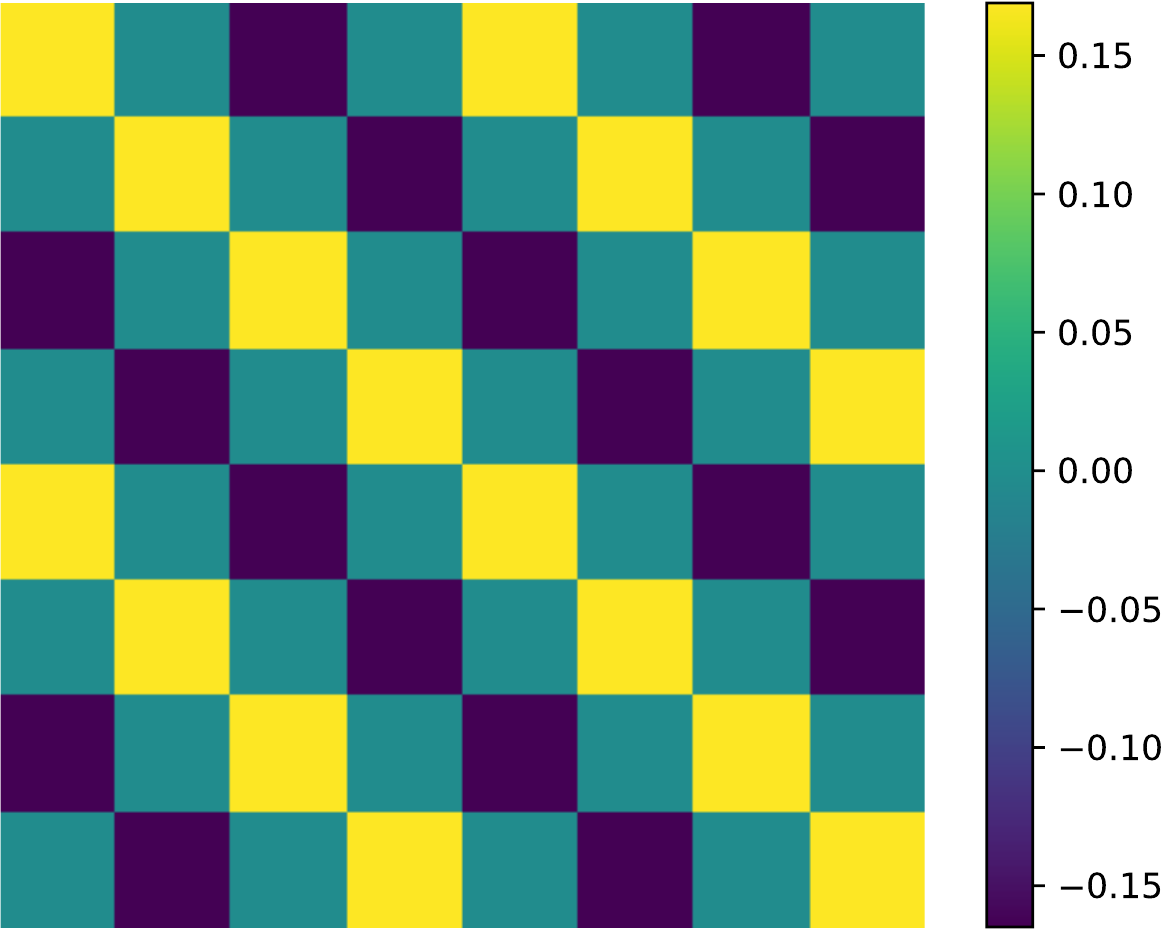}
}
\subfigure[$a_{m}(t)$ for $f_i=1.2$]
{
	\includegraphics[width=0.45\hsize]{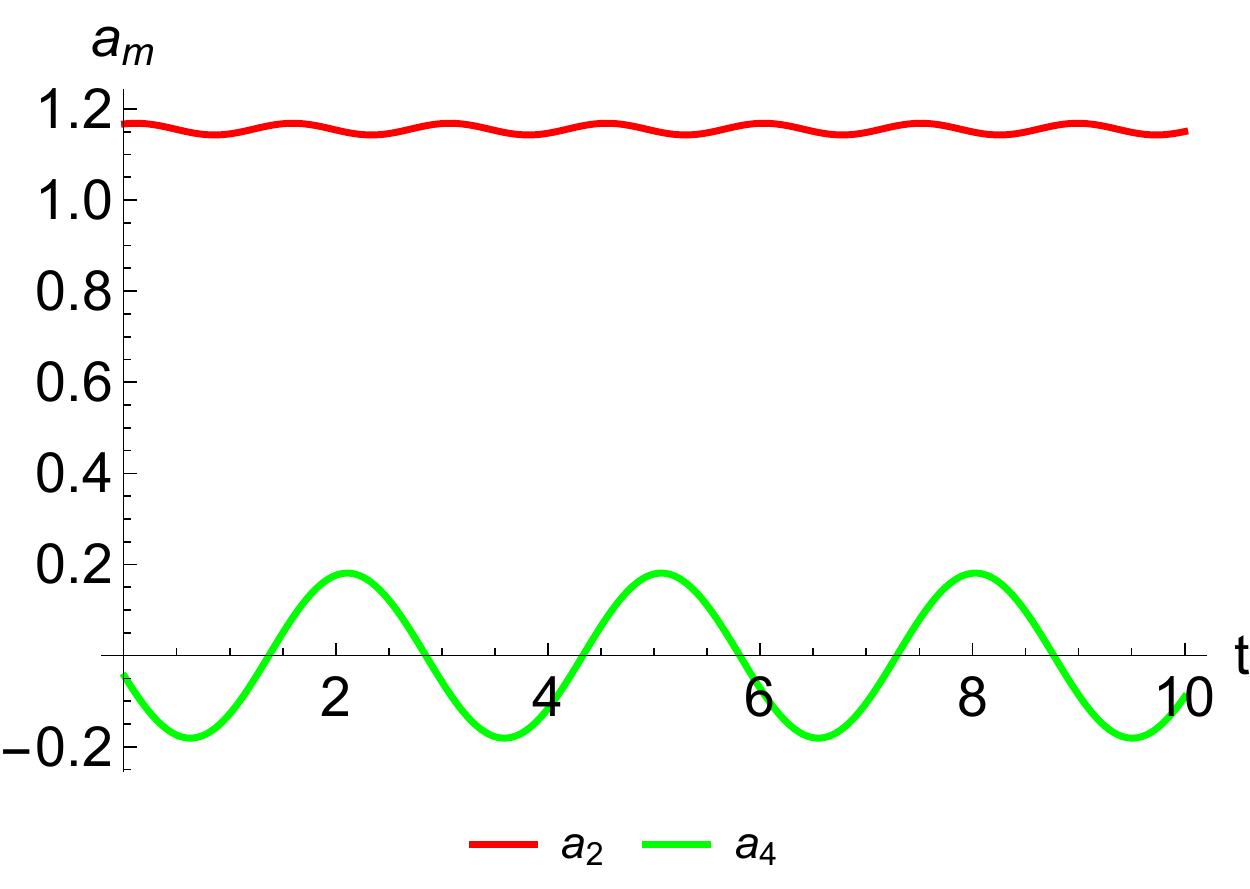}
}
\subfigure[$a_2 \sim a_4$ for $f_i=1.2$]
{
	\includegraphics[width=0.45\hsize]{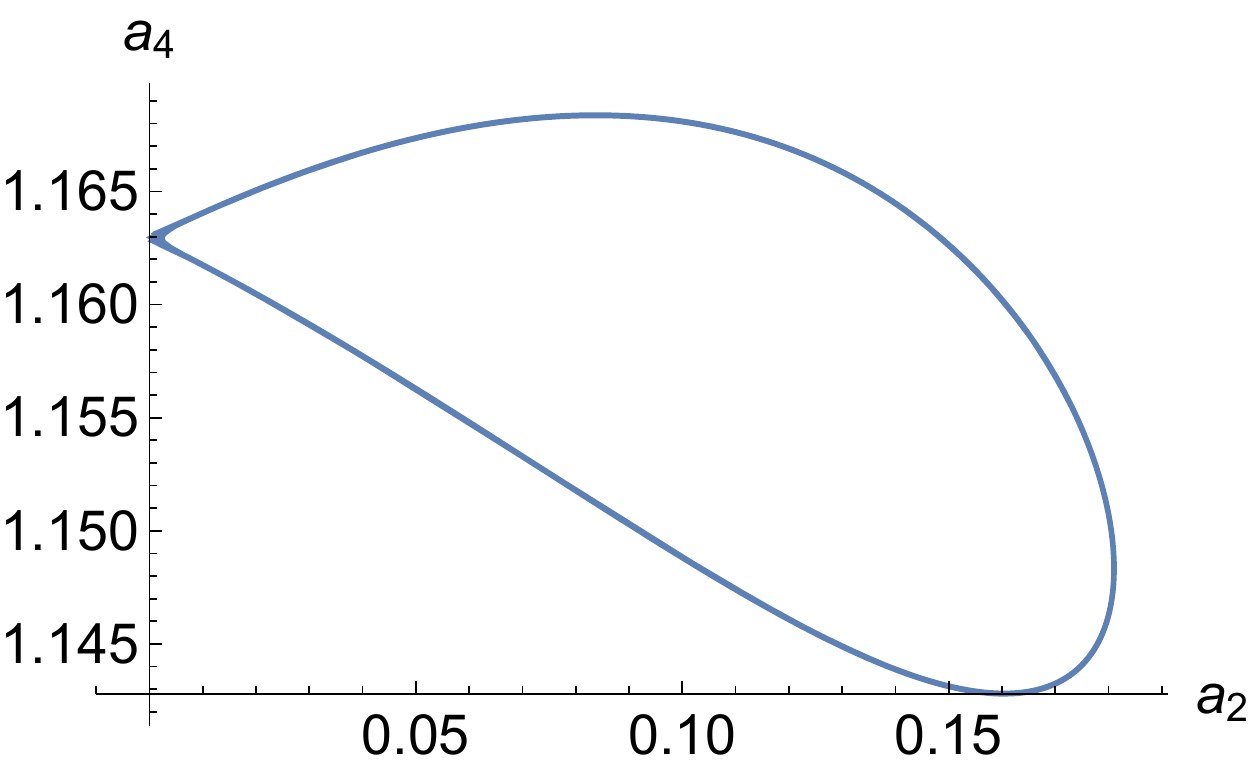}
}
\subfigure[$C_{{\bf x}{\bf x}}$ for $f_i=2$]
{
	\includegraphics[width=0.45\hsize]{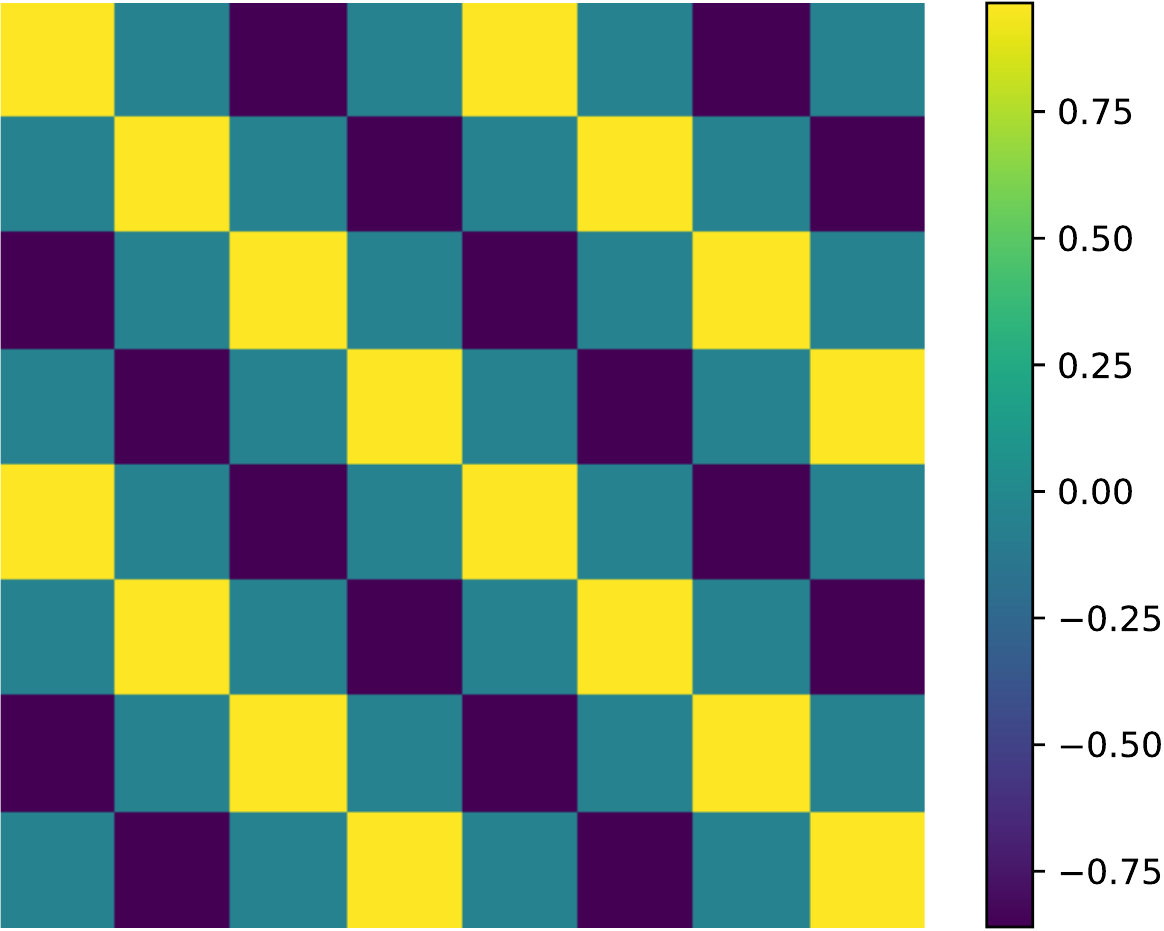}
}
\subfigure[$a_{m}(t)$ for $f_i=2$]
{
	\includegraphics[width=0.45\hsize]{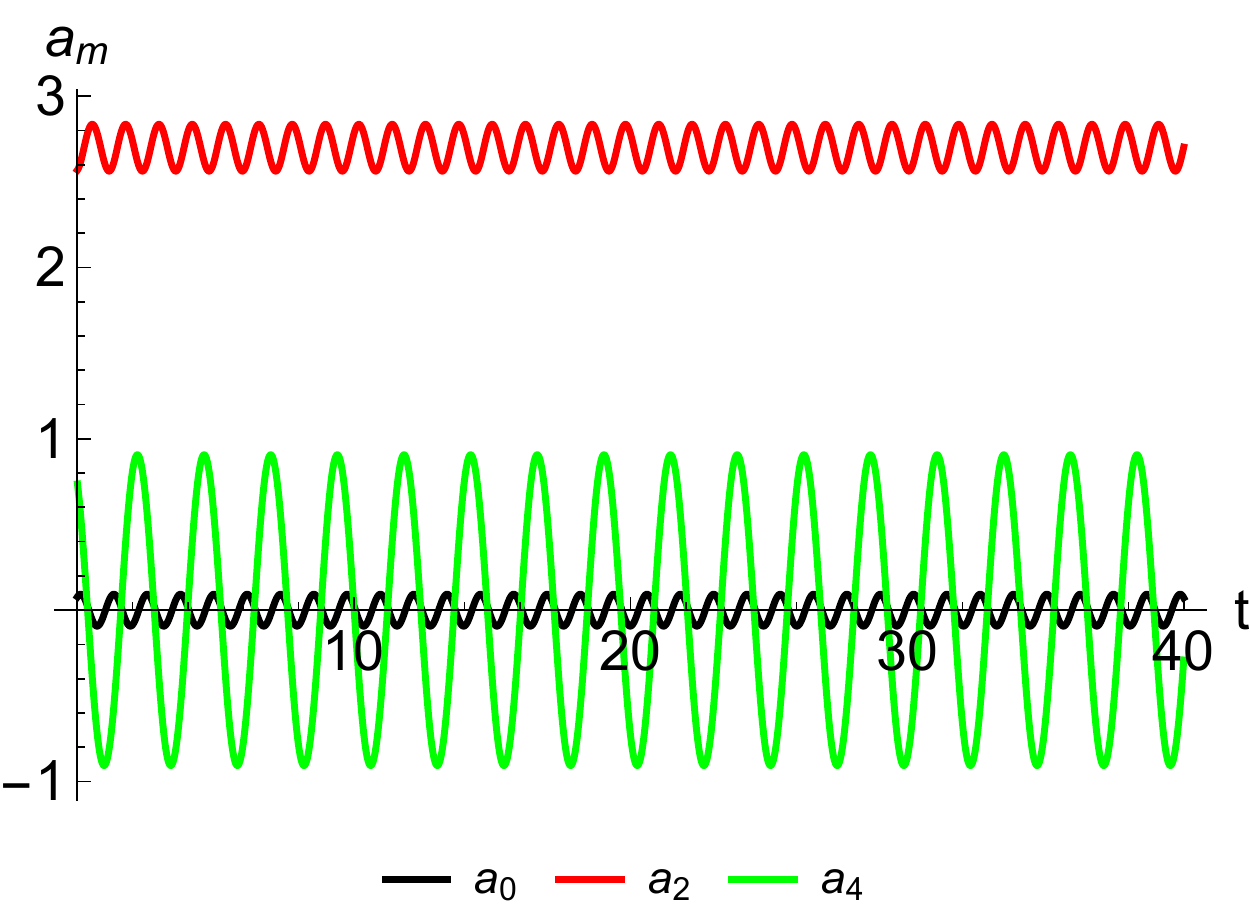}
}
\subfigure[$a_0 \sim a_2 \sim a_4$ for $f_i=2$]
{
	\includegraphics[width=0.45\hsize]{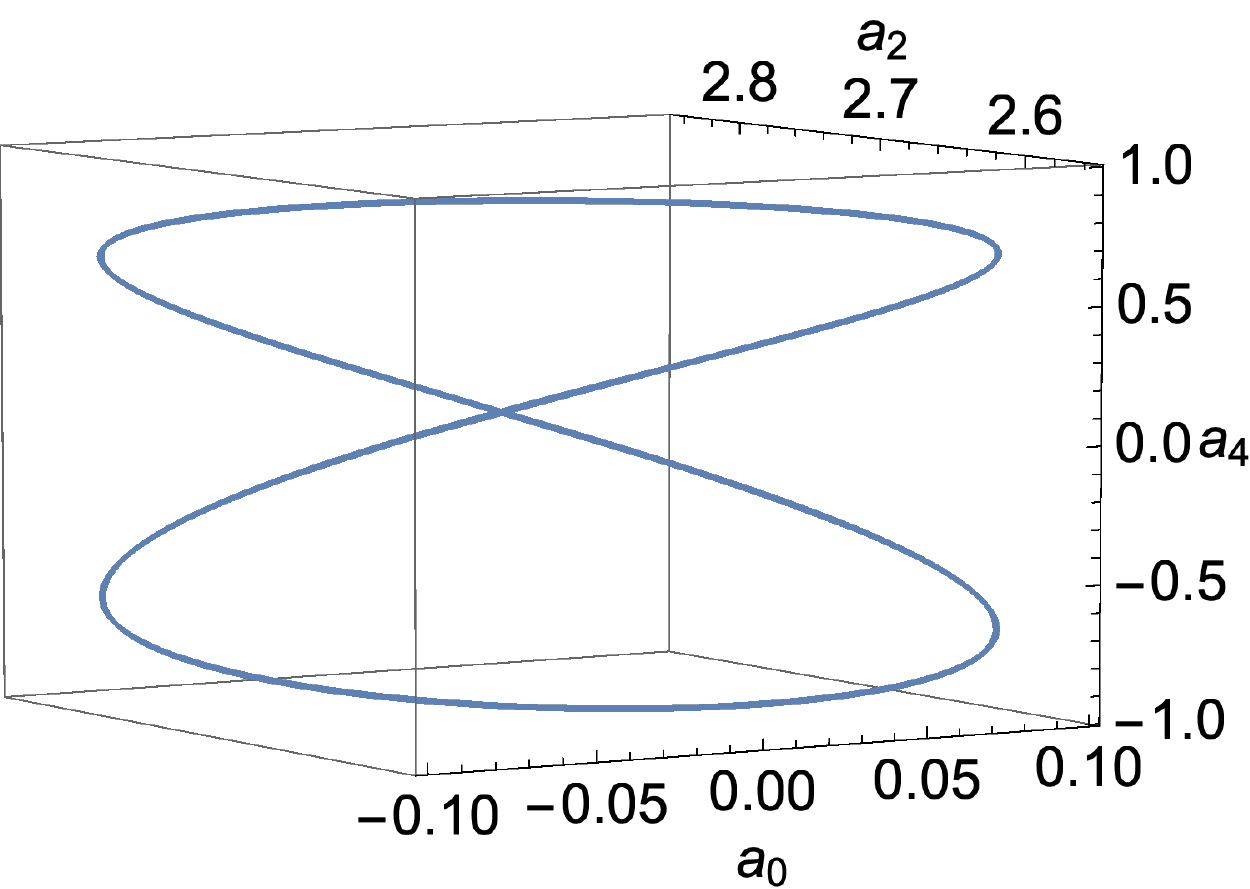}
}
\caption{The plots of the covariance matrix in (a) \& (d) and the nonzero spectral coefficients, $a_m(t)$, as the function of time in (b) \& (e) and in the trajectories of $a_m$ in the phases space in (c) \& (f) for the Lorenz96 system in the periodic state for $f_i=1.2$ and $2$.}
\label{cov_periodic}
\end{figure}

We apply the eigen-decomposition of the covariance matrix, $C_{\bf xx}$, obtained by ensemble averaging the solutions of the dynamical equation in order to analysis the spatial complexity of the periodic oscillations of $x_i$ for different external forcing parameter, $f_i$. As observed in the numerical solution in DNS for the case of $f_i=1.02$, the oscillation pattern of $x_i$ is comprised of a single wave number, which corresponds to the Fourier modes of $m=2$. The eigen-pairs of the covariance matrix is doubly-folded, i.e., the covariance matrix has one nonzero eigenvalue and two distinctive eigenvectors parallel to the sine and cosine components of the Fourier basis of $m=2$. We use the adaptive basis function, ${\cal D}_m({\bf x})$, defined in (\ref{dynbs}) for representing the state vector, ${\bf x}$, in the spectral space, such that the spectral coefficients, $a_m(t)$, is unique in all time, where $m$ is the wave number of the Fourier basis. We obtain $5$ nonzero eigen-pairs for $m=0,2$ and $4$ modes for $f_i=2$. The eigen-pairs for $m=2$ and $4$ are doubly folded and the Fourier modes of $m=0$ and $m=4$ are excited due to the self-interaction of $m=2$ models and oscillate twice as fast as the $m=2$ models. See Fig. (\ref{cov_periodic})
for the test cases of $f_i=1.2$ with $4$ eigen-pairs and $2$ with $5$ eigen-pairs for comparison. For all test cases, the covariance matrices are dominated by the diagonal elements.

We solve the original and the reduced DSS equation in CE2/2.5/3 approximations using strategies, {\it i}) and {\it ii}) and compare the low-order cumulants in DSS with those obtained in DNS. We find that the low-order statistics of the Lorenz96 system in the periodic state can be very accurately approximated by the CE2/2.5 and CE3 approximations. If the third order cumulants are considered in CE2.5/3 approximations, we must introduce the eddy damping parameter, $\tau_d$, to stabilize the numerical integrations, where $\tau_d$ approximates the correlation time of the oscillation of $x_i$ with the typical value within the range from $10^{-2}$ to $10^{-1}$. In our experiments, we find that the optimal $\tau_d$ is in the order of $10^{-1}$, which is approximately $10$ to $100$ times smaller than the periodic oscillation of the state vector, ${\bf x}$. The solution of the DSS equations are listed and compared in Table (\ref{tb1}),
\begin{table*}
\begin{ruledtabular}
\begin{tabular}{l|r|r|r|r|r|r|r|r}
       & $f_i$ & $\tau_d^{-1}$& $C_{x_{i}}$ & $\lambda_{m=0}$ &  $\lambda_{m=1}$    & $\lambda_{m=2}$  & $\lambda_{m=3}$   & $\lambda_{m=4}$      \\ \hline
DNS              & $1.02$&            &  $1.00$ & $0$& $0$  & $6.6\times 10^{-2}$ & $0$   & $1.7\times 10^{-4}$ \\ \hline
CE2.5/CE2.5t &           & $10$    &  $1.00$ & $1.1\times 10^{-4}$& $0$  & $6.5\times 10^{-2}$  & $0$   & $2.0\times 10^{-4}$ \\ \hline
CE2.5r           &           & $10$    &  $1.00$ &        --             &  --  & $6.6\times 10^{-2}$   & --    &    --       \\ \hline
CE3/CE3t       &           & $10$    &  $1.00$ & $1.0\times 10^{-4}$& $0$  & $6.5\times 10^{-2}$  & $0$   & $2.0\times 10^{-4}$ \\ \hline
CE3r              &           & $10$    &  $1.00$ &        --             & --   & $6.6\times 10^{-2}$  & --    &     --       \\ \hline\hline
DNS              & $1.2$  &            &  $1.04$ & $1.0\times 10^{-5}$& $0$  & $6.7\times 10^{-1}$ & $0$   & $1.6\times 10^{-2}$ \\ \hline
CE2.5/CE2.5t &  & $8$   &  $1.04$ & $1.5\times 10^{-2}$& $0$  & $6.7\times 10^{-1}$ & $0$   & $1.0\times 10^{-2}$ \\ \hline
CE2.5r          &  & $8$   &  $1.04$ & $1.5\times 10^{-2}$& --   & $6.7\times 10^{-1}$ & --   & $1.0\times 10^{-2}$ \\ \hline
CE3/CE3t      &  & $15$ &  $1.03$ & $1.5\times 10^{-2}$& $0$  & $7.1\times 10^{-1}$ & $0$   & $0.8\times 10^{-2}$ \\ \hline
CE3r             &  & $15$ &  $1.03$ & $1.5\times 10^{-2}$& --   & $7.1\times 10^{-1}$ & --    & $0.8\times 10^{-2}$ \\ \hline\hline
DNS              & $2$    &            &  $1.19$ & $4.3\times 10^{-3}$& $0$  & $3.7$ & $0$   & $4.1\times 10^{-1}$ \\ \hline
CE2.5/CE2.5t           &  & $10$ &  $1.15$ & $3.6\times 10^{-1}$& $0$  & $3.6$ & $0$   & $2.3\times 10^{-1}$ \\ \hline
CE2.5r         &  & $10$ &  $1.15$ & $3.6\times 10^{-1}$& --   & $3.6$ & --    & $2.3\times 10^{-1}$ \\ \hline
CE3/CE3t      &  & $10$ &  $1.19$ & $4.9\times 10^{-1}$& $0$  & $3.5$ & $0$   & $2.4\times 10^{-1}$ \\ \hline
CE3r             &  & $10$ &  $1.19$ & $4.9\times 10^{-1}$& --   & $3.5$ & --    & $2.4\times 10^{-1}$ \\ \hline
\end{tabular}
\end{ruledtabular}
\caption{The low-order cumulants obtained in DNS and CE2.5/CE3 for the Lorenz96 system in the periodic state for $f_i=1.02, 1.2$ and $2$ with the spatial resolution, ${\xmax n}=8$, where the covariance matrix are represented by the eigenvalue, $\lambda_i$ and Fourier basis functions ${\cal F}_k$. The notations, ``CE2.5r'' and ``CE3r'', stand for the reduced cumulant system with the wave numbers excluded from the covariance matrix, which are noted as ``--'', the test cases of the transformed DSS models are noted as ``CE2.5t'' and ``CE3t'', respectively.}
\label{tb1}
\end{table*}
The notation, ``CE2.5r'' and ``CE3r'', stand for the cumulant equations which are solved by using the strategy, {\it i}), where at each time step, we apply the eigen-decomposition of the covariance matrix and retain the leading order eigen-pairs and ``CE2.5t'' and ``CE3t'', are used to represent the diagonalised cumulant system using strategy {\it ii}). By using the strategies {\it i}) and {\it ii}), the low-order statistics are accurately computed as compared with those obtained in DNS. In this dynamical regime, the strategies {\it i}) and {\it ii}) can be combined, e.g., for $f_i=2$, we can solve the diagonalised cumulant equations for $m=2$ components only and obtain the accurate solutions.

\subsection{The Lorenz96 system in the doubly-periodic state}

\begin{figure}[htp]
\centering
\subfigure[$f_i=3.5$ for $\alpha$ group]
{
	\includegraphics[width=0.45\hsize]{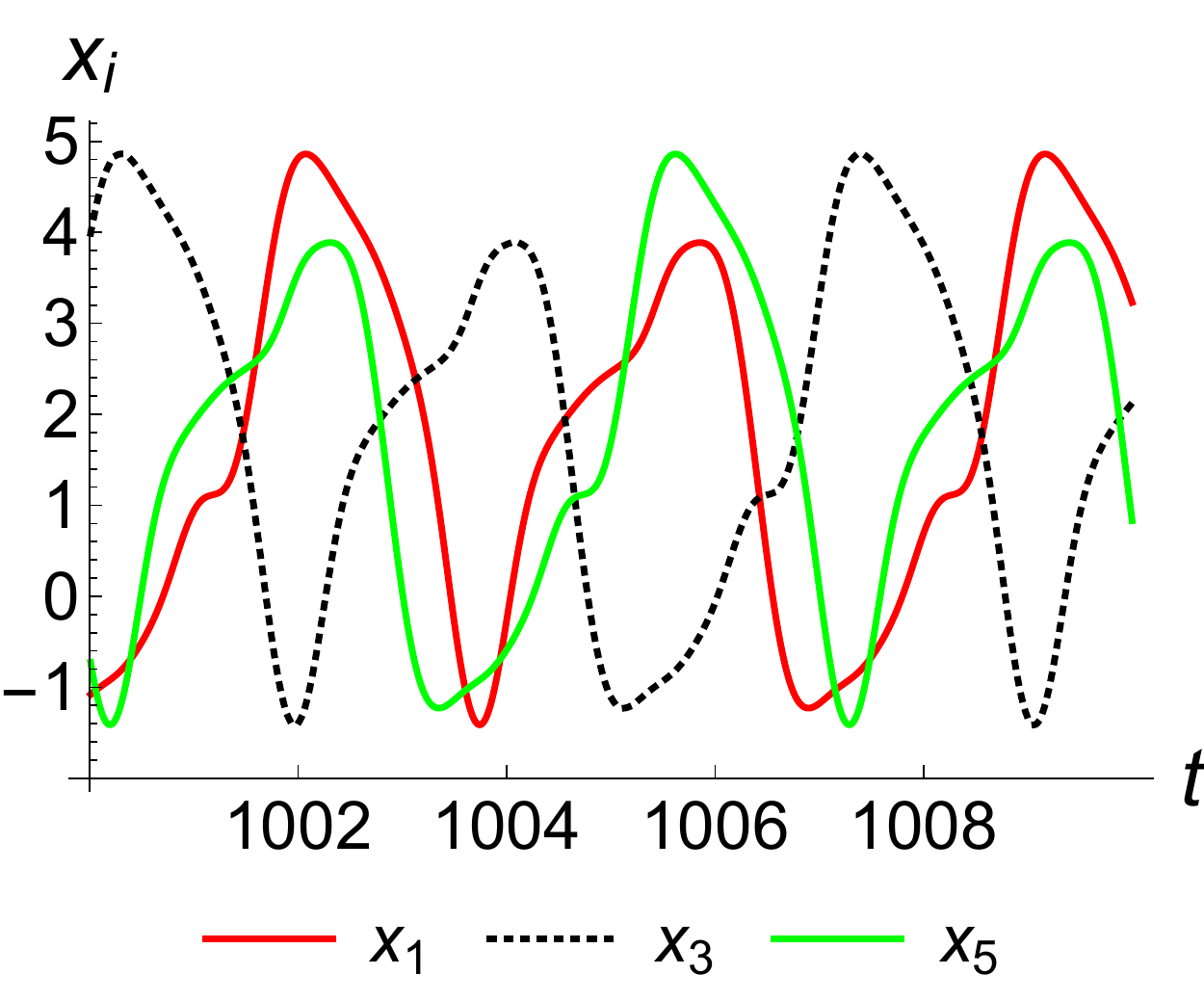}
}
\subfigure[$f_i=3.5$ for $\beta$ group]
{
	\includegraphics[width=0.45\hsize]{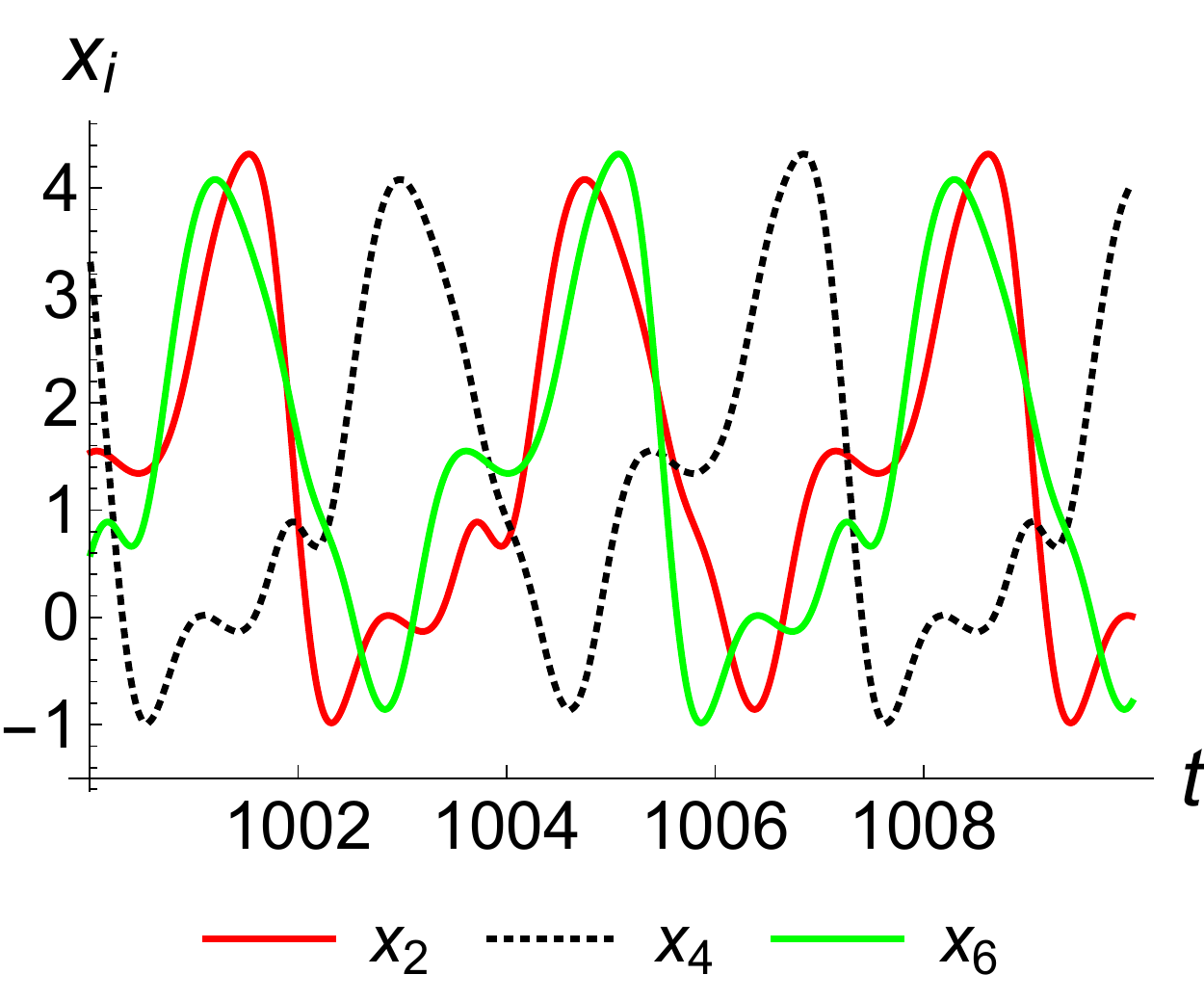}
}
\subfigure[$f_i\sim {\cal N}(3.5,0,01)$]
{
	\includegraphics[width=0.45\hsize]{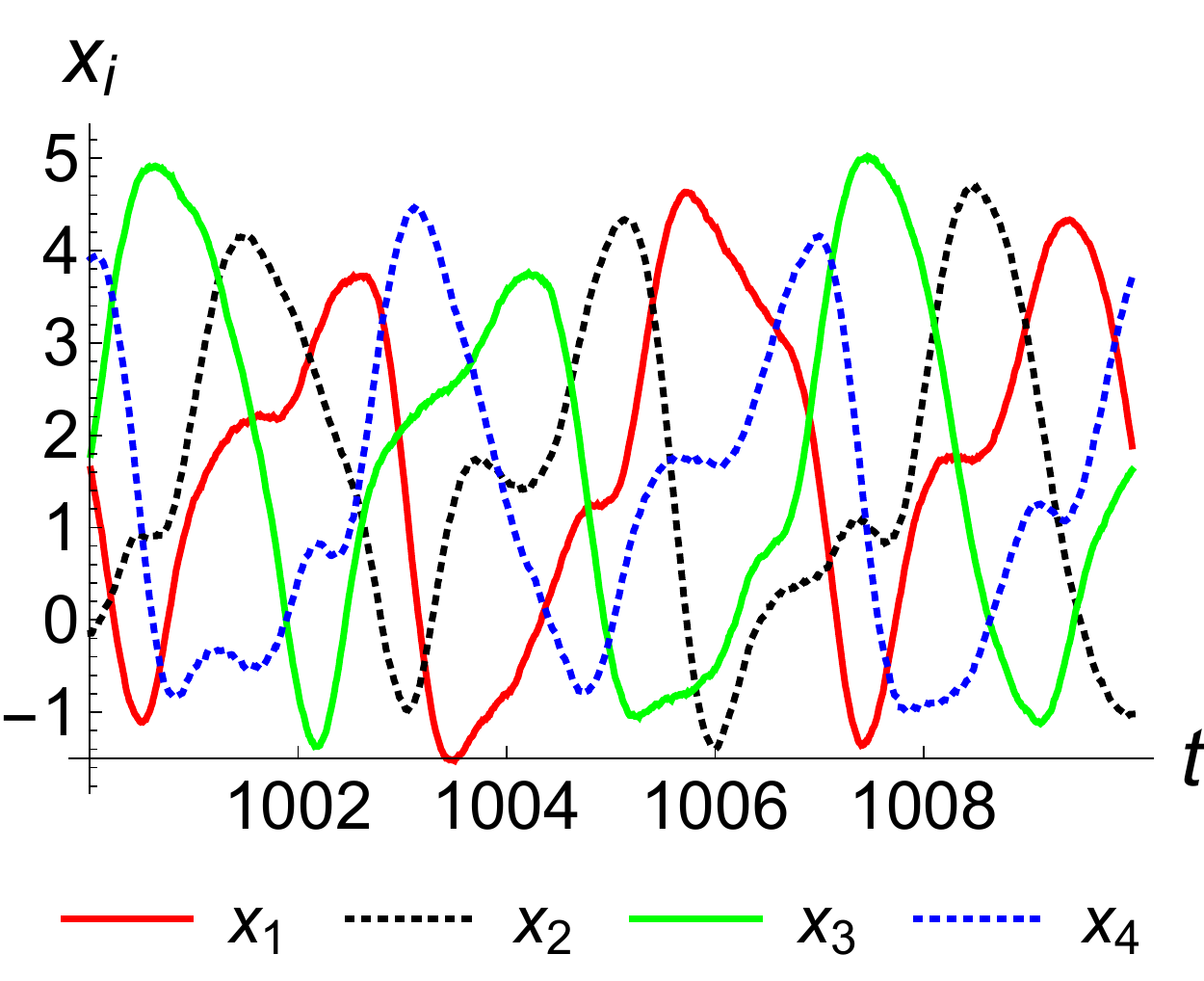}
}
\subfigure[$f_i=3.5$ for $\alpha$ group]
{
	\includegraphics[width=0.45\hsize]{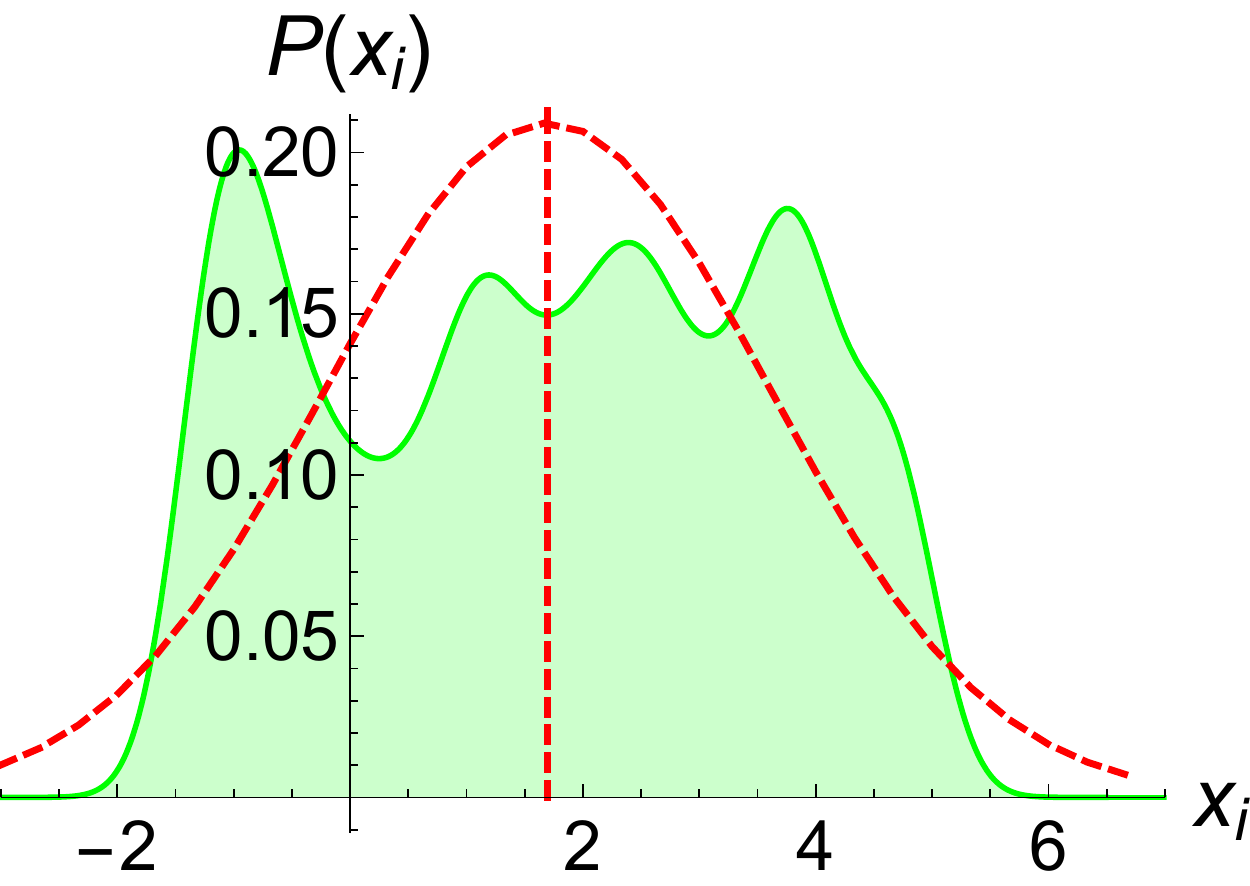}
}
\subfigure[$f_i=3.5$ for $\beta$ group]
{
	\includegraphics[width=0.45\hsize]{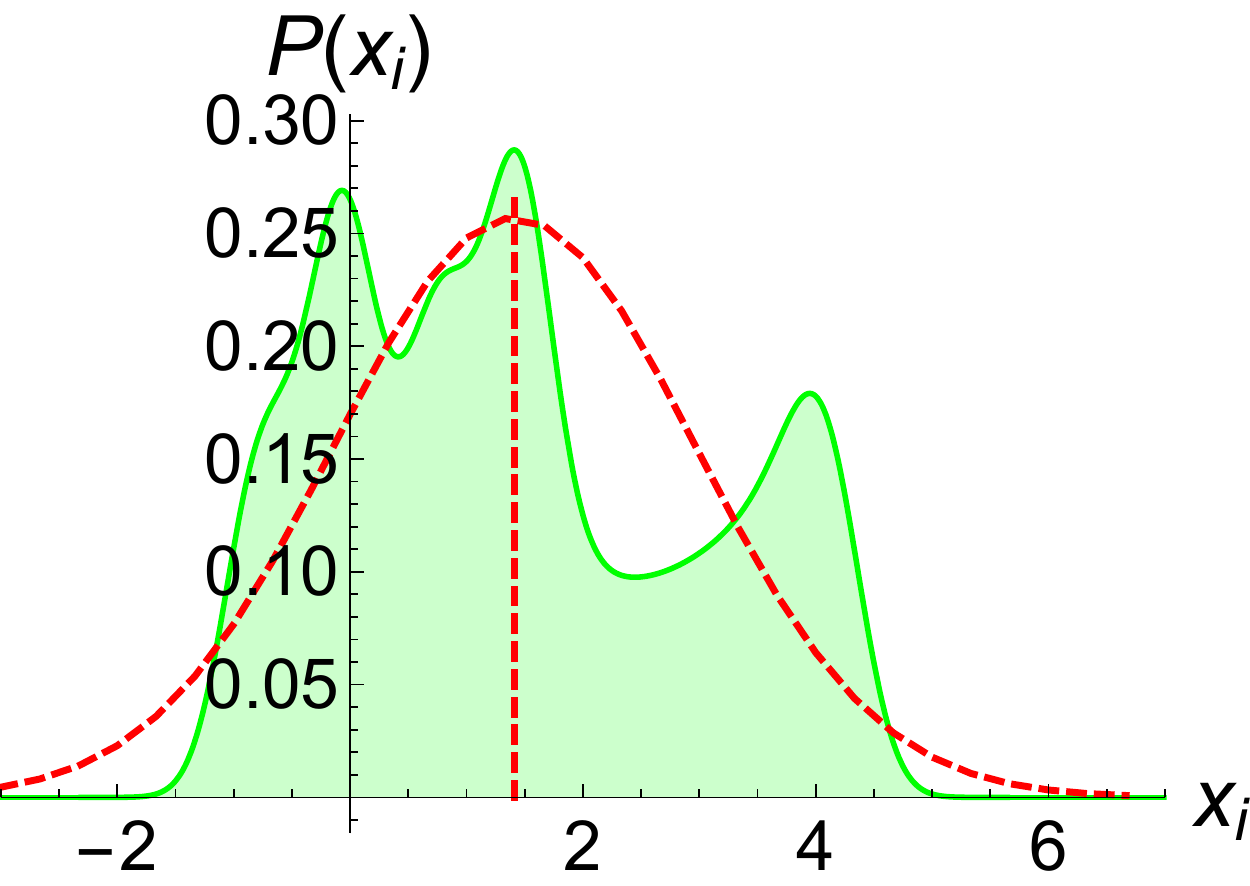}
}
\subfigure[$f_i\sim {\cal N}(3.5,\ 0.01)$]
{
	\includegraphics[width=0.45\hsize]{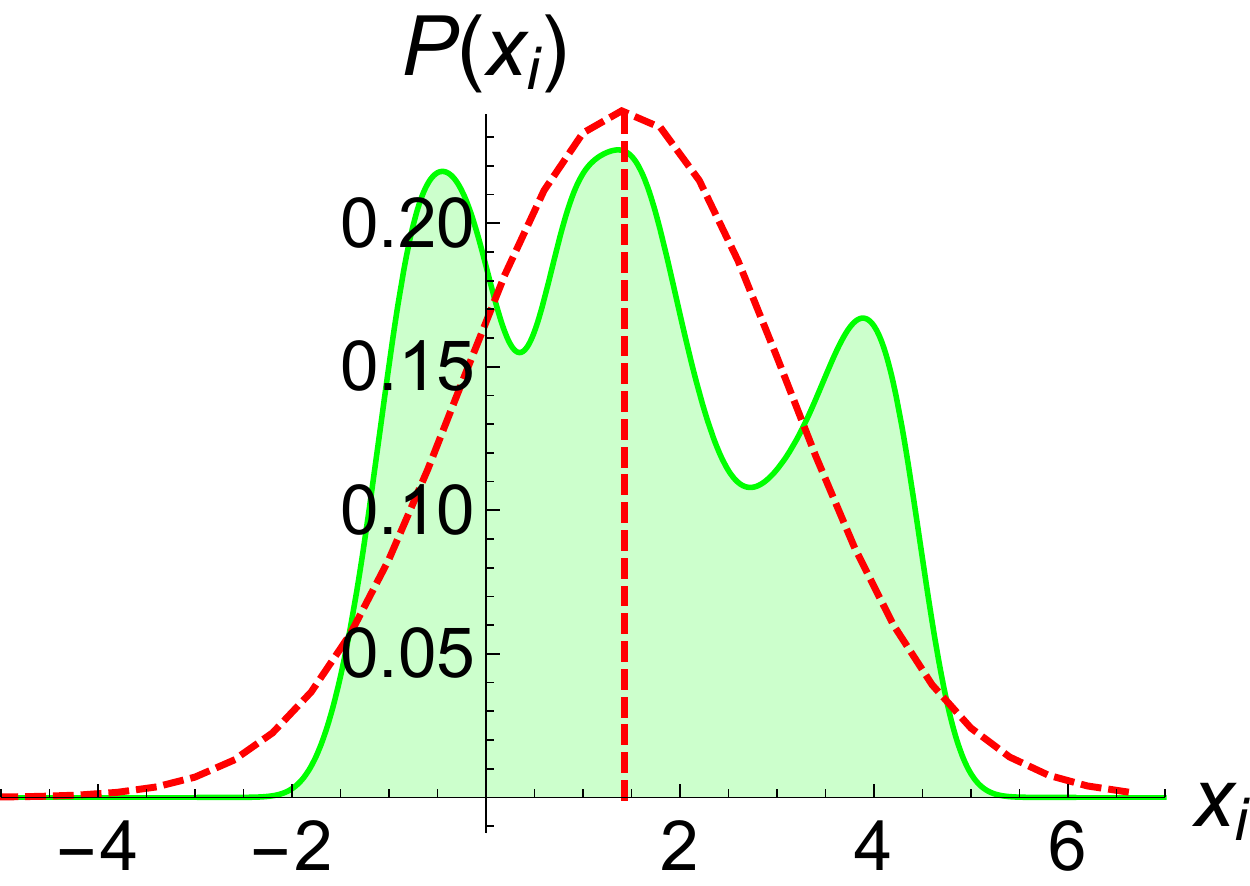}
} 
\subfigure[$f_i=3.5$]
{
	\includegraphics[width=0.9\hsize]{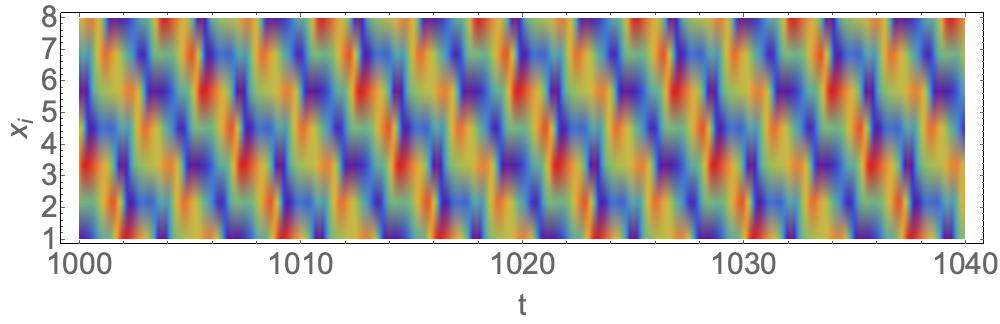}
}
\subfigure[$f_i\sim {\cal N}(3.5, 0.01)$]
{
	\includegraphics[width=0.9\hsize]{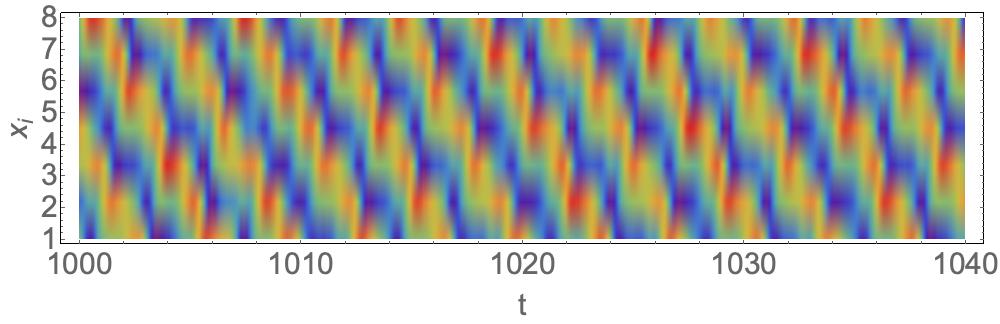}
}
\caption{The plots of the trajectory, PDFs and Hovm{\"o}ller diagram of the Lorenz96 for $f_i=3.5$ and $f_i \sim {\cal N}(3.5, \ 0.01)$.}
\label{plt_periodic2}
\end{figure}

The Lorenz96 system becomes doubly periodic when the external force, $f_i$, passes a  critical value before the solution transitions to spatio-temporal chaos. Here the periodic oscillation of the state vector, ${\bf x}$, is divided into two groups, namely $\alpha$ and $\beta$ group. Each group comprises of the state variable, $x_i$, with purely even or odd indices, $i$. The state variable, $x_i$, in both groups oscillate with the same frequency but in different amplitude and pattern. This special oscillation state is only observed in the Lorenz96 system with the spatial resolution, ${\xmax n}=5, 8, 10, \cdots$, e.g., see \cite{Kekem_2018}. The dynamical system is driven by the equal force, $f_i$, for all nodes. In DNS, the translational symmetry of the wave solution of the dynamical system is preserved and the state variable, $x_i$, can settle in the $\alpha$ or $\beta$ group with equal probability, which depends on the choice of the initial condition. The spatial configuration of $x_{\alpha}$ shown in Fig. (\ref{plt_periodic2}a \& d) is less complex than for the ${\beta}$ group in Fig. (\ref{plt_periodic2}b \& e). The PDFs of $x_i$ have  finite support and are distinctively different from Gaussian. For the $\alpha$ group, ${\cal P}(x_{\alpha})$ has four local maxima whilst for the ${\beta}$ group, ${\cal P}_{{\beta}}$ has only three. Very interestingly, we observe that the bistability of the Lorenz96 system is very sensitive to the stochastic force, even for a weak noise term. By adding a weak noise to the external force, $f_i$, the periodic oscillation of $x_i$ in the $\alpha$ and $\beta$ group immediately converges to a single state, e.g., see Fig. (\ref{plt_periodic2} c, \& f)
for $f_i \sim {\cal N}(3.5, 0.01)$ for comparison.

\begin{figure}[htp]
\centering
\subfigure[$f_i=3.5$ ]
{
	\includegraphics[width=0.45\hsize]{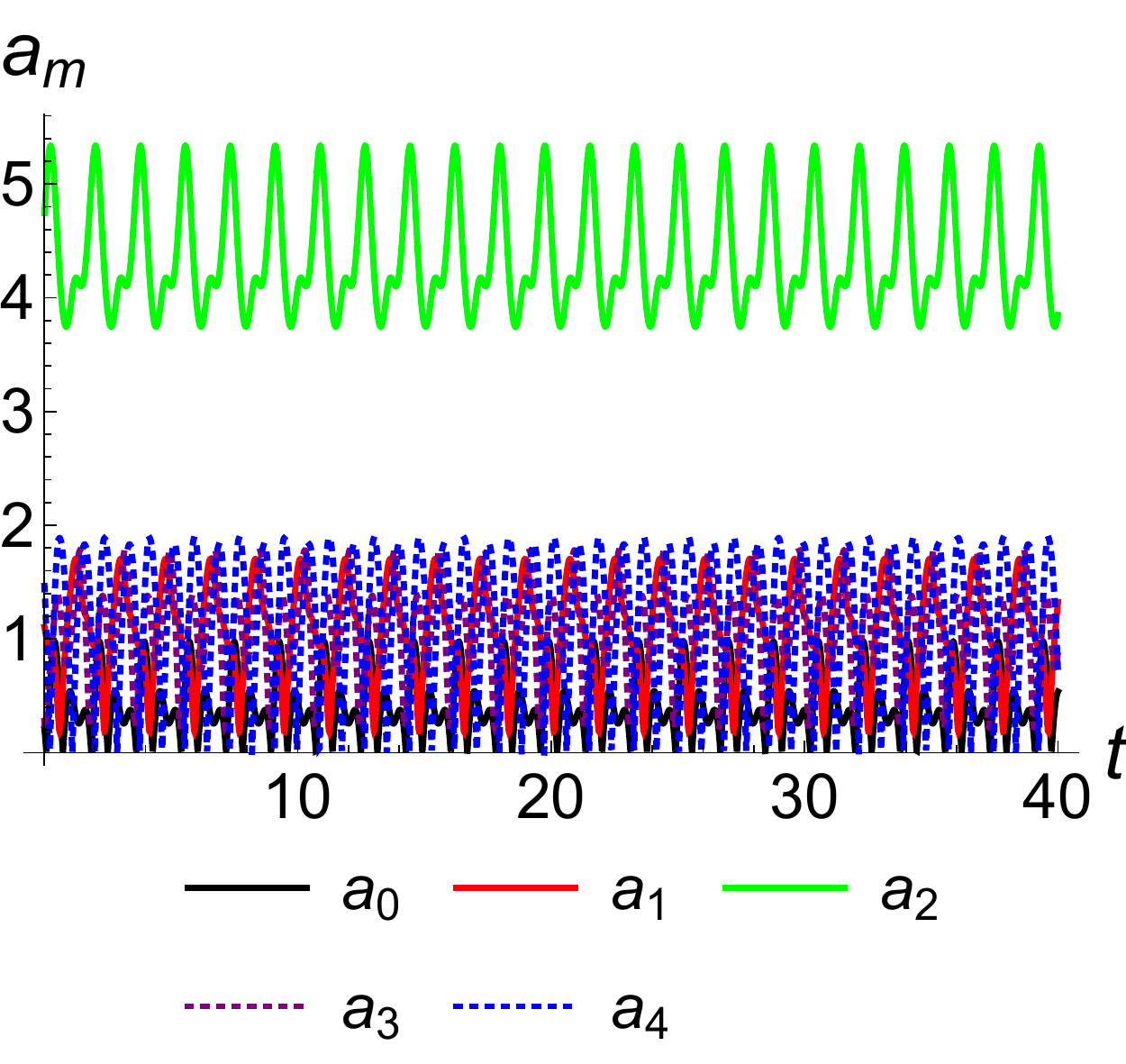}
}
\subfigure[$f_i\sim {\cal N}(3.5,0.01)$]
{
	\includegraphics[width=0.45\hsize]{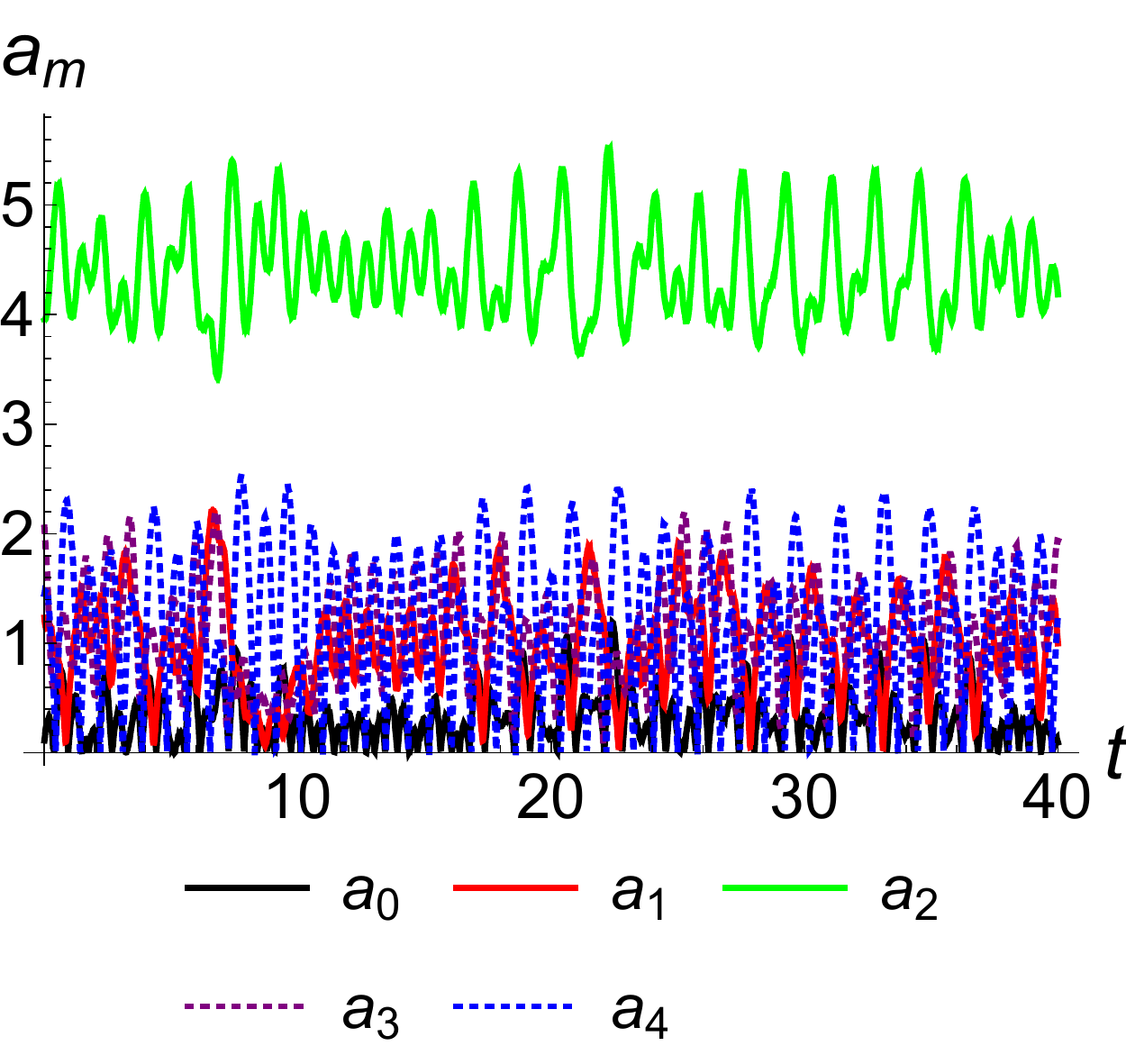}
}
\subfigure[$f_i\sim {\cal N}(3.5,\ 1)$]
{
	\includegraphics[width=0.45\hsize]{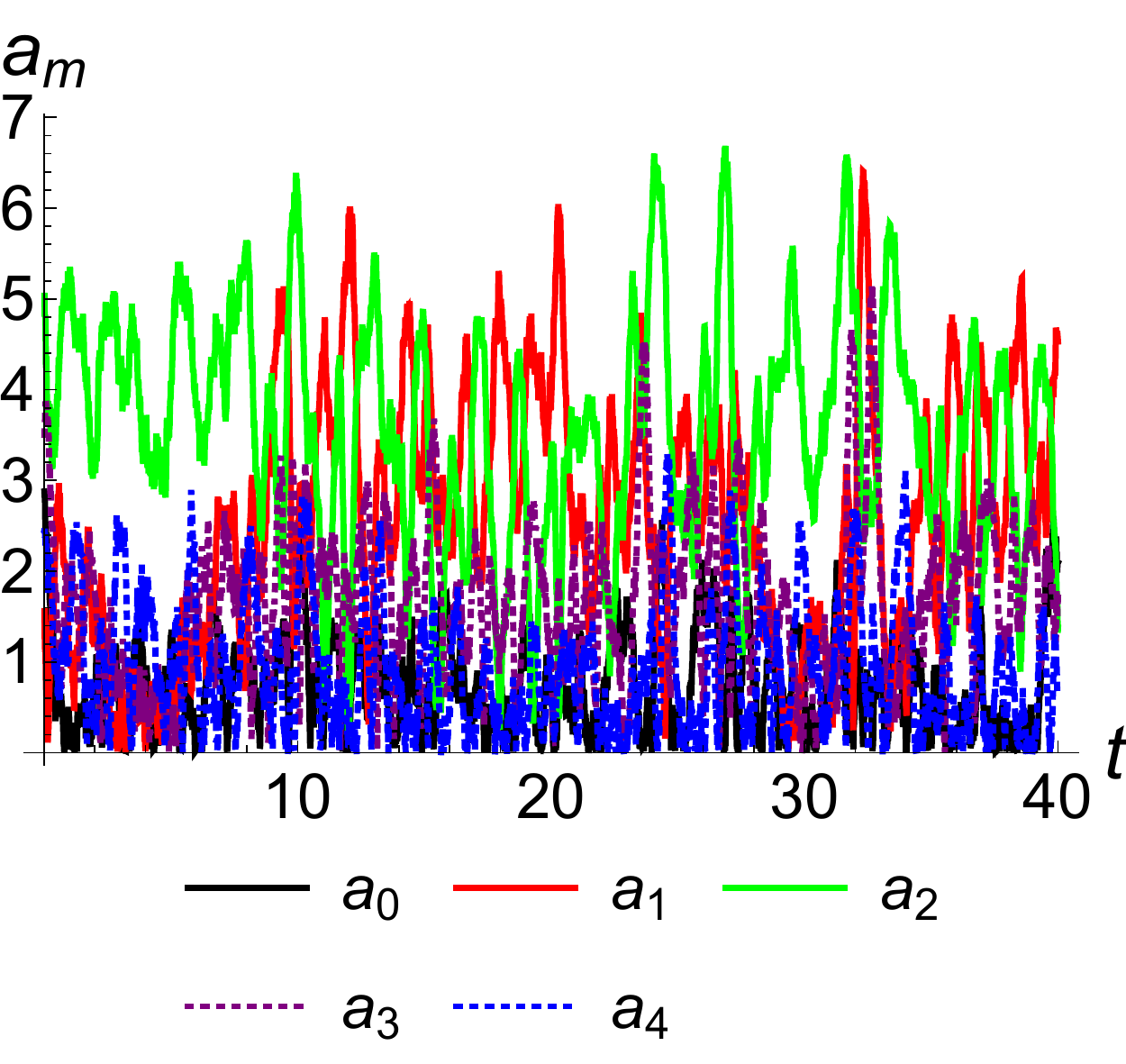}
}
\caption{The spectral coefficients, $a_m$, as the function of time, $t$, for the Lorenz96 for $f_i=3.5$, $f_i \sim {\cal N}(3.5, \ 0.01)$ and ${\cal N}(3.5, \ 1)$.}
\label{plt_periodic3}
\end{figure}

\begin{figure}
\centering
\subfigure[$f_i=3.5$ in DNS]
{
	\includegraphics[width=0.45\hsize]{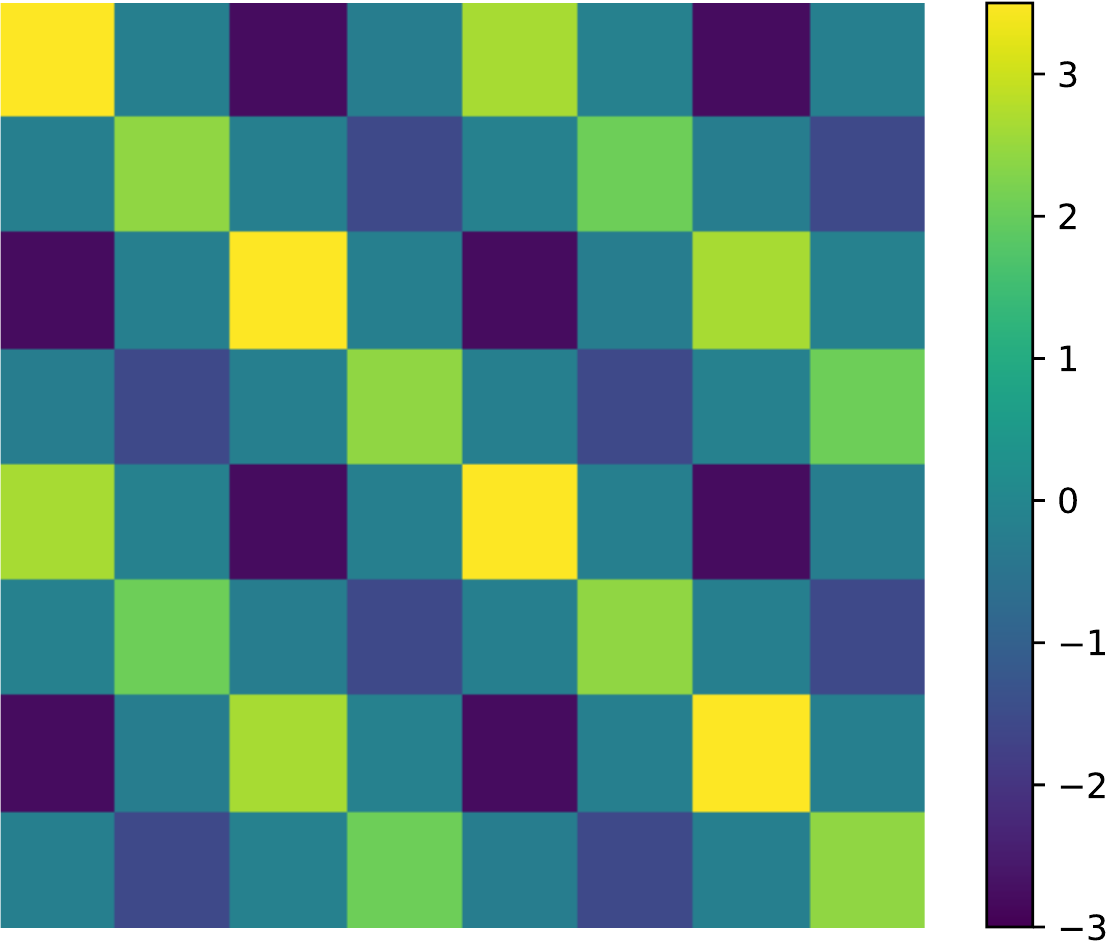}
}
\subfigure[$f_i\sim{\cal N}(3.5, 10^{-2})$ in DNS]
{
	\includegraphics[width=0.45\hsize]{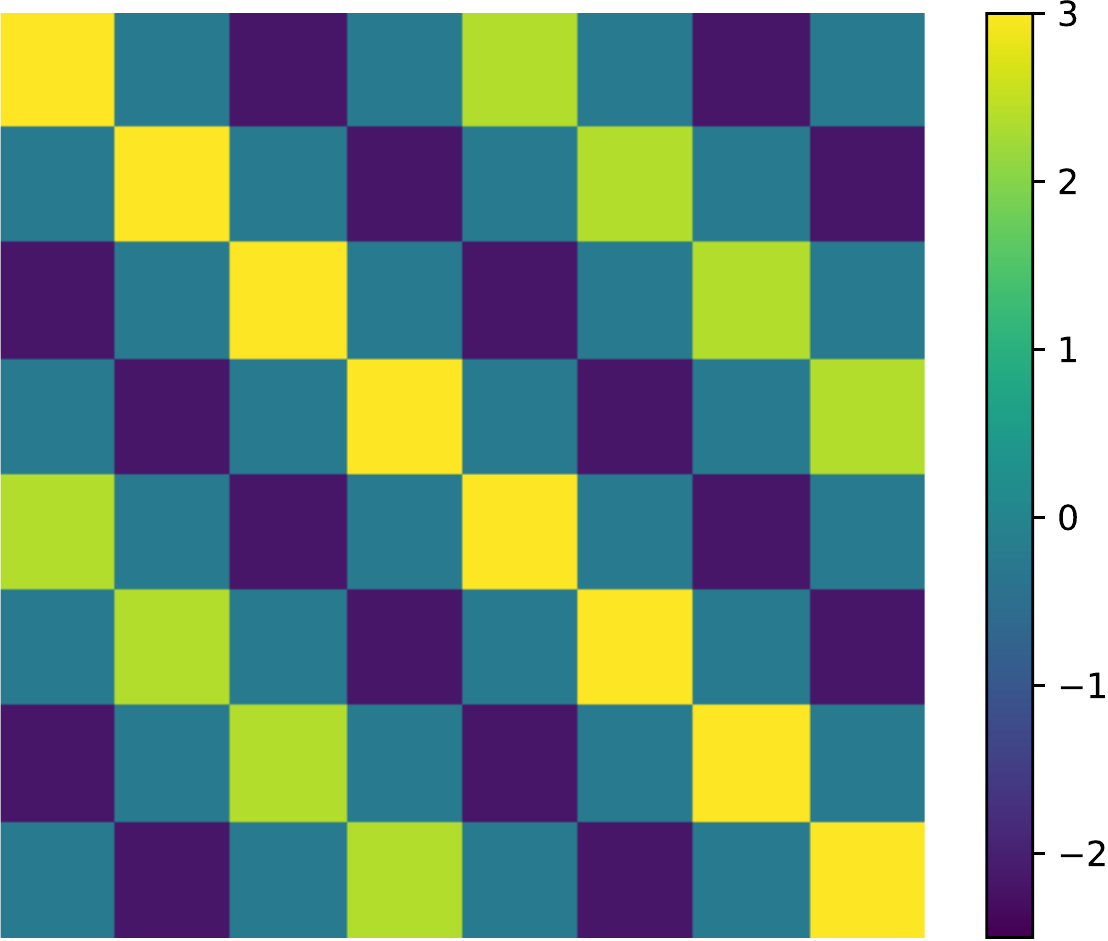}
}
\subfigure[$f_i\sim{\cal N}(3.5, 1)$ in DNS]
{
	\includegraphics[width=0.45\hsize]{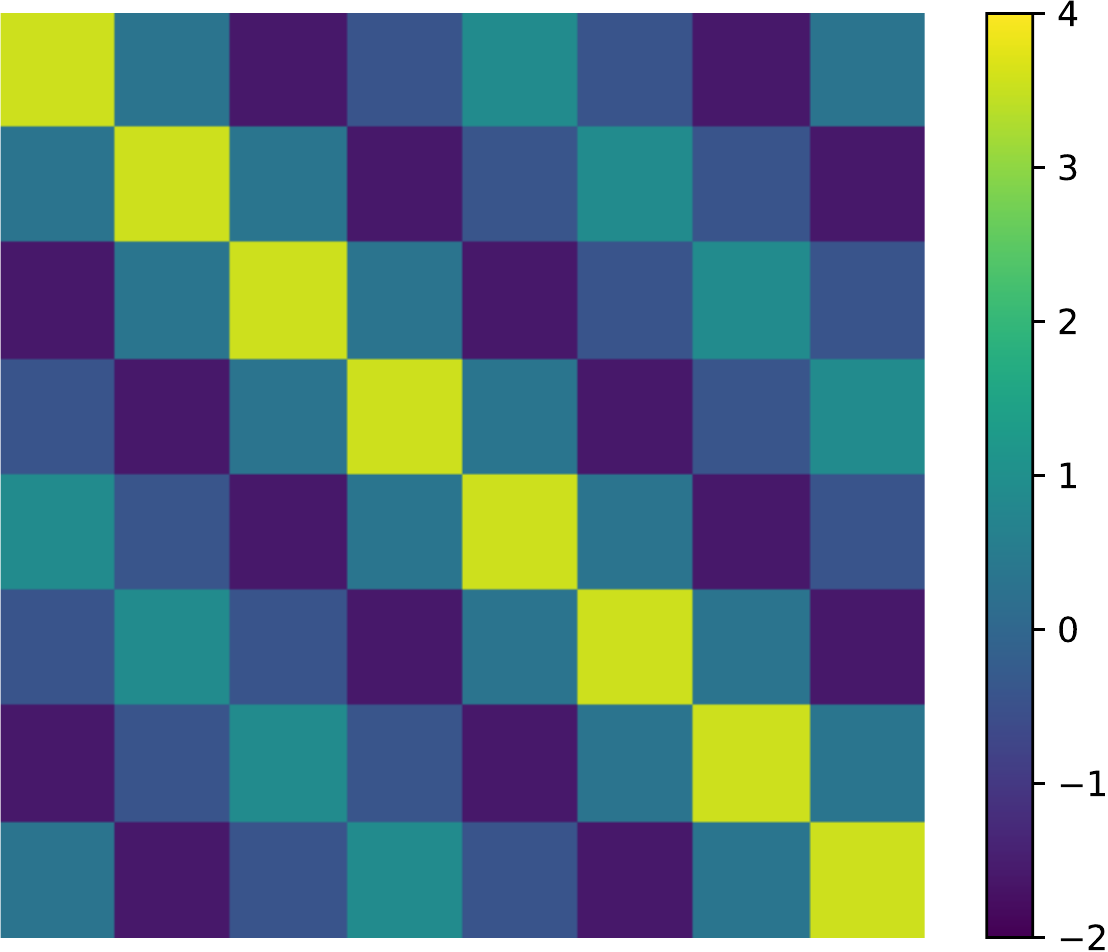}
}
\subfigure[$f_i=3.5$ in CE2.5 for $\tau_d=10$]
{
	\includegraphics[width=0.45\hsize]{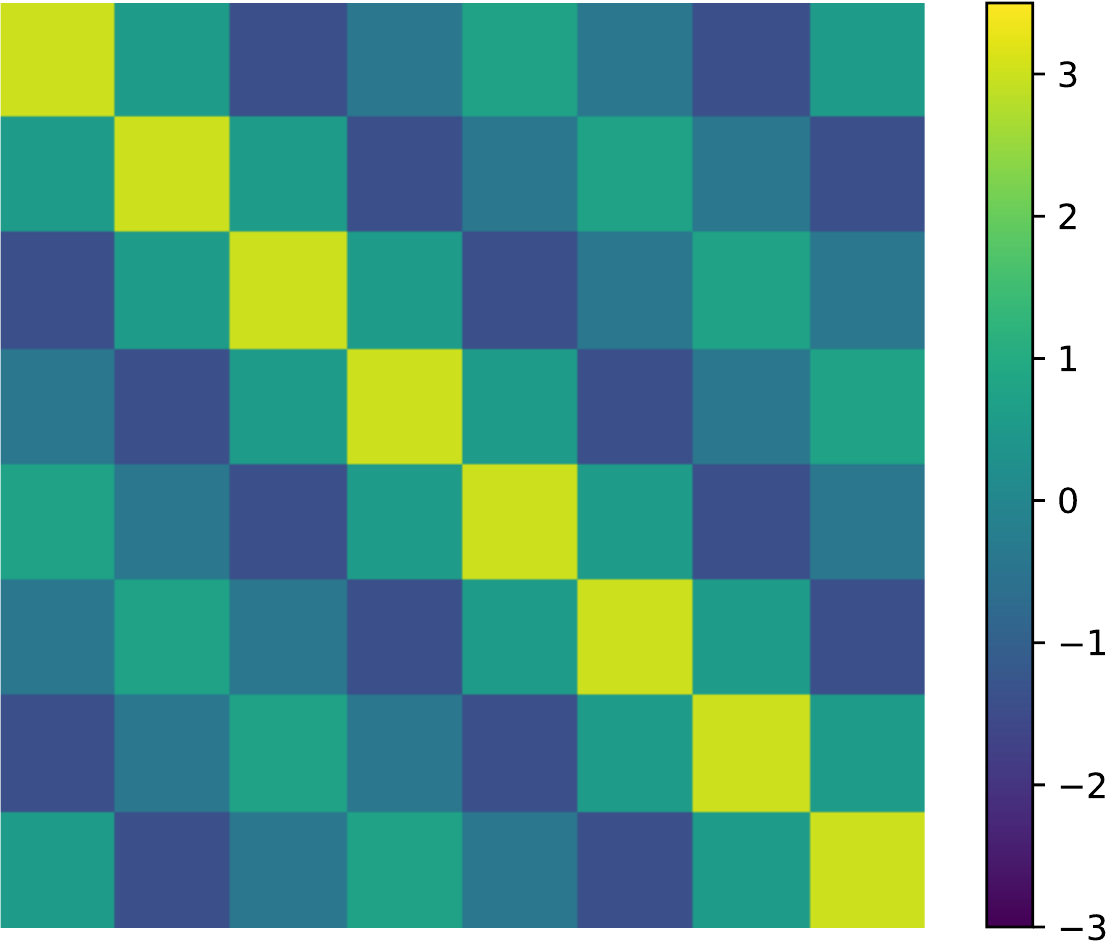}
}
\subfigure[$f_i\sim{\cal N}(3.5, 10^{-2})$ in CE3 for $\tau_d=10$]
{
	\includegraphics[width=0.45\hsize]{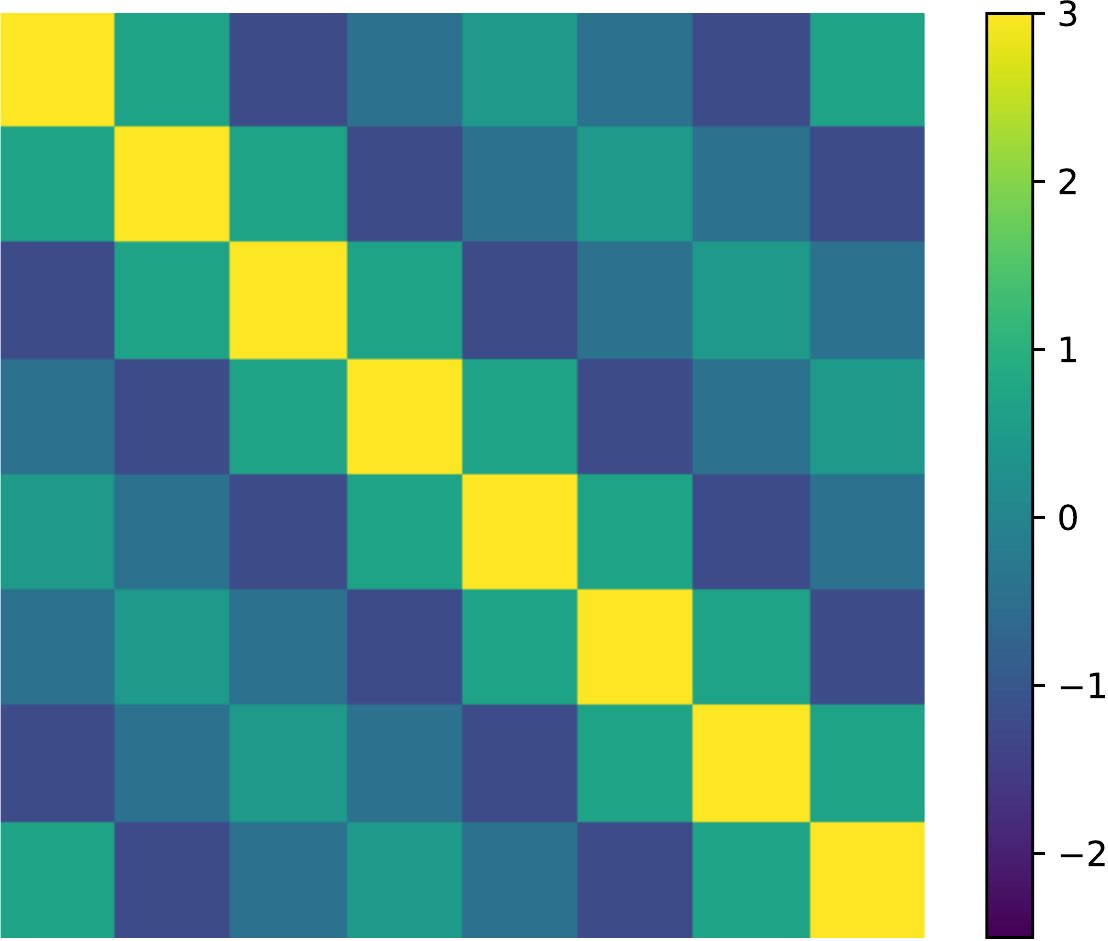}
}
\subfigure[$f_i\sim{\cal N}(3.5, 1)$ in CE2.5 for $\tau_d=8$]
{
	\includegraphics[width=0.45\hsize]{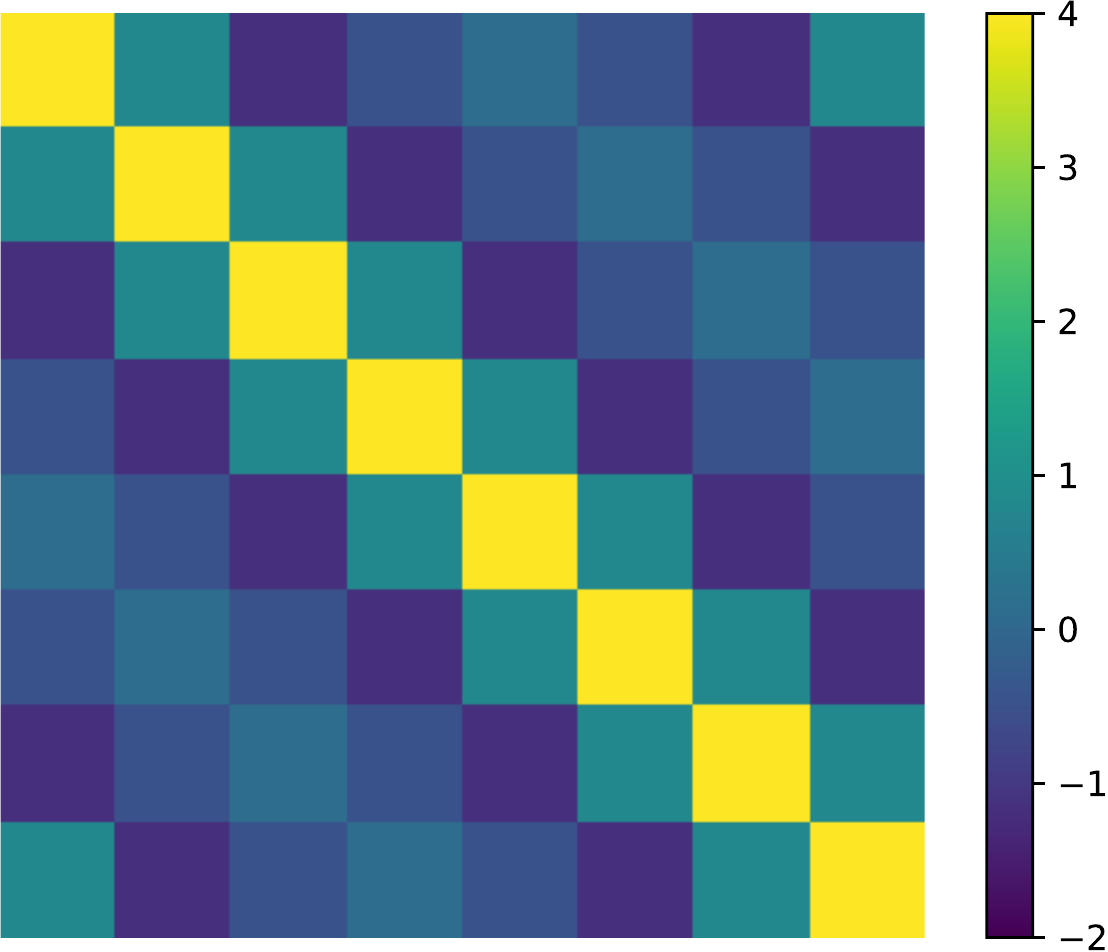}
}
\caption{The covariance matrix, $C_{\bf xx}$, obtained in DNS and DSS for $f_i=3.5$ with and without noise.}
\label{cov_dperiodic}
\end{figure}

In this dynamical regime, the spatial oscillation of the state vactor, ${\bf x}$, becomes more complex than the cases in the periodic regime, e.g., see Fig (\ref{plt_periodic3})
for the spectral coefficients, $a_m$, as a function of time and Fig. (\ref{cov_dperiodic})
for the plots of the covariance matrix, $C_{\bf xx}$ obtained in DNS and DSS for $f_i=3.5$, ${\cal N}(3.5, 10^{-2})$ and ${\cal N}(3.5, 1)$. The covariance matrix for all test cases is dominated by the diagonal elements. In this dynamical regimes, all eigenvectors, which are parallel to the Fourier basis function, are doubly folded for $m>1$ with the associated eigenvalues significantly greater than zero.

We solve the CE2.5 and CE3 equations to obtain the low-order statistics of the Lorenz96 system and we compare with those obtained in DNS for a variety of external forcing, i.e. for the cases $f_i=3.5$, $f_i \sim {\cal N}(3.5, \ 0.01), {\cal N}(3.5, \ 0.1)$ and ${\cal N}(3.5, \ 1)$. Among all test cases with or without noise, the CE2.5 and CE3 approximations are numerically stable for the eddy damping parameter selected to be in the range of $\tau_d^{-1} \in [0.1, 100]$; whist the CE2 approximation is only stable but not accurate for the system with strong random noise, i.e. $f_i\sim{\cal N}(3.5,1)$. Within DNS, for the test case of $f_i=3.5$ with no noise, the dynamics of each node converges to the $\alpha$ and $\beta$ group with equal probability. Very interestingly, we find that the ensemble CE2.5 and CE3 approximations compute the statistical average of the ``superpositioned'' oscillation of the $\alpha$ and $\beta$ groups for all state variables, $x_i$, e.g., see Table (\ref{tb2}),
\begin{table*}
\begin{ruledtabular}
\begin{tabular}{l|c|r|r|r|r|r||r|r|r|r}
          & $f_i$  & $\tau_d^{-1}$   & $C_{x_{\alpha}}$ & $C_{x_{\alpha}x_{\alpha}}$ & $C_{x_{\alpha}x_{\alpha+1}}$ & $C_{x_{\alpha}x_{\alpha+2}}$ & $C_{x_{\beta}}$ & $C_{x_{\beta}x_{\beta}}$ & $C_{x_{\beta}x_{\beta+1}}$ & $C_{x_{\beta}x_{\beta+2}}$ \\\hline
DNS    & $3.5$&  & $1.71$      & $3.60$      & $-0.19$       & $-2.78$   & $1.42$      & $2.43$      & $-0.20$       & $-1.56$  \\\hline
CE2.5  & & $10$ & $1.51$ &  $3.01$ & $0.58$      & $-1.41$ & $1.51$ &  $3.01$ & $0.58$      & $-1.41$        \\\hline
CE2.5  & & $20$ & $1.36$ &  $2.91$ & $0.38$      & $-1.76$ & $1.36$ &  $2.91$ & $0.38$      & $-1.76$        \\\hline
CE3    & & $10$ & $1.59$ &  $3.04$ & $0.68$      & $-1.22$ & $1.59$ &  $3.04$ & $0.68$      & $-1.22$        \\\hline
CE3    & & $20$ & $1.38$ &  $2.92$ & $0.43$      & $-1.68$ & $1.38$ &  $2.92$ & $0.43$      & $-1.68$        \\\hline
\end{tabular}
\end{ruledtabular}
\caption{The low-order cumulants obtained in DNS and DSS for the Lorenz96 system for $f_i=3.5$ without noise for the spatial resolution, ${\xmax n}=8$.}
\label{tb2}
\end{table*}

However, the solution of the low-order cumulants for this noiseless case is found to be very sensitive to the choice of $\tau_d$. For all other cases in the presence of noise, the PDF of each node converges to an unique distribution, owing  to the presence of random forcing. Here the low-order cumulants obtained in DSS agree well with those in DNS for the the second order cumulants (covariance matrix) written in terms of the eigen-pairs for the test cases, as shown in Table (\ref{tb3})
\begin{table}[htp]
\centering
\begin{tabular}{l|c|r|r|r|r|r}
          & $f_i$  & $\tau_d^{-1}$   & $C_{x_i}$ & $C_{x_ix_i}$ & $C_{x_ix_{i+1}}$ & $C_{x_ix_{i+2}}$ \\\hline
DNS    & ${\cal N}(3.5,10^{-2})$ &   & $1.57$      & $3.08$      & $-0.21$       & $-2.16$     \\\hline
CE2.5  & & $10$  & $1.59$      & $3.05$      & $0.63$       & $-1.23$     \\\hline
CE3  &    & $10$  & $1.59$      & $3.05$      & $0.69$       & $-1.22$     \\\hline\hline
DNS    & ${\cal N}(3.5,10^{-1})$ &   & $1.55$      & $3.04$      & $-0.24$       & $-2.17$     \\\hline
CE2.5  & & $10$  & $1.58$      & $3.13$      & $0.66$       & $-1.26$     \\\hline
CE3  &    & $10$  & $1.59$      & $3.13$      & $0.71$       & $-1.22$     \\\hline\hline
DNS    & ${\cal N}(3.5,1)$ &      & $1.57$      & $3.55$      & $0.32$       & $-1.61$     \\\hline
CE2.5  & & $8$                        & $1.53$     & $4.01$      & $0.80$       & $-1.17$     \\\hline
CE3     & & $8$                        & $1.51$     & $4.00$      & $0.85$       & $-1.13$     
\end{tabular}
\caption{The low-order cumulants obtained in DNS and DSS for the Lorenz96 system for $f_i\sim{\cal N}(3.5,\ 0.01), {\cal N}(3.5,\ 0.1)$ and ${\cal N}(3.5,\ 1)$.}
\label{tb3}
\end{table}
for comparison, except for some off-diagonal elements in the covariance matrix, which are approximately ten times smaller than the diagonal elements and become less accurate, e.g., see $C_{x_i x_{i+1}}$, in Table (\ref{tb3}). In this dynamical regime, all eigenvectors are excited with the associated eigenvalues significantly greater than zero. We further find that the eigen-decomposition of the covariance matrix is sensitive to the sub- and super-diagonal elements, e.g., see Table (\ref{tb4a})
\begin{table*}
\begin{ruledtabular}
\begin{tabular}{l|c|c|c|c|c|c|c|c}
       & $f_i$ & $\tau_d^{-1}$& $C_{x_{i}}$ & $\lambda_{m=0}$ &  $\lambda_{m=1}$    & $\lambda_{m=2}$  & $\lambda_{m=3}$   & $\lambda_{m=4}$      \\ \hline
DNS & ${\cal N}(3.5,10^{-2})$& &  $1.57$ & $0.18$& $0.64$  & $9.76$ & $0.72$   & $1.95$ \\ \hline
CE2.5/CE2.5t && $10$ &  $1.59$ & $1.31$& $4.13$  & $6.07$ & $1.03$   & $0.65$ \\ \hline
CE2.5r           && $5$   &  $1.57$ & -- & $0.04$  & $9.23$ & --   & $1.52$ \\ \hline
CE3/CE3t  && $10$ &  $1.59$ & $1.56$& $4.17$  & $5.96$ & $0.98$   & $0.59$ \\ \hline\hline
DNS & ${\cal N}(3.5,10^{-1})$& &  $1.55$ & $0.18$& $0.64$  & $9.77$ & $0.72$   & $1.95$ \\ \hline
CE2.5/CE2.5t && $10$ &  $1.59$ & $1.40$& $4.30$  & $6.10$ & $1.10$   & $0.68$ \\ \hline
CE3/CE3t  && $10$ &  $1.59$ & $1.65$& $4.33$  & $6.00$ & $1.06$   & $0.63$ \\ \hline\hline
DNS & ${\cal N}(3.5, 1)$& &  $1.57$ & $0.99$& $3.71$  & $7.67$ & $1.58$   & $1.44$ \\ \hline
CE2               &&       &  $0.89$ & $1.00$& $2.71$  & $9.31$ & $0.61$   & $0.36$ \\ \hline
CE2.5/CE2.5t && $8$ &  $1.53$ & $2.46$& $5.67$  & $6.51$ & $2.05$   & $1.17$ \\ \hline
CE3/CE3t  && $8$ &   $1.53$ & $2.80$& $5.60$  & $6.50$ & $1.97$   & $1.09$ \\ \hline
CE3r         && $8$ &   $1.43$ & $2.19$& $4.05$  & $5.09$ & --   & --  \\ \hline
CE3r         && $8$ &   $1.54$ & $2.51$& $5.14$  & $5.75$ & $1.70$   & -- 
\end{tabular}
\end{ruledtabular}
\caption{An alternative representations of the low-order statistics for the test cases shown in Table (\ref{tb3}), where ``CE2.5r/CE3r'' and ``CE2.5t/CE3t'' stand for the solutions obtained by solving the reduced and transformed cumulant equations using startegies, {\it i}) and {\it ii}), respectively.}
\label{tb4a}
\end{table*}

In this dynamical regime, the computational expense of the DSS equations can be reduced via strategy {\it i}). But as we increase the randomness of the Gaussian noise, more energy is pumped into the incoherent part of the dynamical system, and this significantly increases the importance of the eigenvectors for the covariance matrix, especially for those eigenvectors with smaller eigenvalues, e.g., for $f_i \sim {\cal N}(3.5, 0.01)$, the smallest eigenvalue is $0.18$ for $m=0$ mode and for $f_i \sim {\cal N}(3.5, 1)$ the value is increased to $0.99$. We find that for the Lorenz96 system driven by the greater random force, more eigenvectors must therefore be retained in order to obtain stable numerical solutions. The diagonalised cumulant approximation via  strategy, {\it ii}), is found to be numerically stable for any initial condition and the solutions are identical to those obtained by solving the full original cumulant equations.

\subsection{The Lorenz96 system in the chaotic states}

\begin{figure}[htp]
\centering
\subfigure[$f_i=5$]
{
	\includegraphics[width=0.45\hsize]{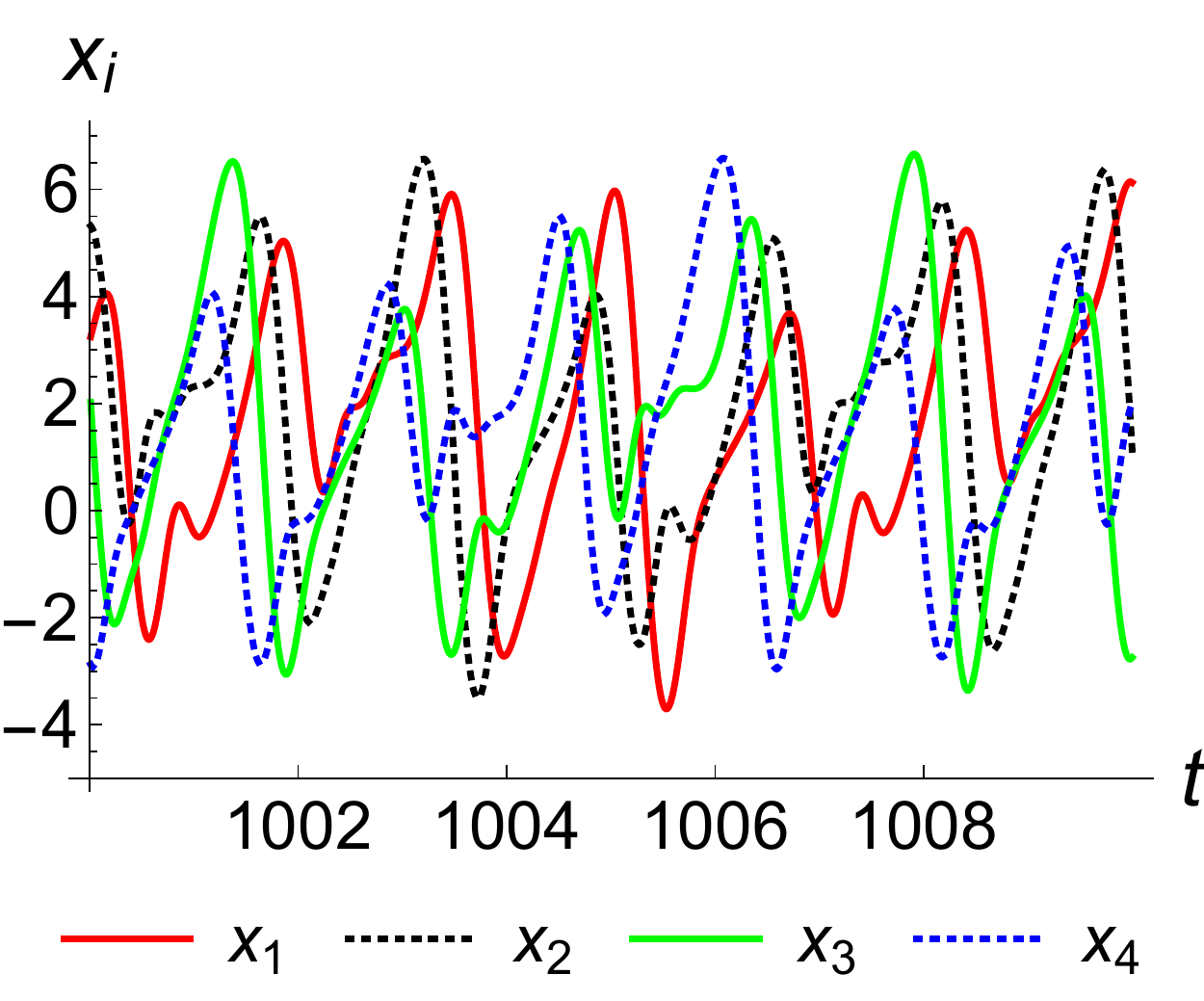}
}
\subfigure[$f_i=5$]
{
	\includegraphics[width=0.45\hsize]{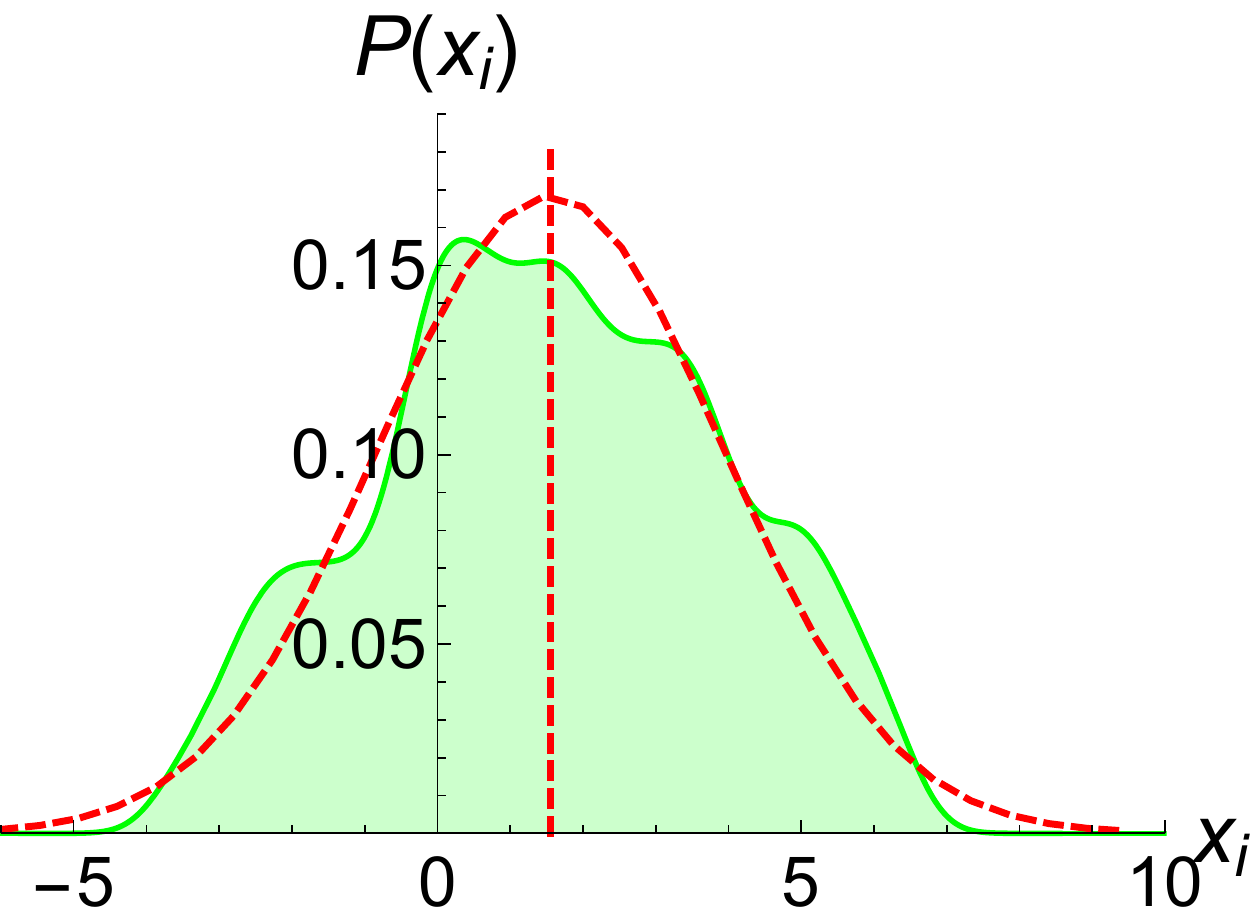}
}
\subfigure[$f_i=20$]
{
	\includegraphics[width=0.45\hsize]{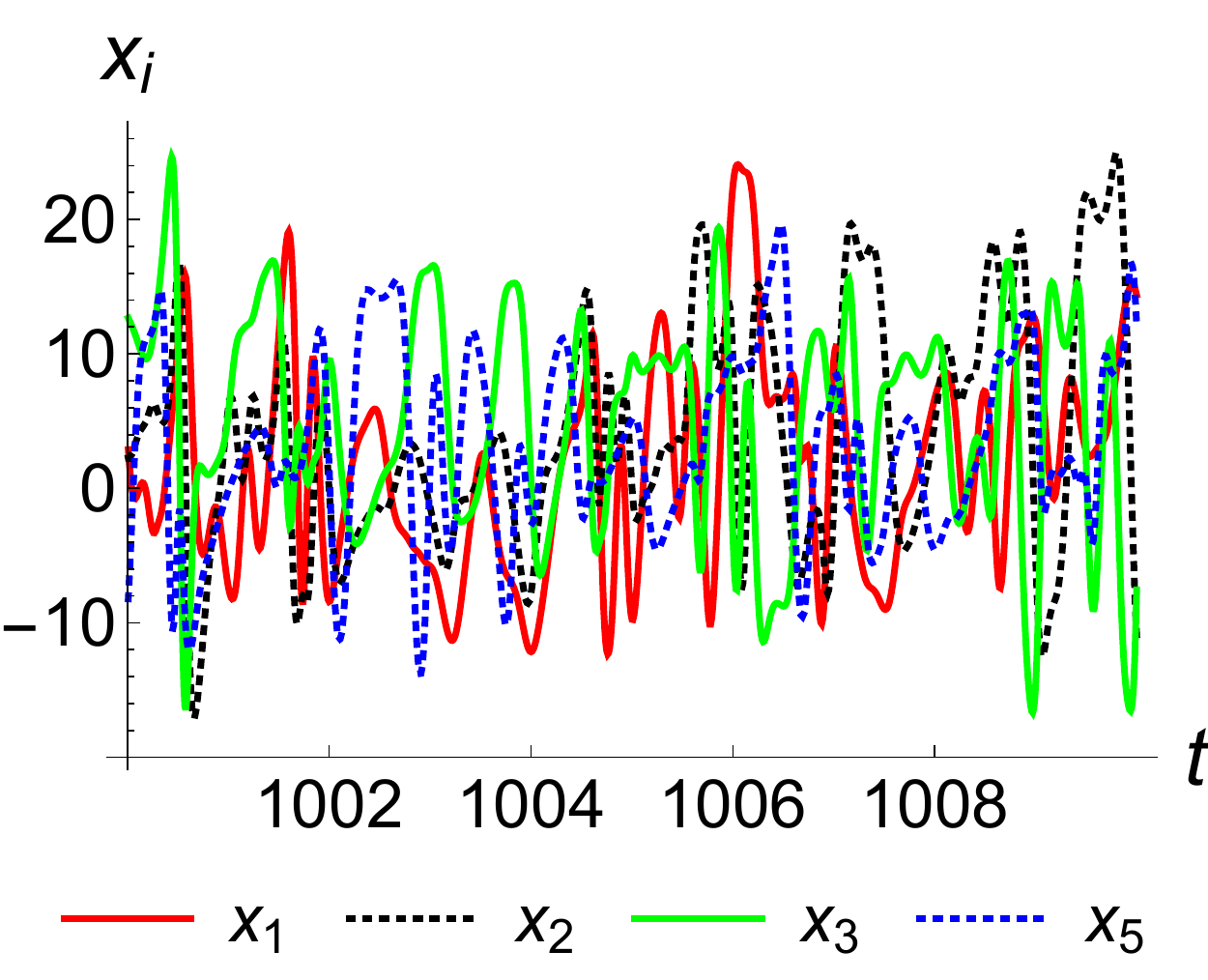}
}
\subfigure[$f_i=20$]
{
	\includegraphics[width=0.45\hsize]{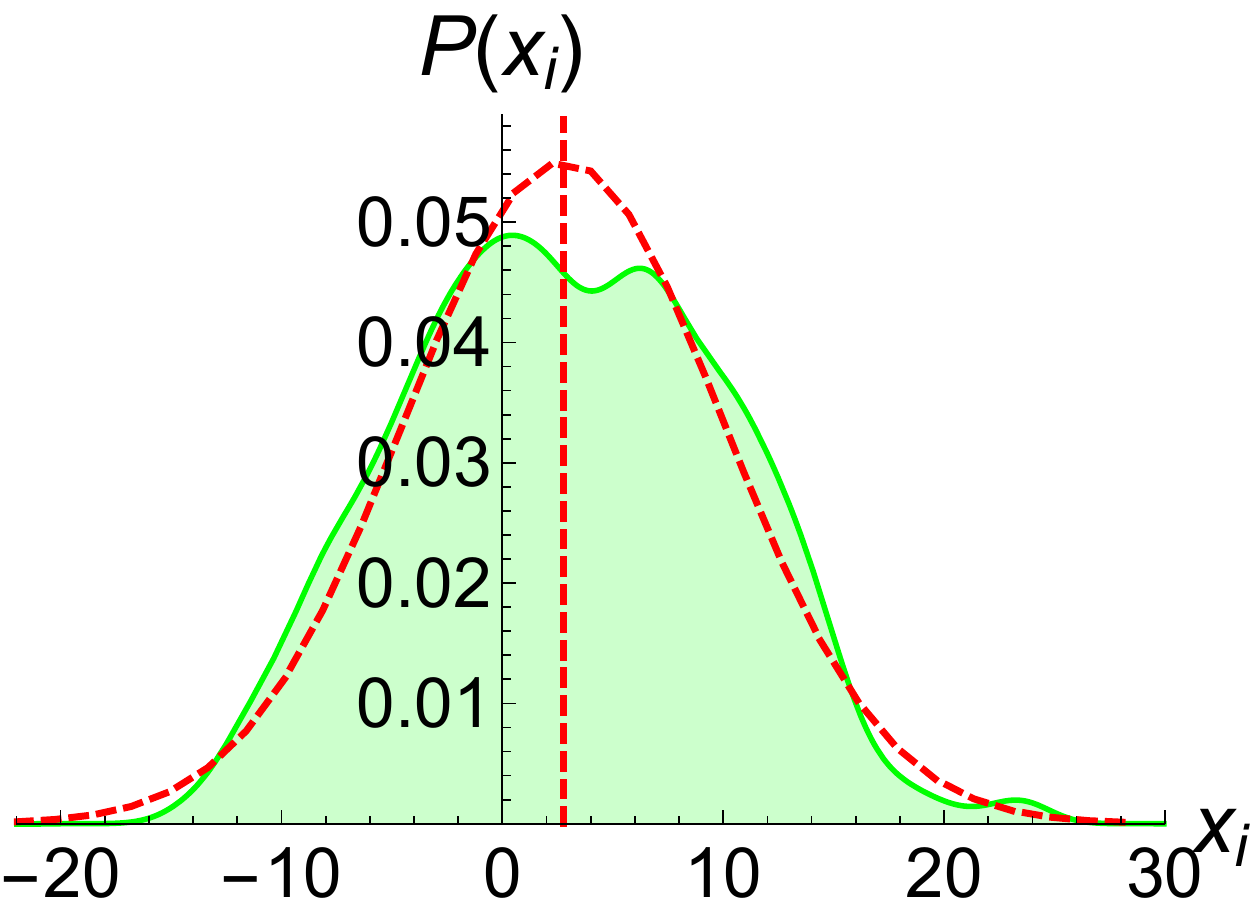}
}
\subfigure[$a_{m}(t)$ for $f_i=5$]
{
	\includegraphics[width=0.45\hsize]{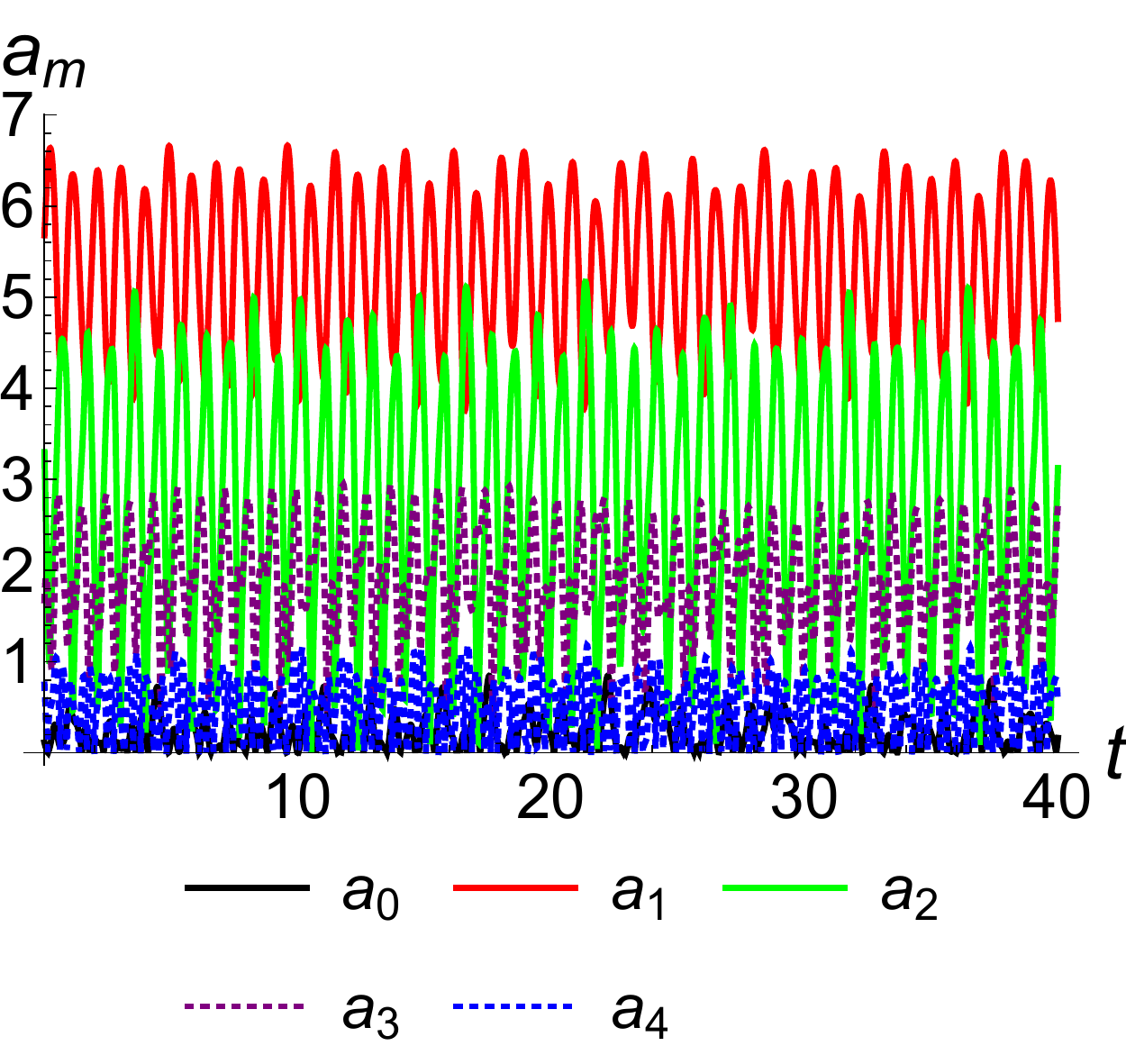}
}
\subfigure[$a_1 \sim a_2 \sim a_4$ for $f_i=5$]
{
	\includegraphics[width=0.45\hsize]{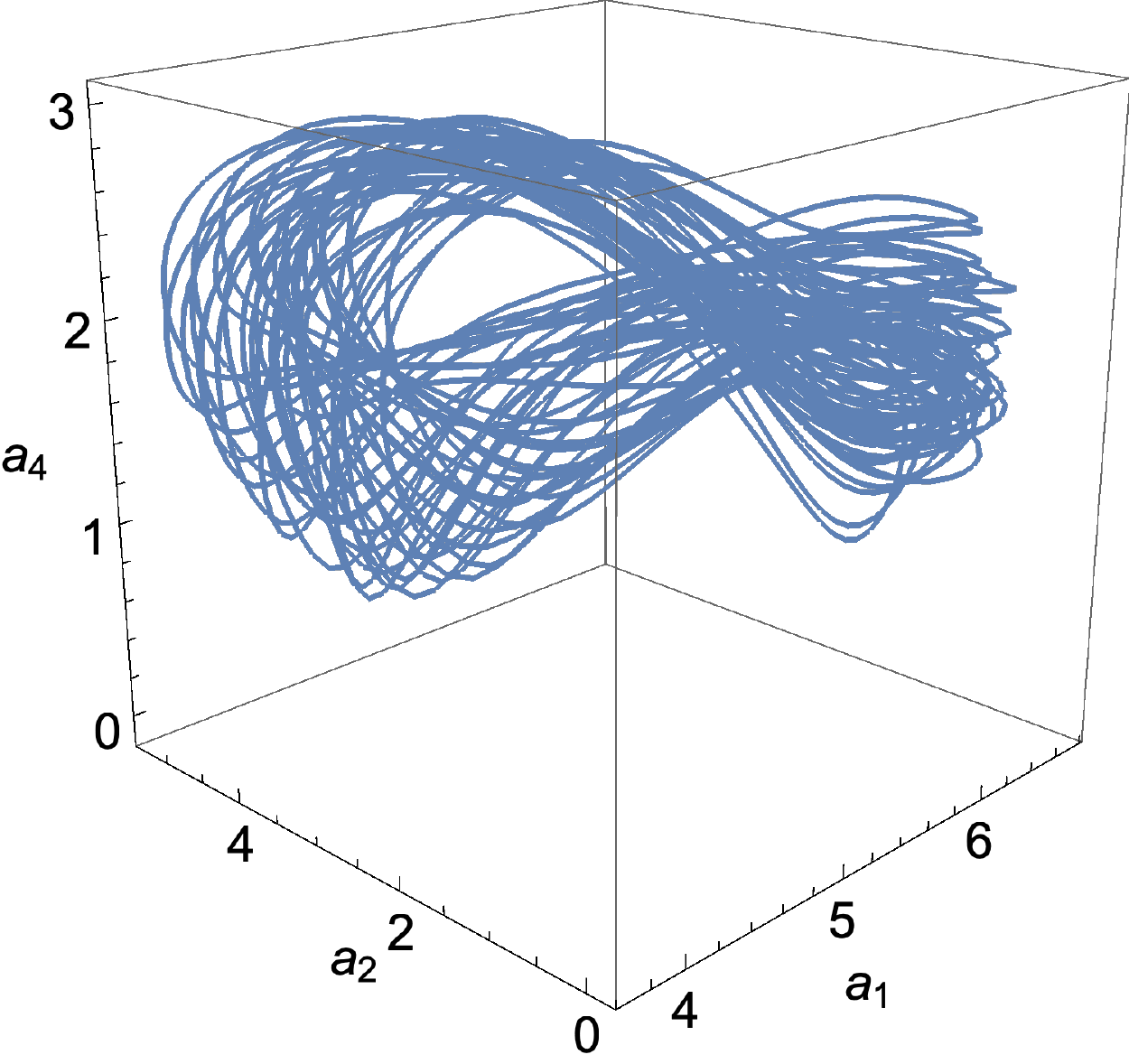}
}
\subfigure[$a_{m}(t)$ for $f_i=20$]
{
	\includegraphics[width=0.45\hsize]{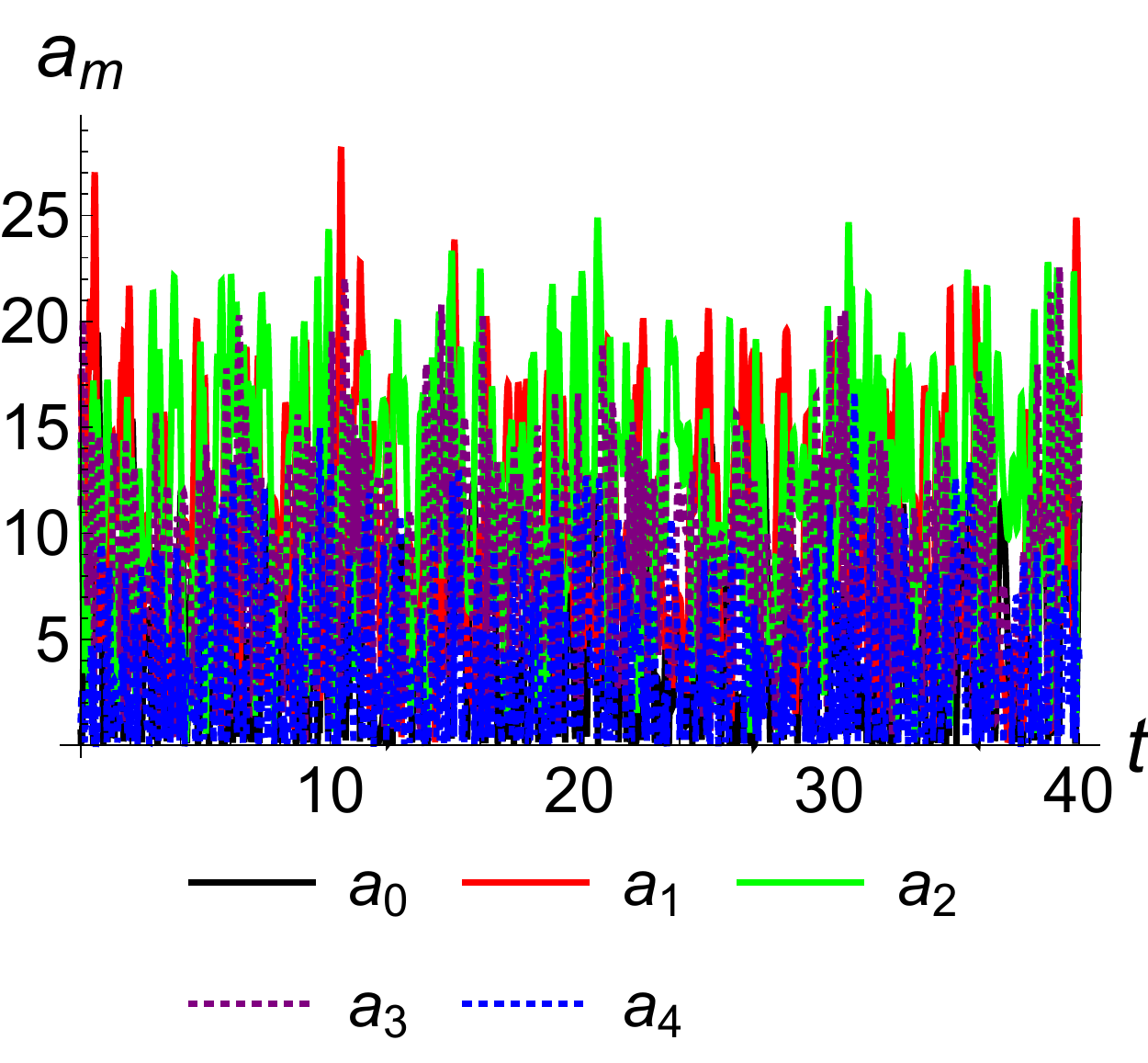}
}
\subfigure[$a_1 \sim a_2 \sim a_4$ for $f_i=20$]
{
	\includegraphics[width=0.45\hsize]{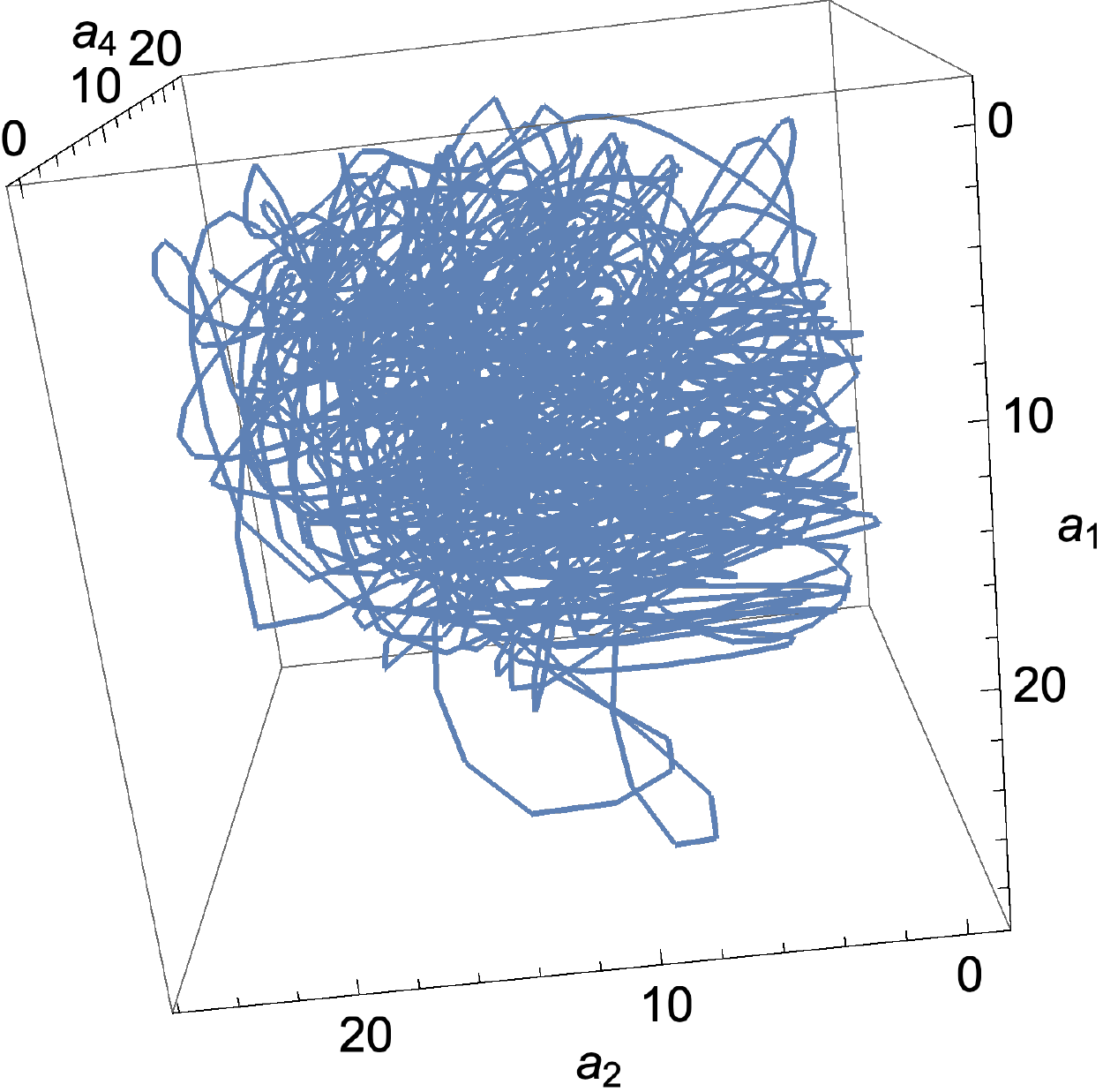}
}
\subfigure[$f_i=5$]
{
	\includegraphics[width=0.9\hsize]{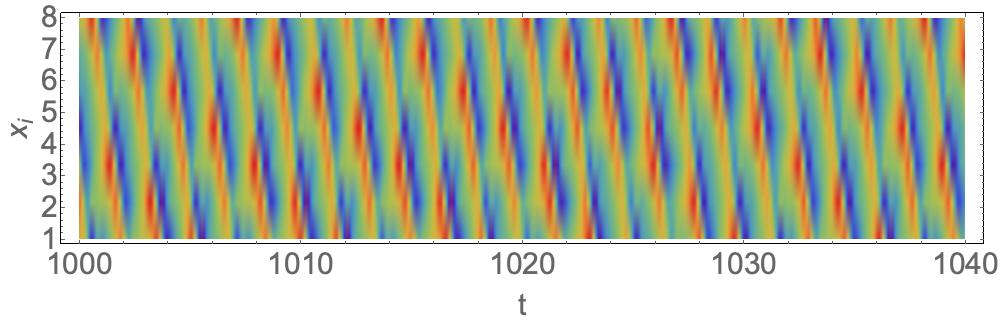}
}
\subfigure[$f_i=20$]
{
	\includegraphics[width=0.9\hsize]{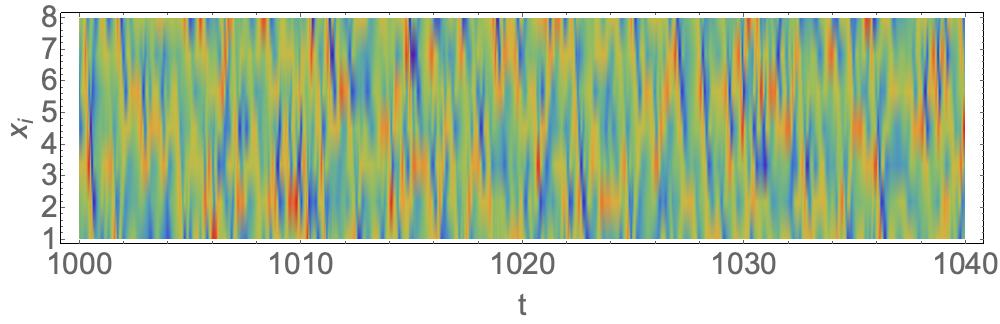}
}
\caption{The illustration of the Lorenz96 in the chaotic state for $f_i=5$ and $20$ in DNS.}
\label{plt_chaotic}
\end{figure}

When the external force becomes large, i.e., $f_i \geq 4.5$, the Lorenz96 system settles into a chaotic state where the spatio-temporal complexity of the dynamical system subsequently increases, if the external forcing, $f_i$, becomes large, e.g., see Fig. (\ref{plt_chaotic})
for the typical solutions of the Lorenz96 system obtained in DNS in the weakly chaotic state for $f_i = 5$ and strongly chaotic state for $f_i=20$, where in the weakly chaotic regime for $f_i=5$, the state vector, ${\bf x}$, oscillates quasi periodically about a unique mean value; whilst the oscillation of ${\bf x}$ become more chaotic about the mean value as the external force is increased to $f_i=20$. In all cases, the PDFs of the state variable, ${\cal P}(x_i)$ does not depend on $i$ and has compact support. Furthermore, if we increase the chaoticity of the Lorenz96 system, the PDFs of $x_i$ begin to converge to Gaussian but this convergence is slow and the high order cumulant are not negligible.

\begin{figure}
\centering
\subfigure[DNS]
{
	\includegraphics[width=0.45\hsize]{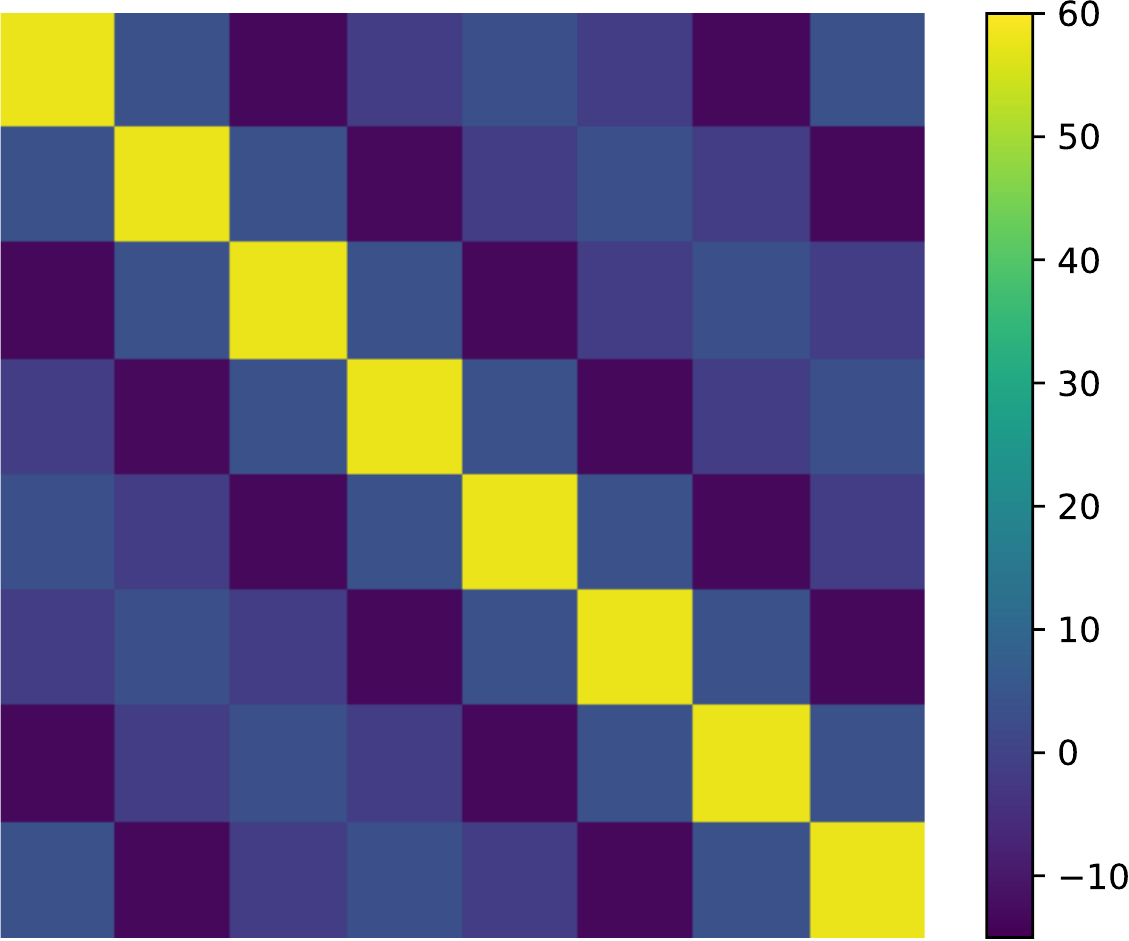}
}
\subfigure[CE2.5]
{
	\includegraphics[width=0.45\hsize]{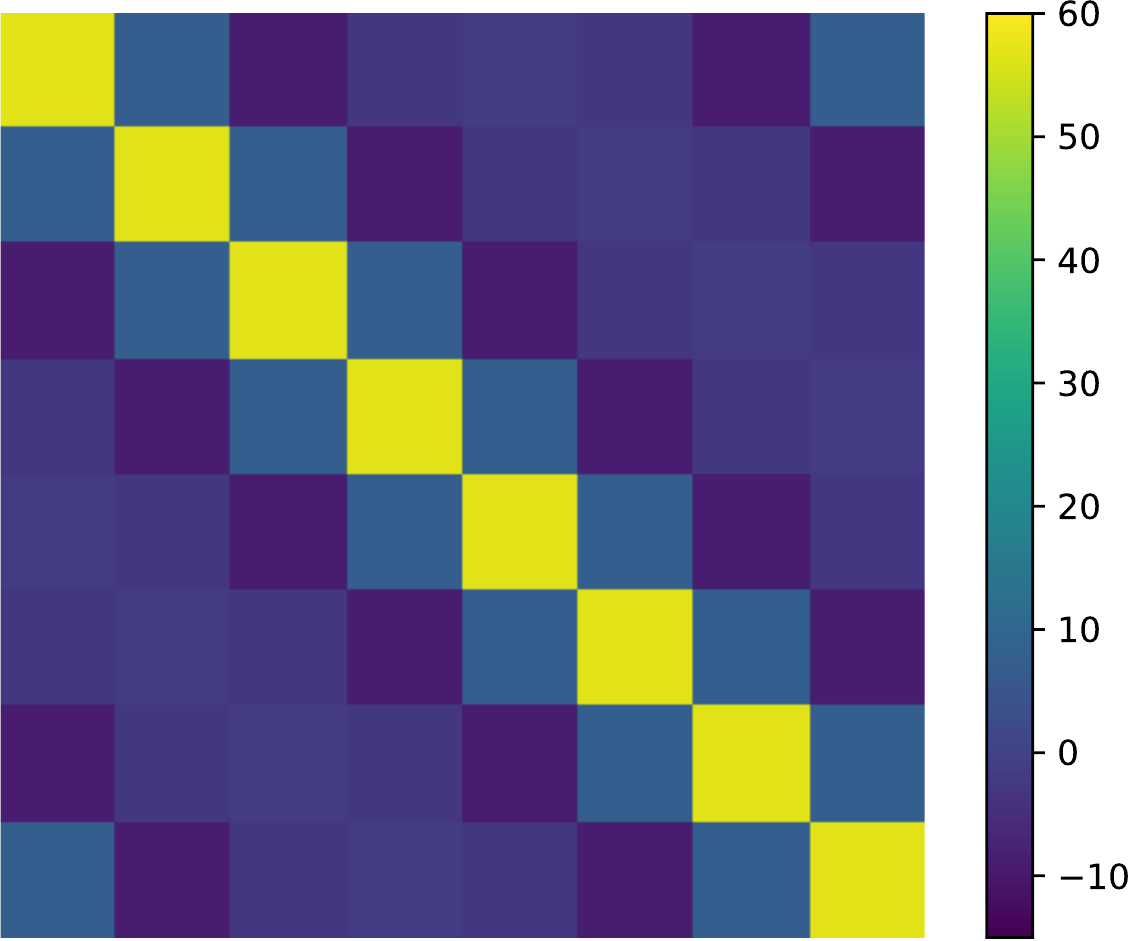}
}
\caption{The covariance matrix, $C_{\bf xx}$, obtained in DNS and CE2.5 with $\tau_d=20$ for $f_i=20$.}
\label{cov_chaotic}
\end{figure}

We forward evolve the dynamical and cumulant equations for long enough time to obtain the statistical equilibrium of the Lorenz96 system in the chaotic state and compare the low-order cumulants obtained in DNS and DSS. The CE2.5 and CE3  equations are stable for numerical computations with the eddy damping parameter, $\tau_d^{-1}$, in the range of ${\cal O}(10^{-1}) \sim {\cal O}(10^2)$, where the optimal $\tau_d^{-1}$ is found to be ${\cal O}(10)$ for all test cases; whilst the CE2 approximation is found to be numerically unstable for all cases. The low-order statistics for all test cases in DNS and DSS are listed and compared in Table (\ref{tb4b}),
\begin{table*}
\begin{ruledtabular}
\begin{tabular}{l|c|r|r|r|r|r||c|c|c|c|c}
          & $f_i$  & $\tau_d^{-1}$   & $C_{x_i}$ & $C_{x_ix_i}$ & $C_{x_ix_{i+1}}$ & $C_{x_ix_{i+2}}$ & $\lambda_{m=0}$ &  $\lambda_{m=1}$    & $\lambda_{m=2}$  & $\lambda_{m=3}$   & $\lambda_{m=4}$\\\hline
DNS    & $5$   &                        & $1.58$      & $5.47$      & $2.18$       & $-1.23$  & $0.09$& $14.54$  & $5.16$ & $1.98$   & $0.38$   \\\hline
CE2.5/CE2.5t &          &            $20$      & $1.60$     & $5.44$      & $1.13$      & $-2.27$   & $2.33$ & $7.39$    & $10.82$ & $1.81$ & $1.16$    \\\hline
CE3/CE3t    &          &            $20$     & $1.61$      & $5.46$      & $1.18$       & $-2.21$  & $2.56$ & $7.51$    & $10.65$ & $1.83$ & $1.10$   \\\hline\hline
DNS    & $20$ &                        & $3.28$      & $57.89$      & $3.48$       & $-1.29$  & $39.10$& $61.32$  & $87.71$ & $47.84$   & $30.35$   \\\hline
CE2.5/CE2.5t &                       & $20$   & $3.42$      & $56.72$      & $7.30$       & $-9.27$  & $45.23$& $72.90$  & $73.45$ & $44.16$   & $27.45$    \\\hline
CE3/CE3t    &                       & $20$   & $3.25$      & $54.49$      & $8.11$       & $-8.63$  & $48.59$& $69.66$  & $70.85$ & $41.11$   & $24.05$   \\\hline
\end{tabular}
\end{ruledtabular}
\caption{The low-order cumulants obtained in DNS and DSS for the  Lorenz96 system in the chaotic state with the spatial resolution, $\xmax{n}=8$, for the external force, $f_i=5$ and $20$, where $\lambda_m$ is the eigenvalue of the covariance matrix and the associated eigenfunctions are given by the Fourier basis. The notations, ``CE2.5t'' and ``CE3t'', stand for the test cases using the diagonalised cumulant system in CE2.5 and CE3.}
\label{tb4b}
\end{table*}
where a selection of the covariance matrix entries is listed on the left hand side of the table and the matrix represented by the eigen-pairs is shown on the right hand side. The eigensystem of the covariance matrix are doubly-folded for all cases, owing to the translational invariance. We find that the mean trajectory, $C_{x_i}$, and the diagonal elements of the covariance matrix, $C_{x_ix_i}$ of the Lorenz96 system are very accurately approximated by the CE2.5 and CE3 systems, i.e., the trace of the covariance matrix is conserved in all DSS models, see e.g. Fig. (\ref{cov_chaotic}).
However, as for  the test cases in the doubly-periodic state, the off-diagonal elements, e.g., $C_{x_i x_{i+1}}$, are found to be less accurate, which reduces the accuracy of the eigen-decomposition of the covariance matrix. In this dynamical regime, the model reduction strategy {\it i}) cannot be used to reduce the complexity of the covariance matrix, as all eigenvalues of the covariance matrix are significantly greater than zero. We further apply the strategy {\it ii}) to diagonalise the second cumulants of the DSS equations and solve the transformed cumulant equations in CE2.5/3 approximations. The solutions of the transformed systems agree perfectly with those obtained by solving the original cumulant equations.

\subsection{The Lorenz96 system driven by inhomogeneous forcing}

\begin{figure}
\centering
\subfigure[]
{
	\includegraphics[width=0.45\hsize]{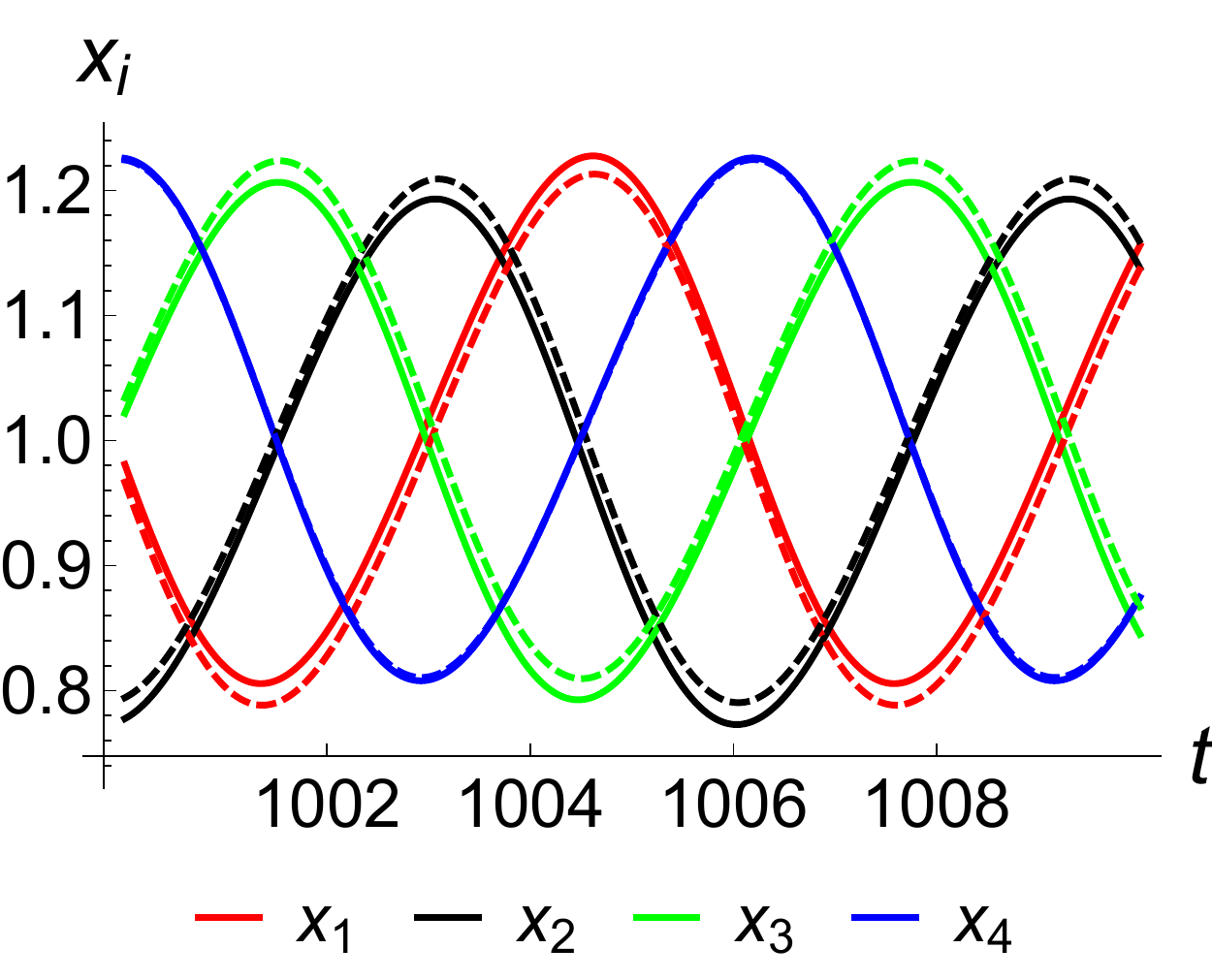}
}
\subfigure[]
{
	\includegraphics[width=0.45\hsize]{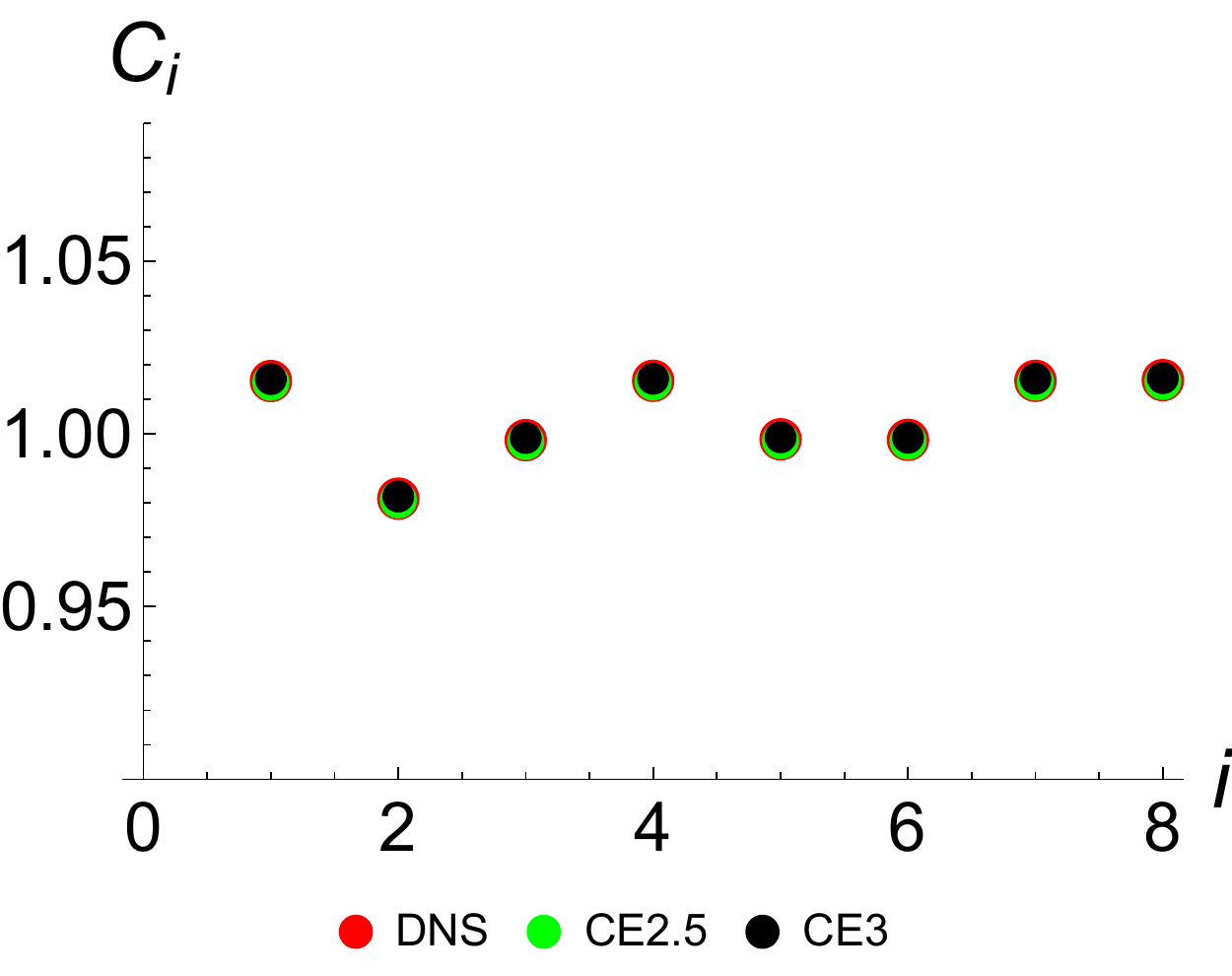}
}
\subfigure[]
{
	\includegraphics[width=0.45\hsize]{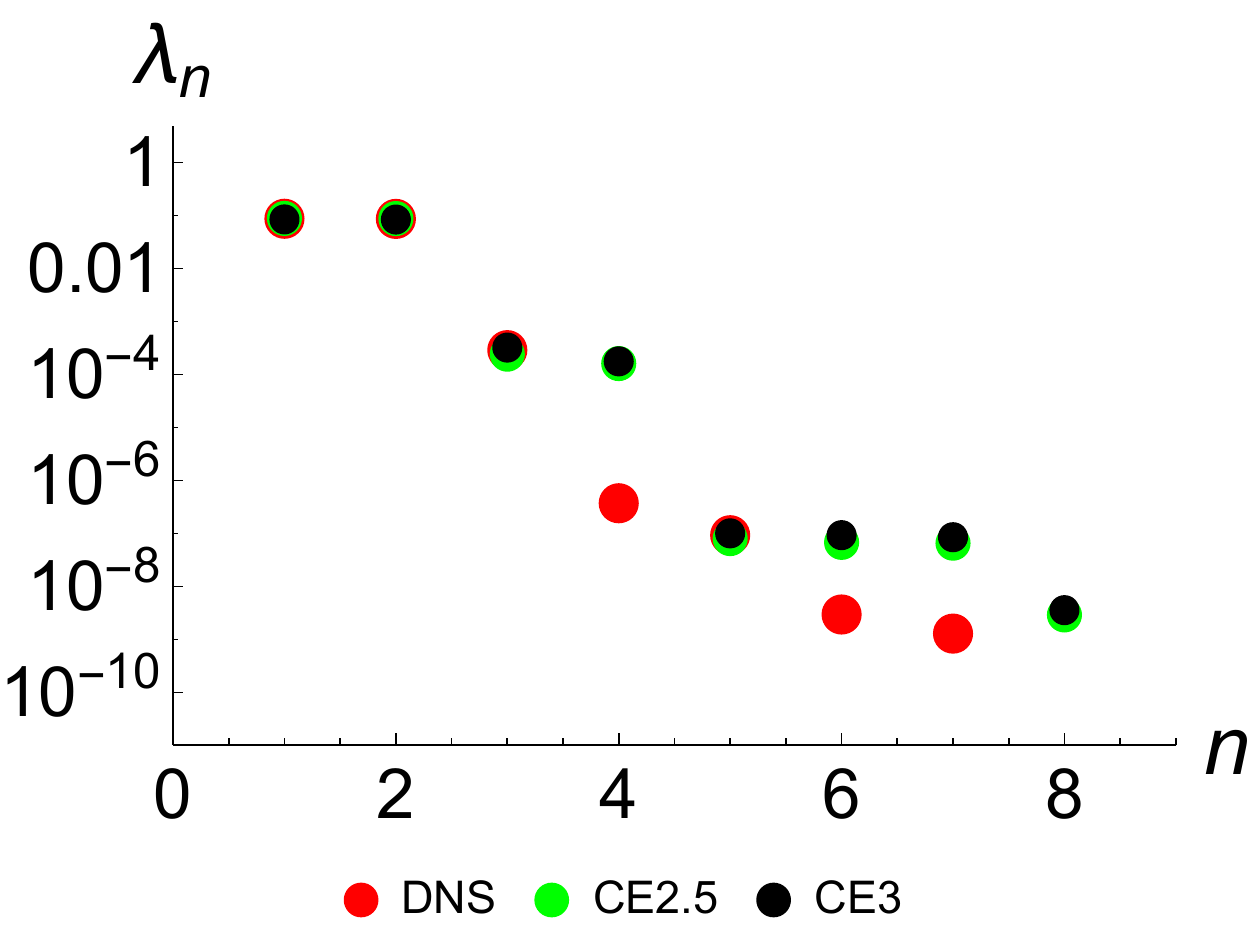}
}
\subfigure[]
{
	\includegraphics[width=0.45\hsize]{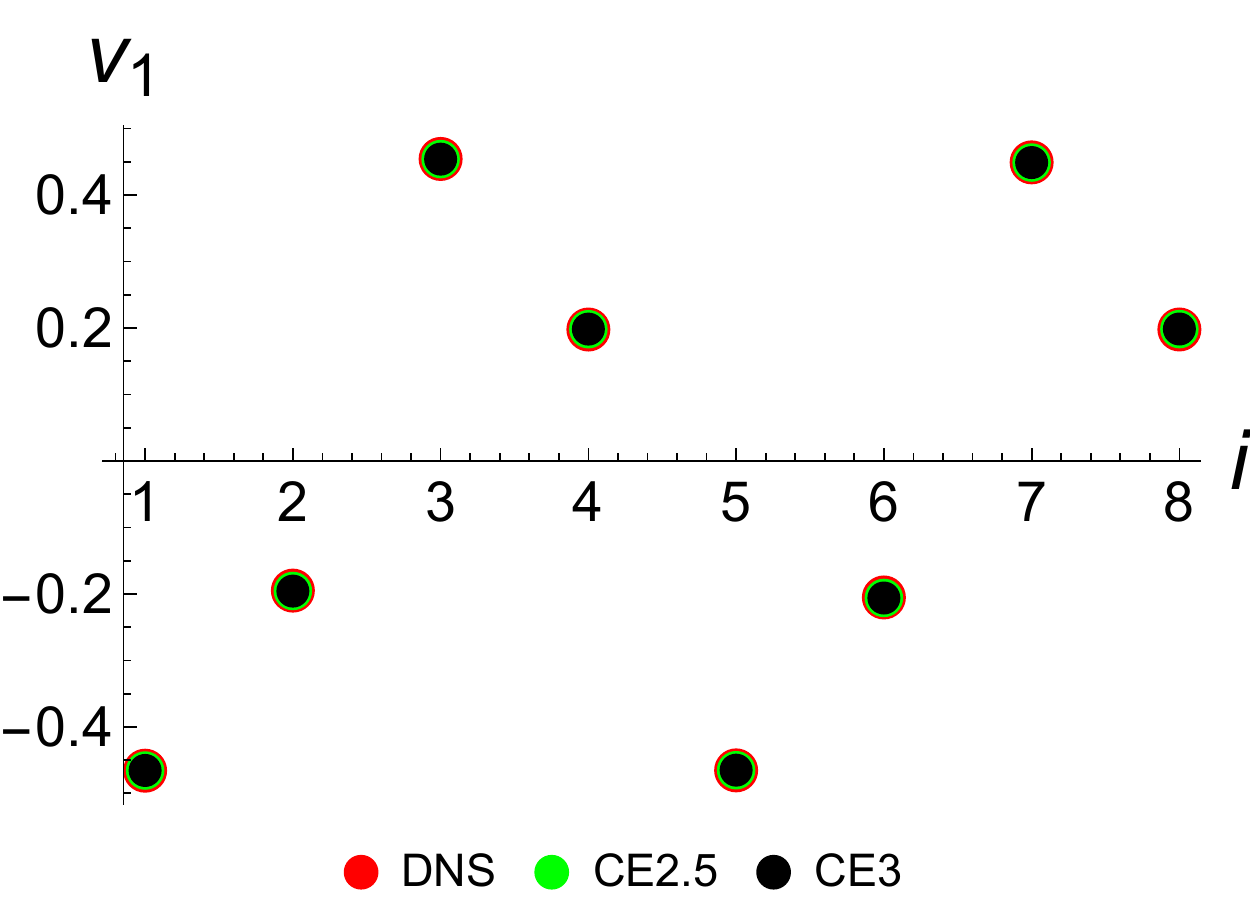}
}
\subfigure[]
{
	\includegraphics[width=0.45\hsize]{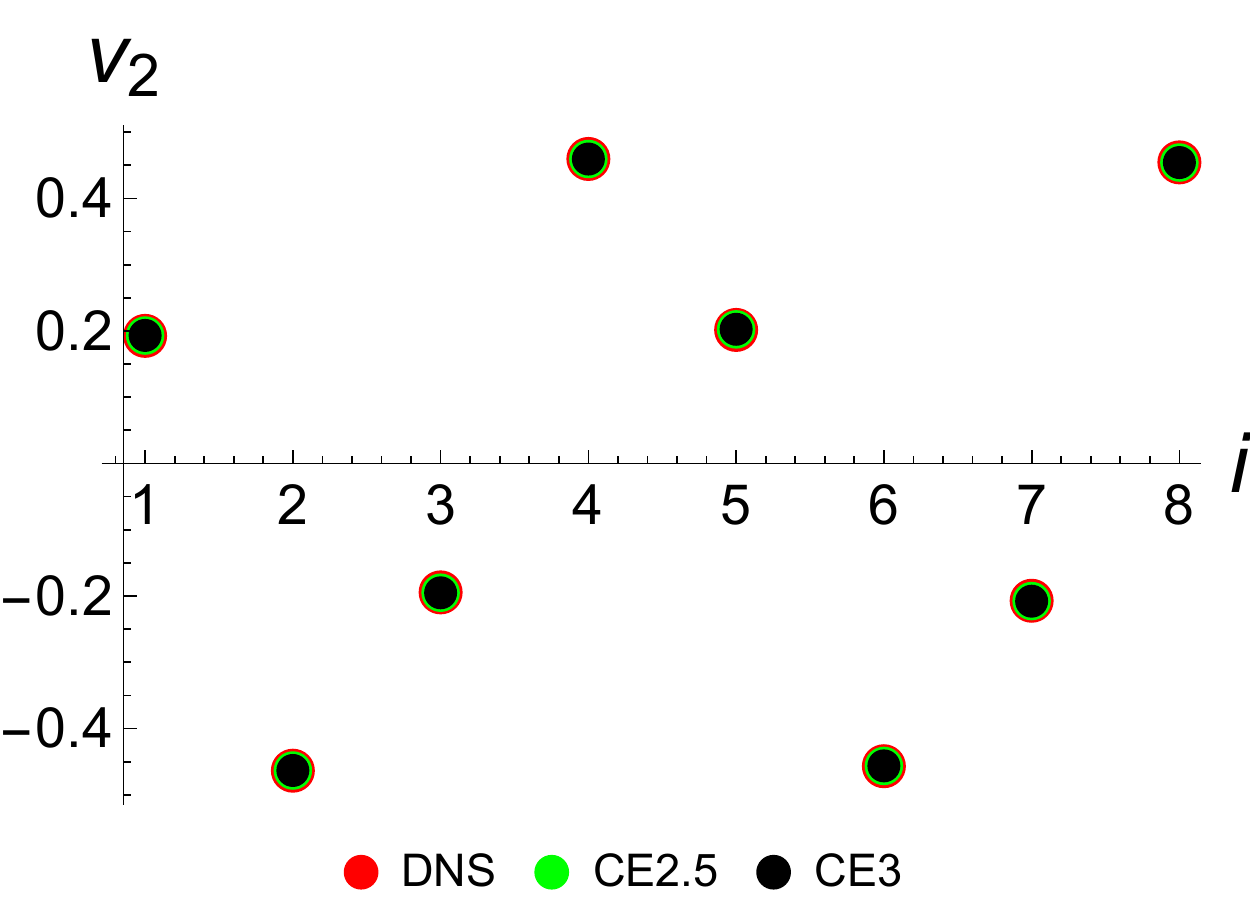}
}
\caption{The illustration of a variety of dynamical and statistical properties of the Lorenz96 system in the periodic state for $c =1.05$ and $f_{i}=1.02$ for $i=2,3,\cdots 8$, where the time series of the dynamical model in DNS is shown in (a), the first cumulants, $C_{x_i}$ obtained in DNS and DSS are plotted in (b), the eigenvalues of the covariance matrix in DNS and DSS are in (c) and the plots of the first two eigenvectors of the covariance in DNS and DSS are illustrated in (d)--(e).}
\label{inho1}
\end{figure}
\begin{figure}
\centering
\subfigure[]
{
	\includegraphics[width=0.45\hsize]{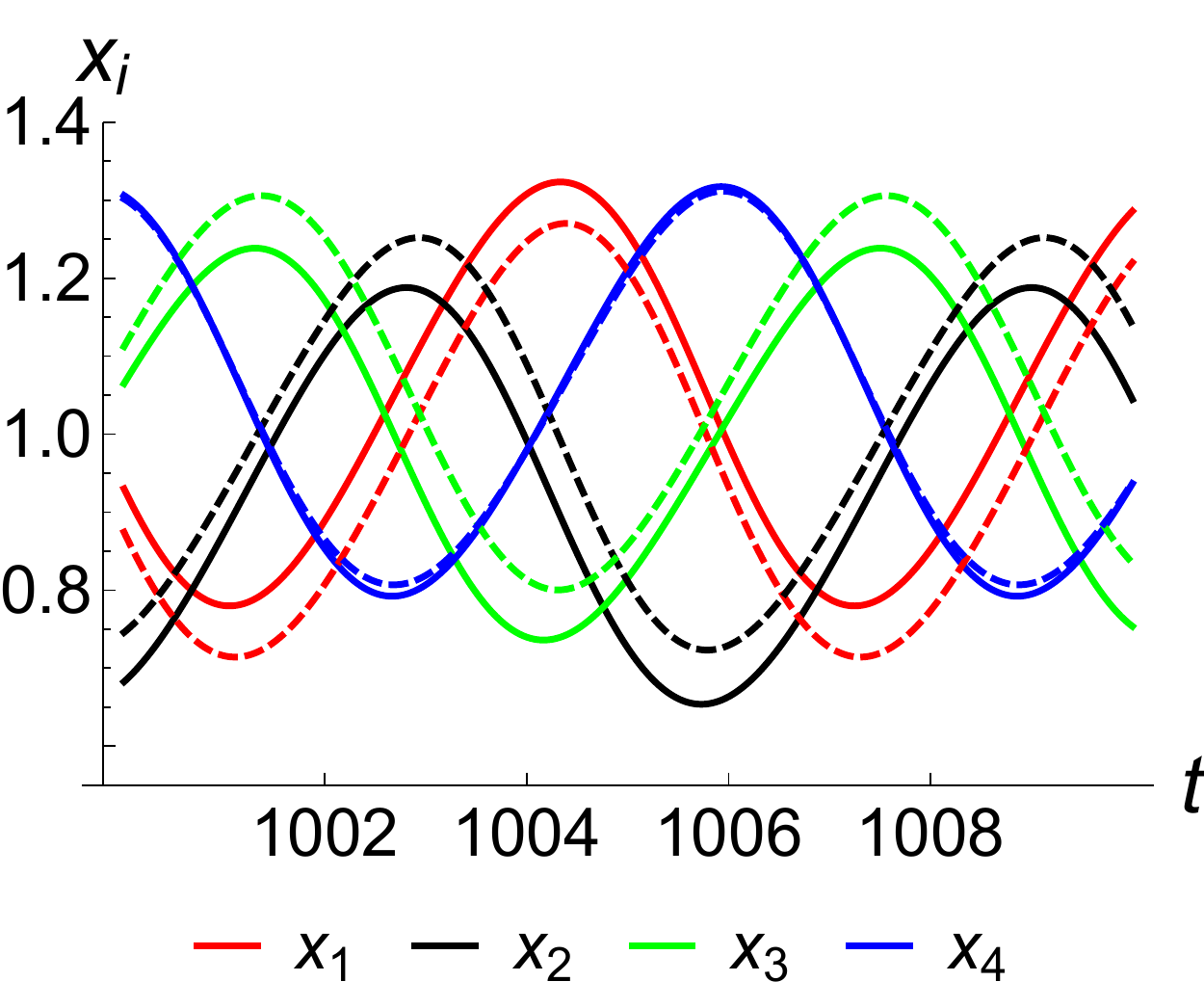}
}
\subfigure[]
{
	\includegraphics[width=0.45\hsize]{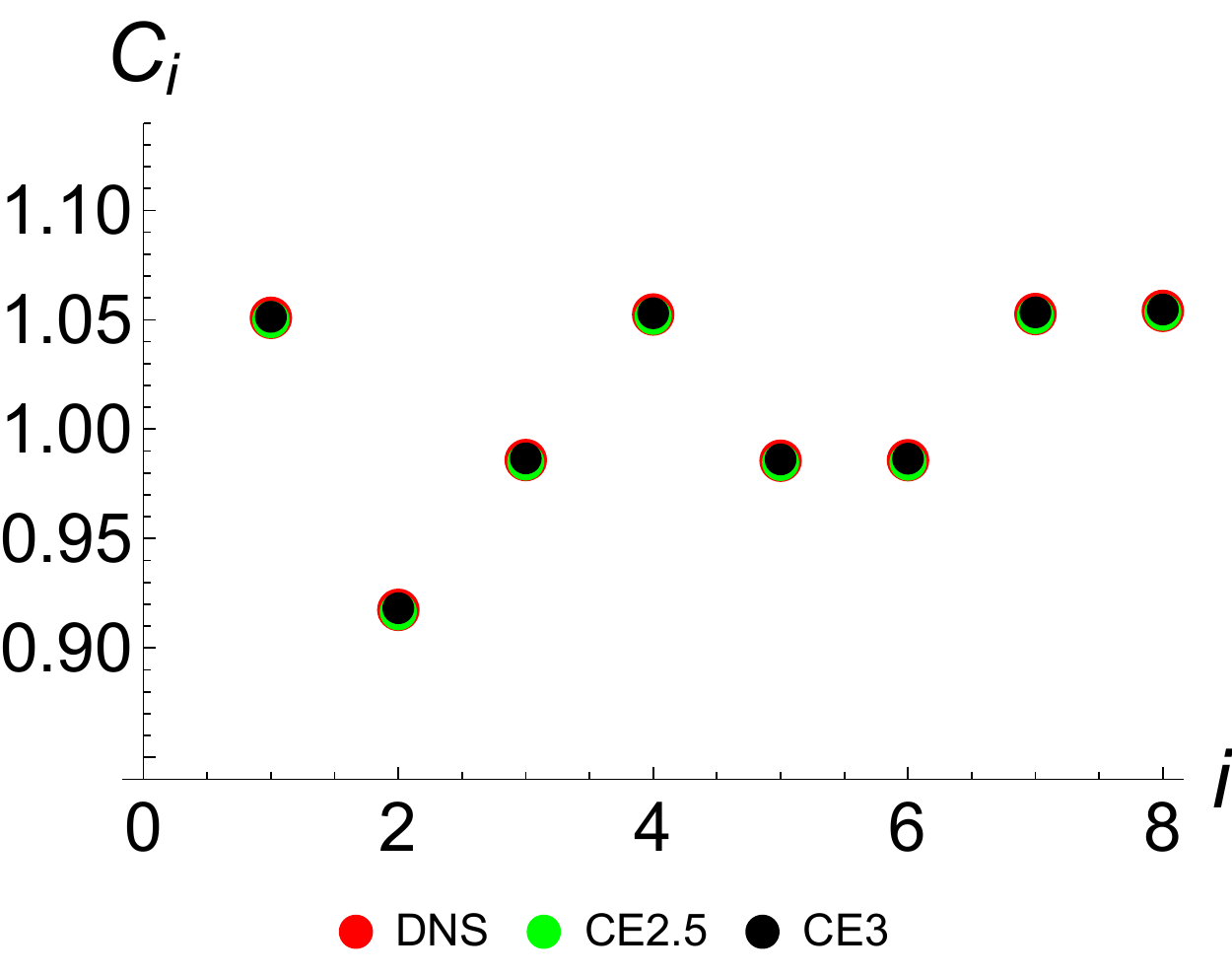}
}
\subfigure[]
{
	\includegraphics[width=0.45\hsize]{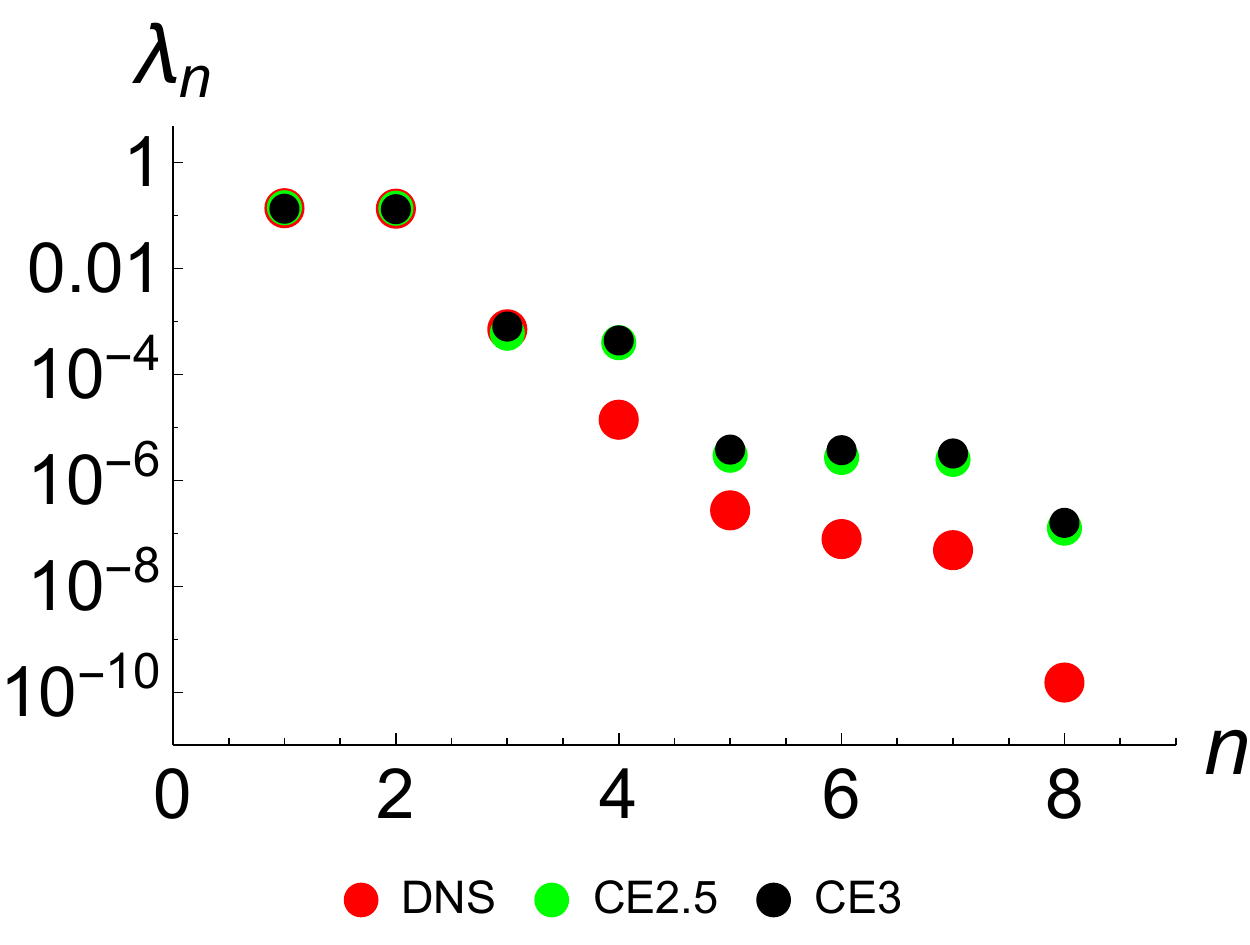}
}
\subfigure[]
{
	\includegraphics[width=0.45\hsize]{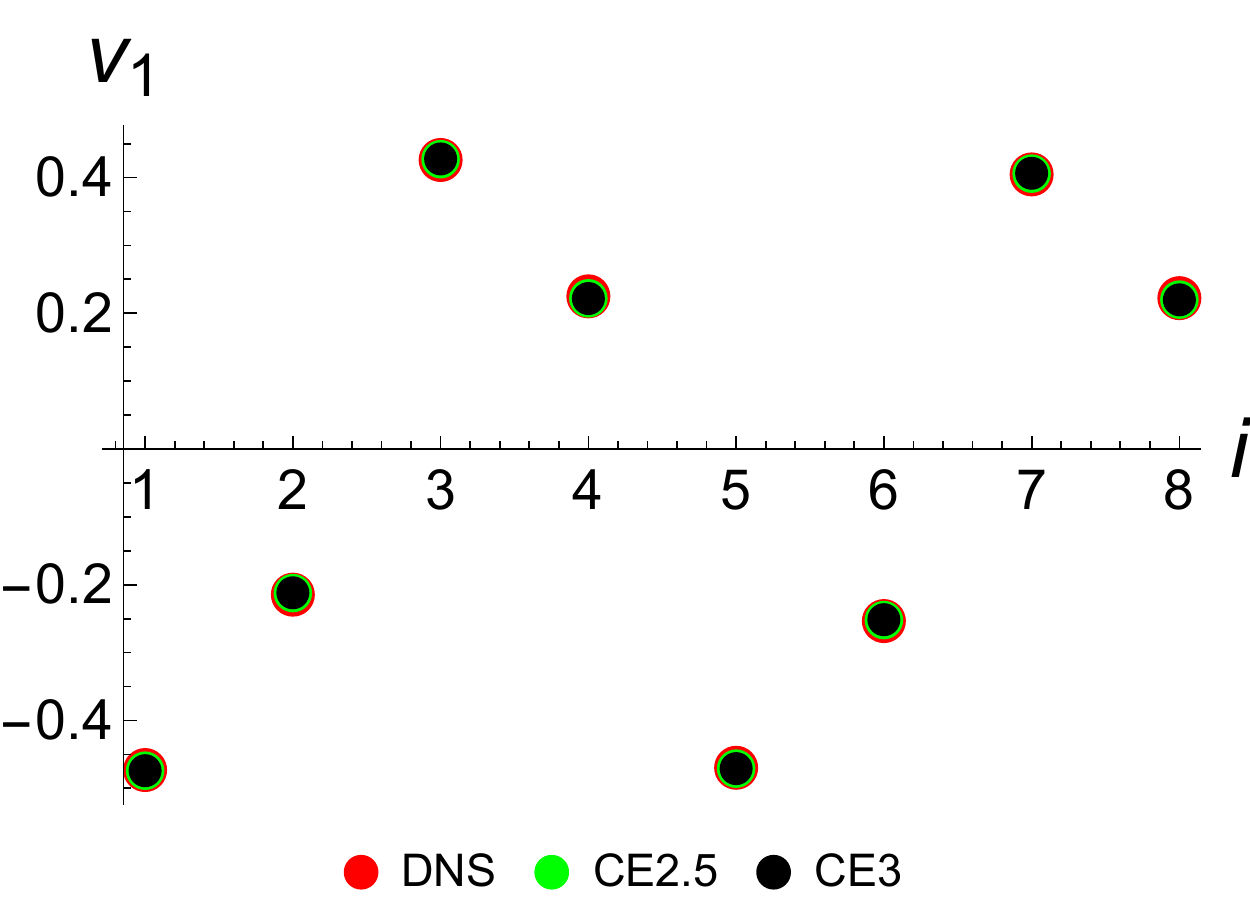}
}
\subfigure[]
{
	\includegraphics[width=0.45\hsize]{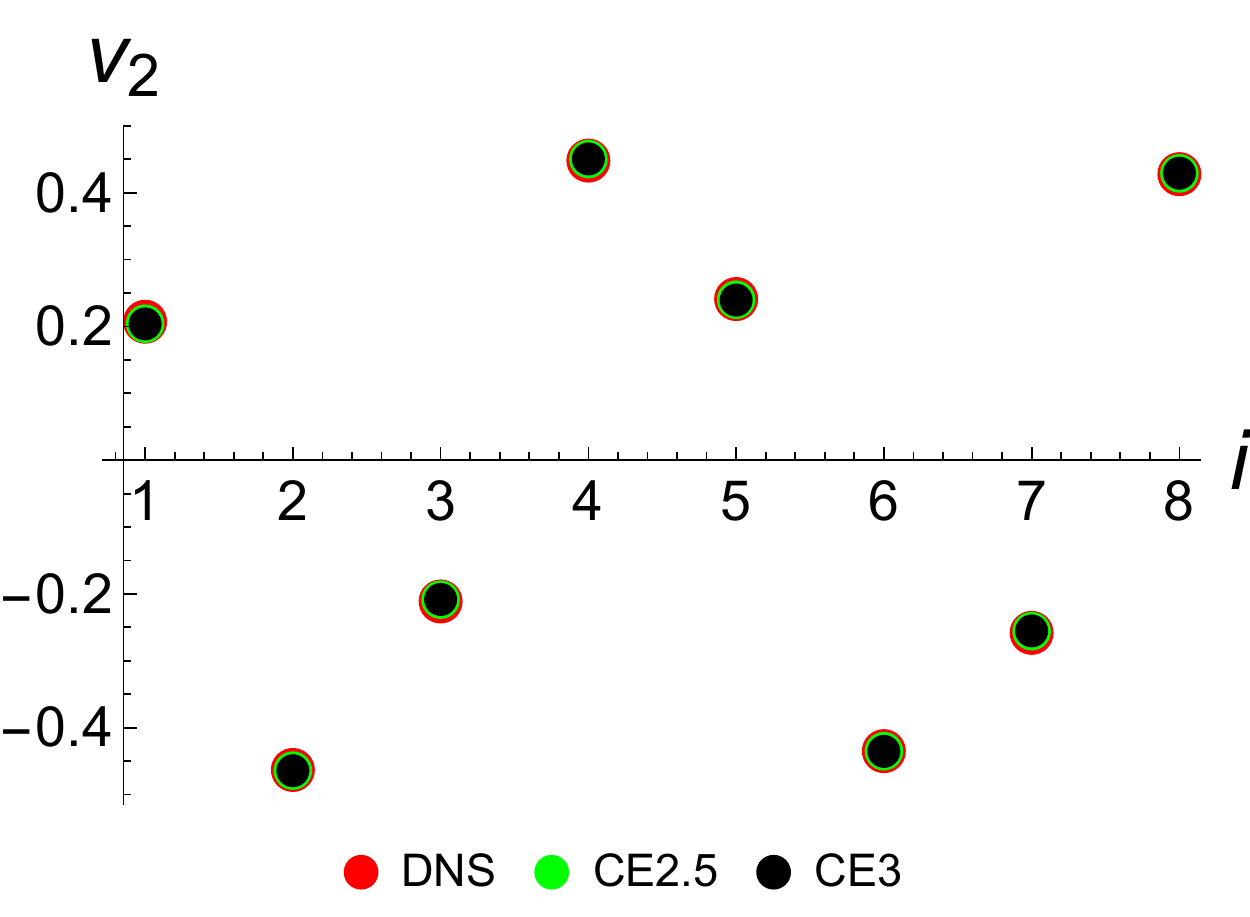}
}
\caption{The illustration of a variety of dynamical and statistical properties obtained in DNS and DSS for the Lorenz96 system in the periodic state for $c =1.2$ and $f_{i}=1.02$ for $i=2,3,\cdots 8$, where the eddy damping parameter is chosen to be $\tau_d=0.1$ for both CE2.5 and CE3 computation. The time series of the dynamical model in DNS is shown in (a), the first cumulants, $C_{x_i}$ obtained in DNS and DSS are plotted in (b), the eigenvalues of the covariance matrix in DNS and DSS are in (c) and the plots of the first two eigenvectors of the covariance in DNS and DSS are illustrated in (d)--(e).}
\label{inho2}
\end{figure}
\begin{figure}
\centering
\subfigure[]
{
	\includegraphics[width=0.45\hsize]{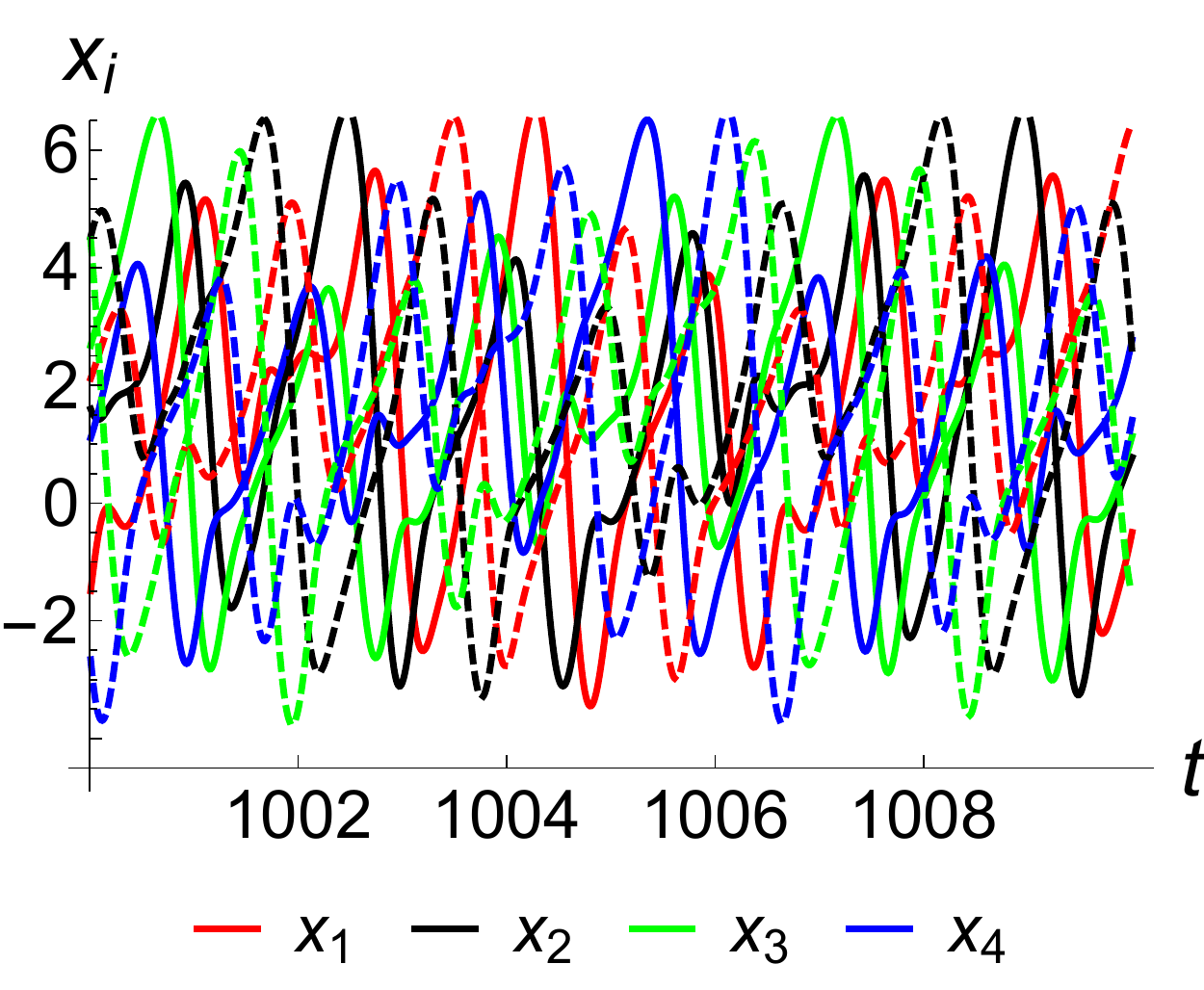}
}
\subfigure[]
{
	\includegraphics[width=0.45\hsize]{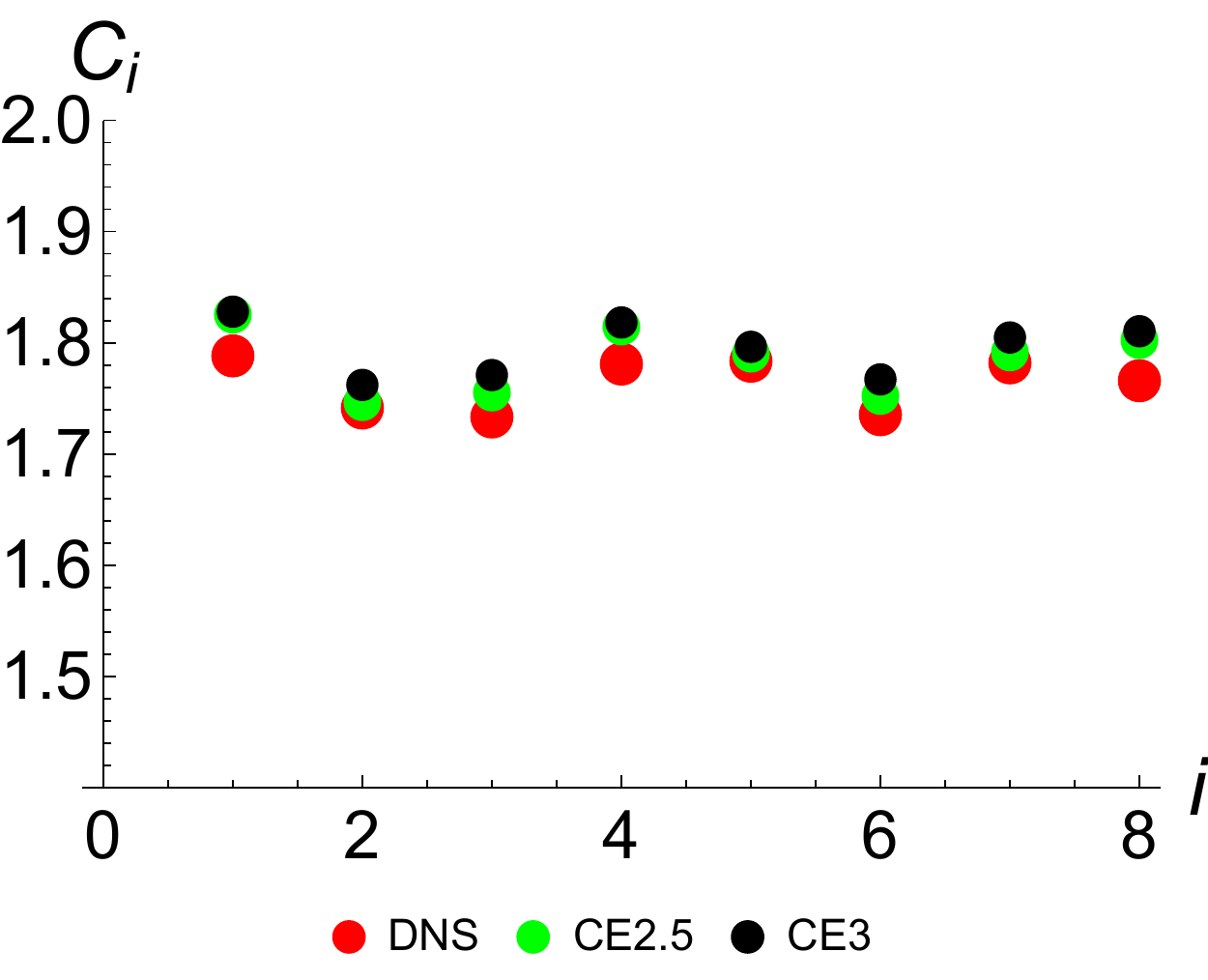}
}
\subfigure[]
{
	\includegraphics[width=0.45\hsize]{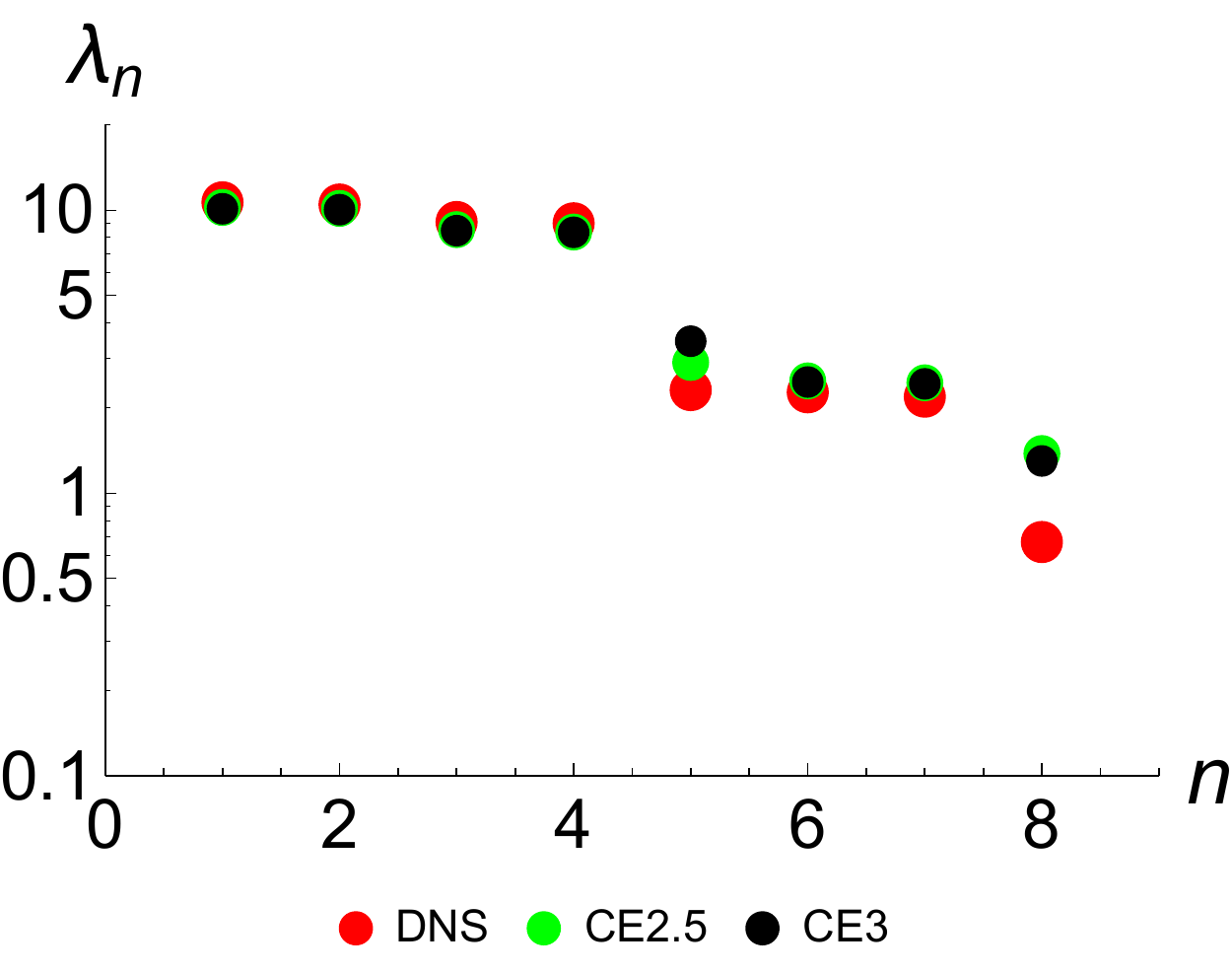}
}
\subfigure[]
{
	\includegraphics[width=0.45\hsize]{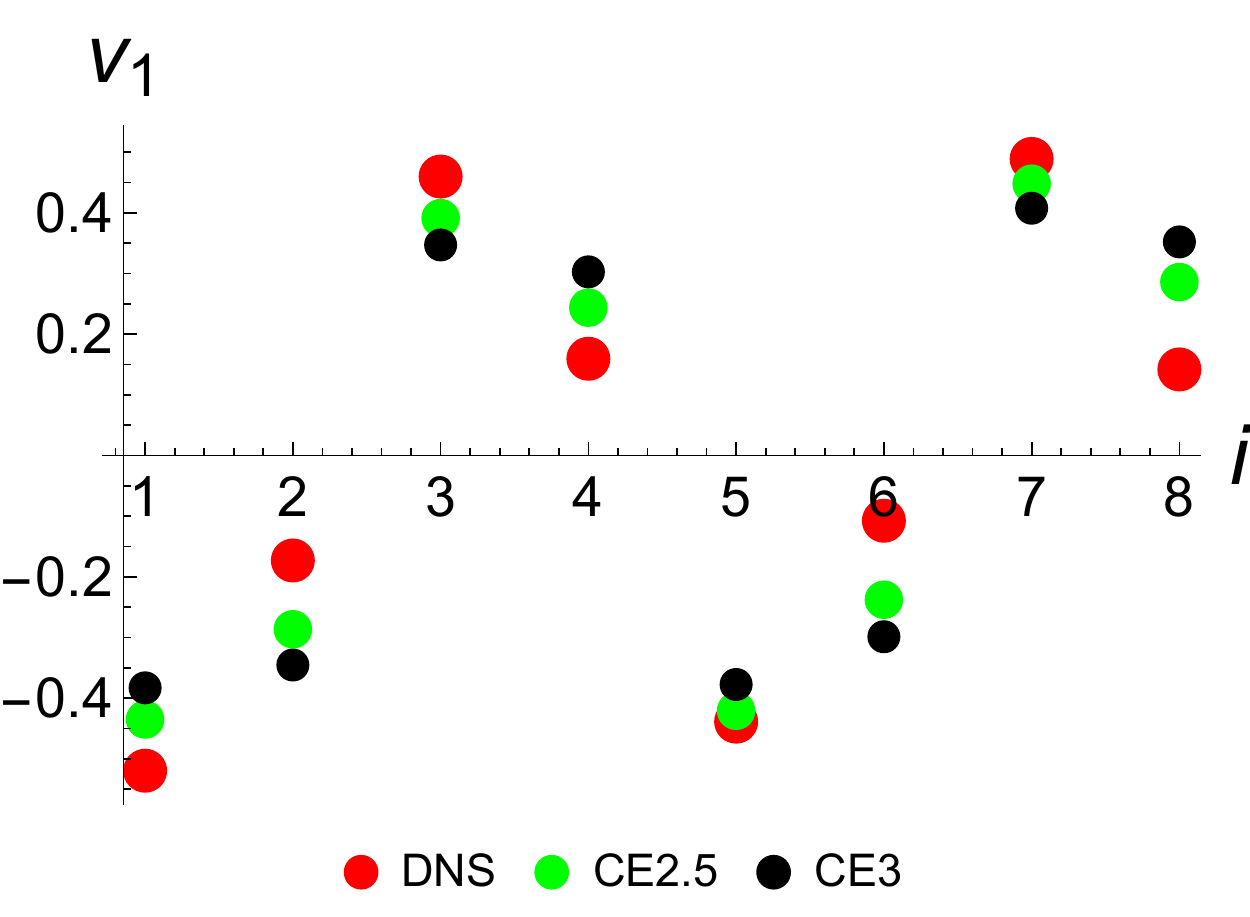}
}
\subfigure[]
{
	\includegraphics[width=0.45\hsize]{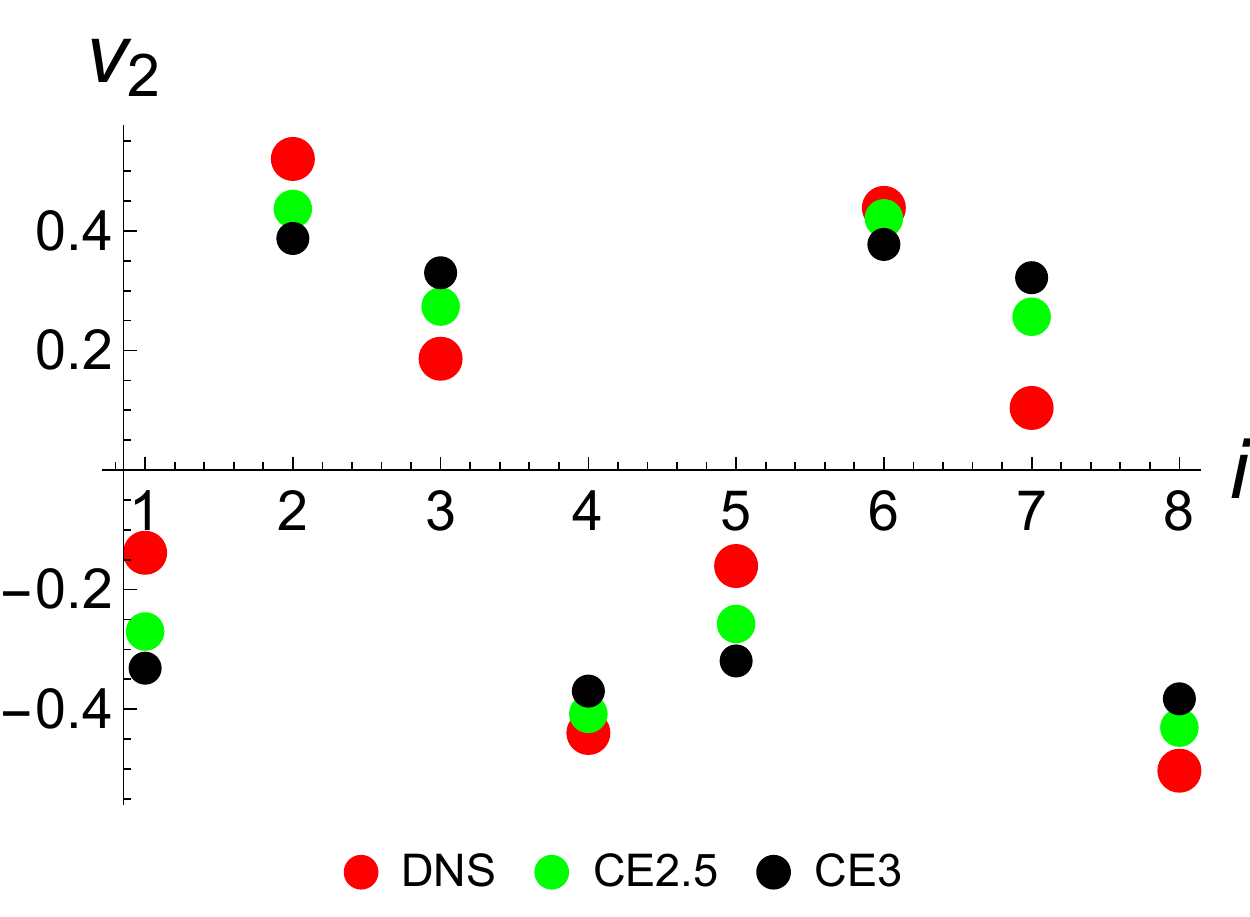}
}
\subfigure[]
{
	\includegraphics[width=0.45\hsize]{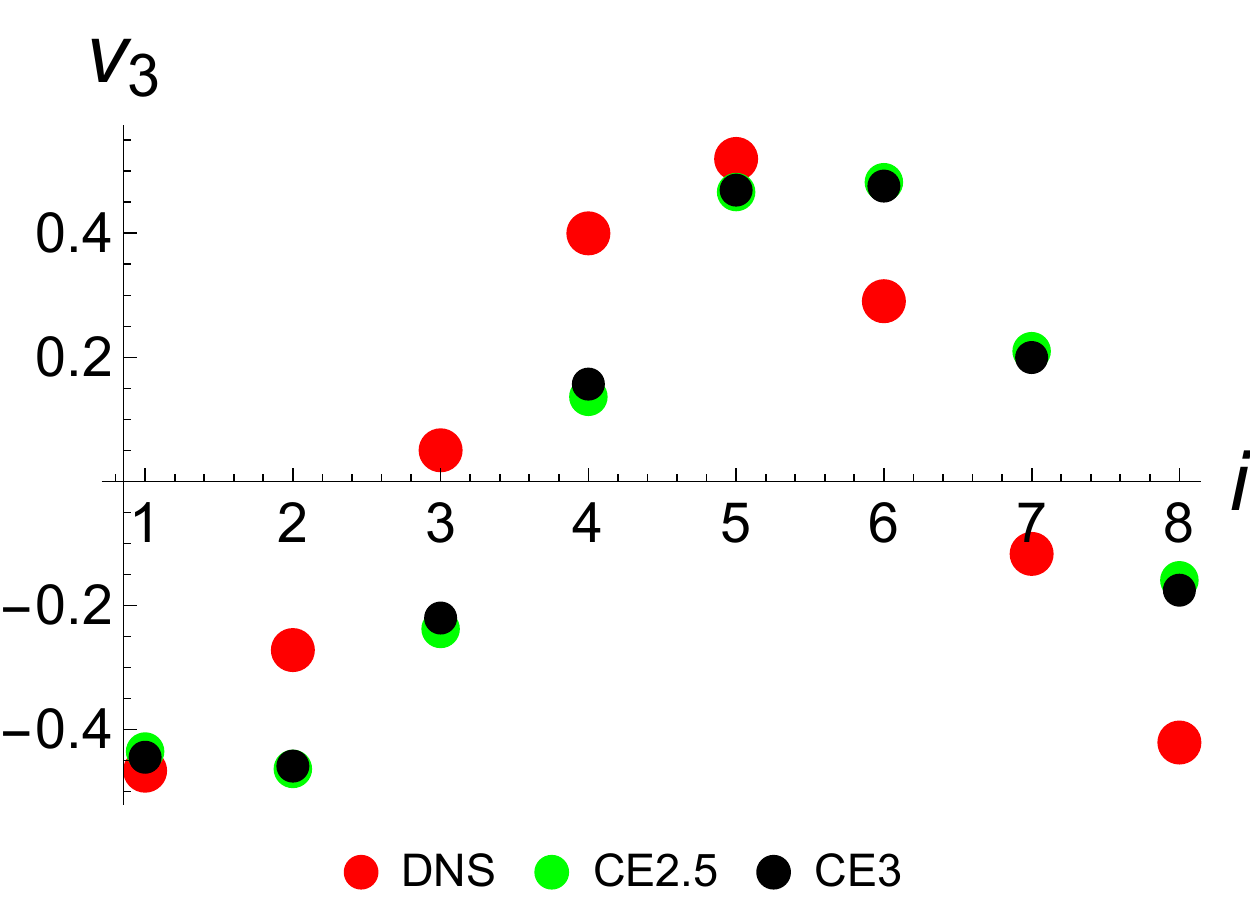}
}
\caption{The illustration of a variety of dynamical and statistical properties obtained in DNS and DSS for the Lorenz96 system in the periodic state for $c =1.2$ and $f_{i}=5$ for $i=2,3,\cdots 8$, where the eddy damping parameter is chosen to be $\tau_d^{-1}=10$ for both CE2.5 and $\tau_d^{-1}=12.5$ for CE3 computation. The time series of the dynamical model in DNS is shown in (a), the first cumulants, $C_{x_i}$ obtained in DNS and DSS are plotted in (b), the eigenvalues of the covariance matrix in DNS and DSS are in (c) and the plots of the first three eigenvectors of the covariance in DNS and DSS are illustrated in (d)--(f).}
\label{inho3}
\end{figure}
\begin{figure}
\centering
\subfigure[]
{
	\includegraphics[width=0.45\hsize]{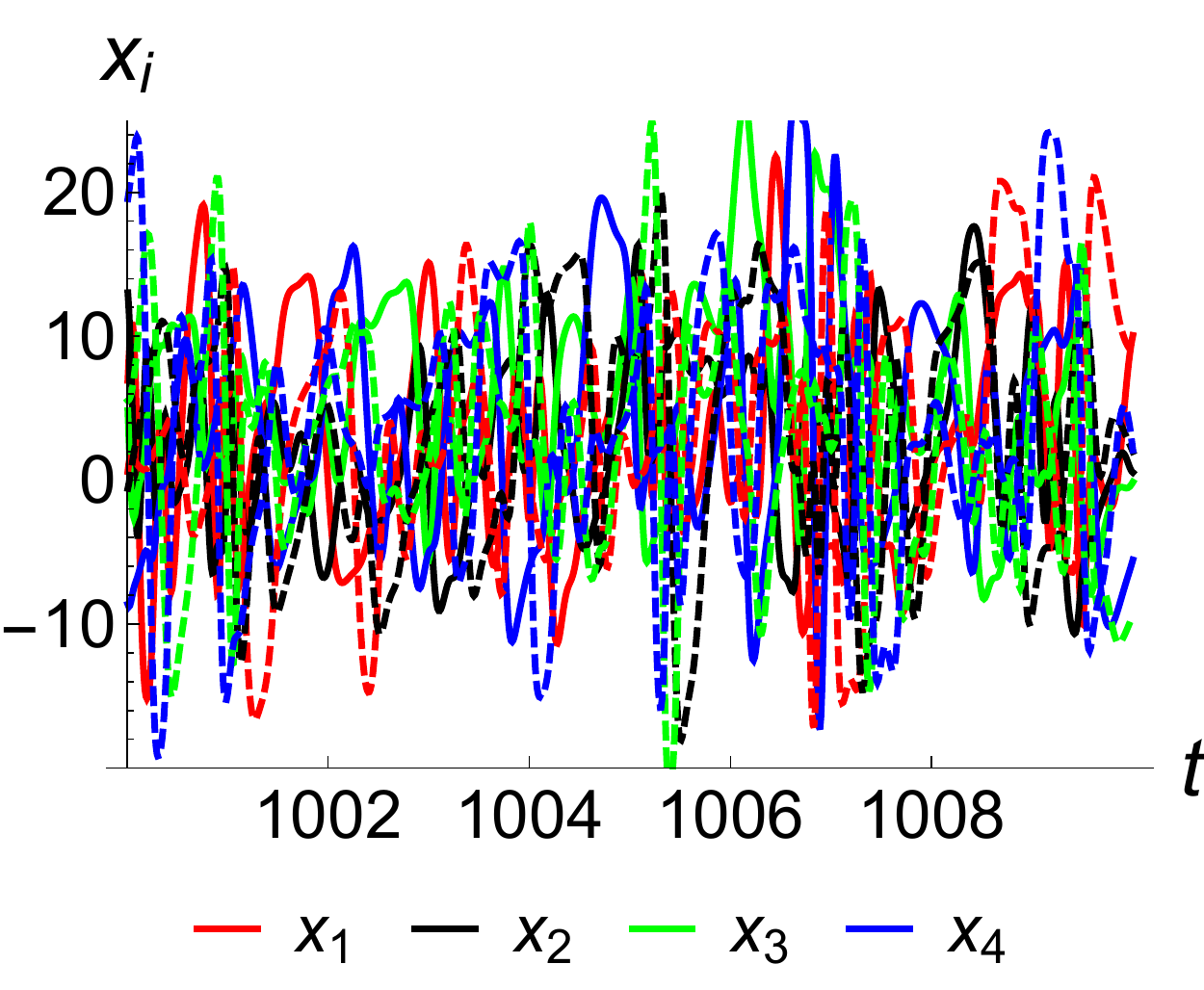}
}
\subfigure[]
{
	\includegraphics[width=0.45\hsize]{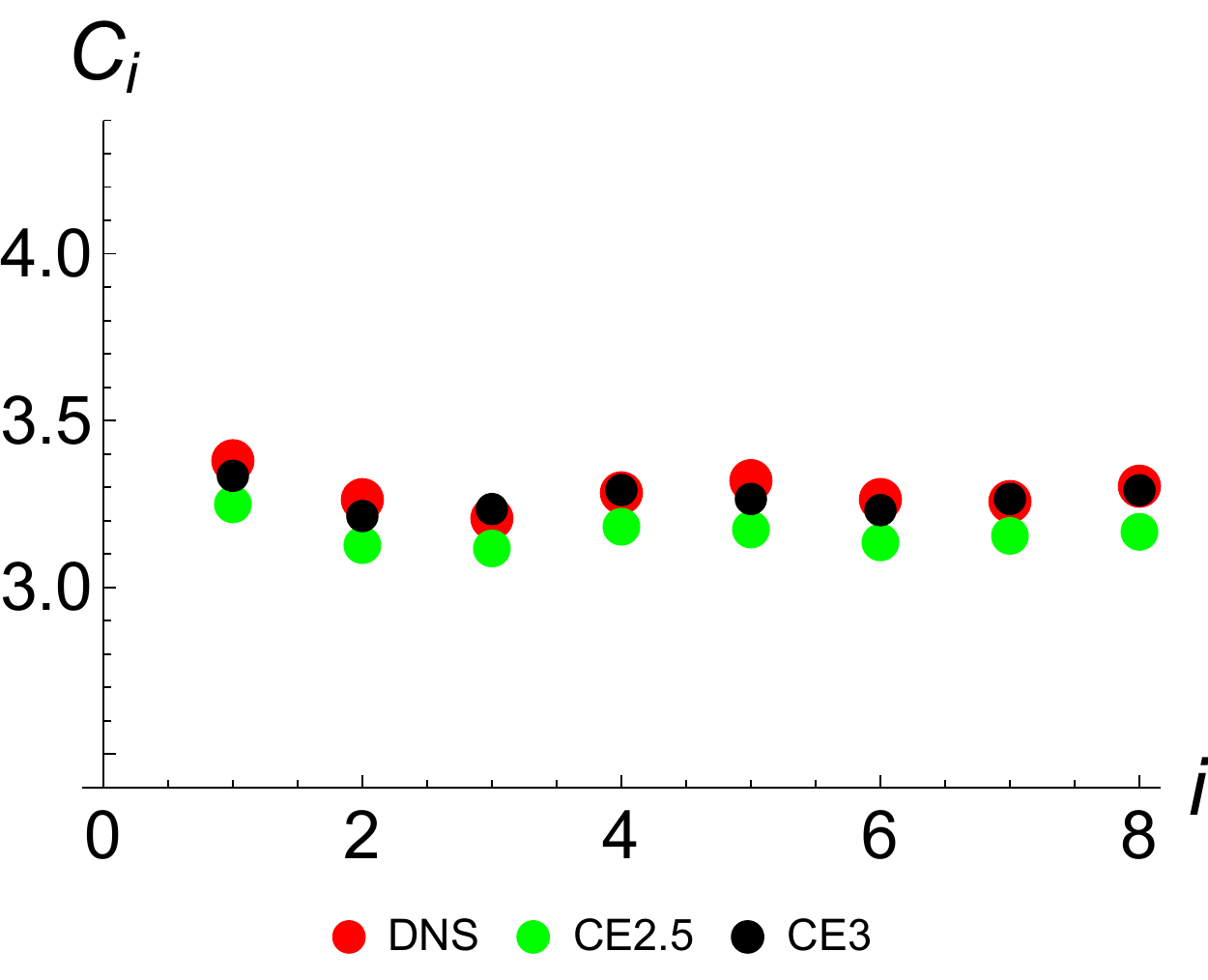}
}
\subfigure[]
{
	\includegraphics[width=0.45\hsize]{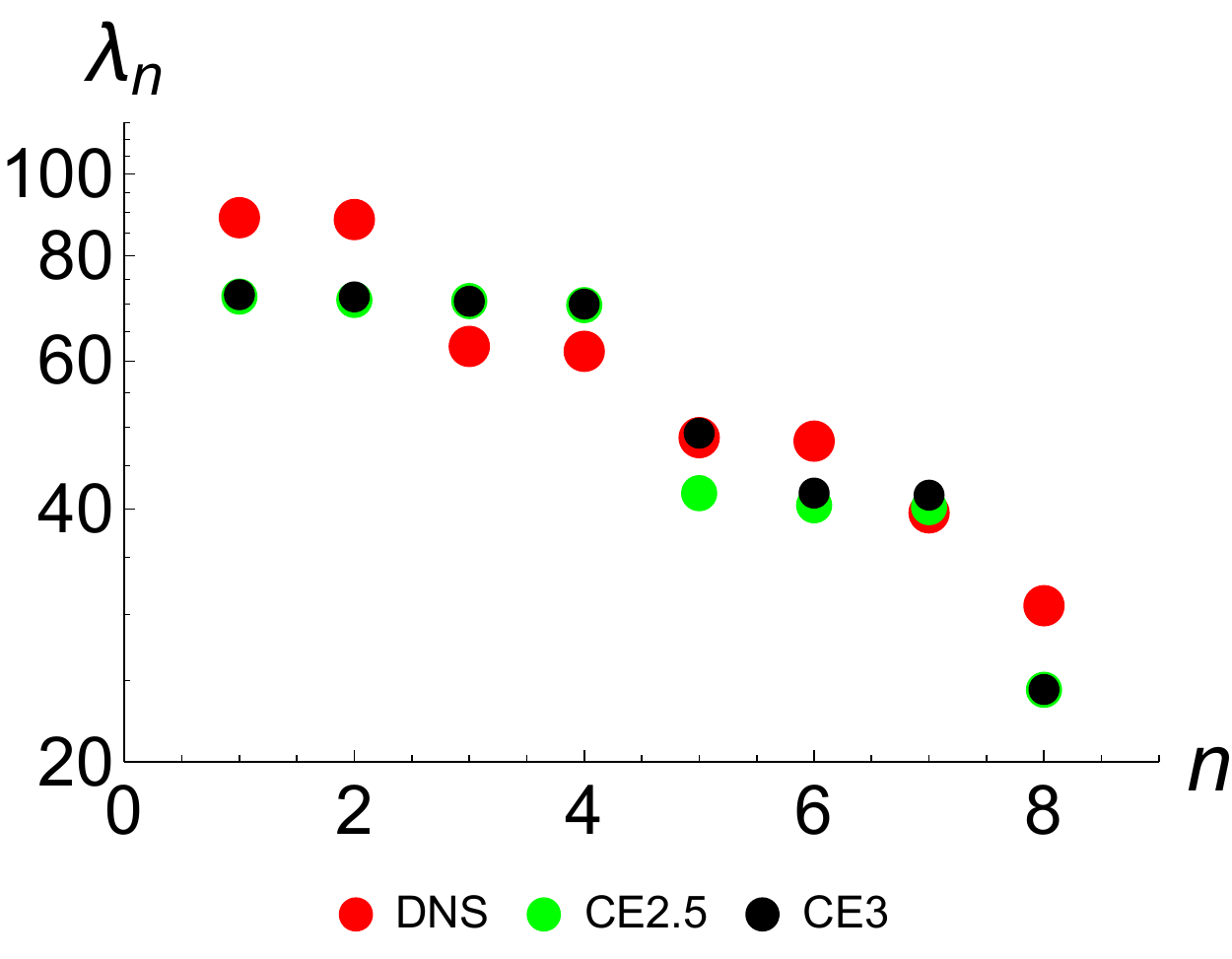}
}
\subfigure[]
{
	\includegraphics[width=0.45\hsize]{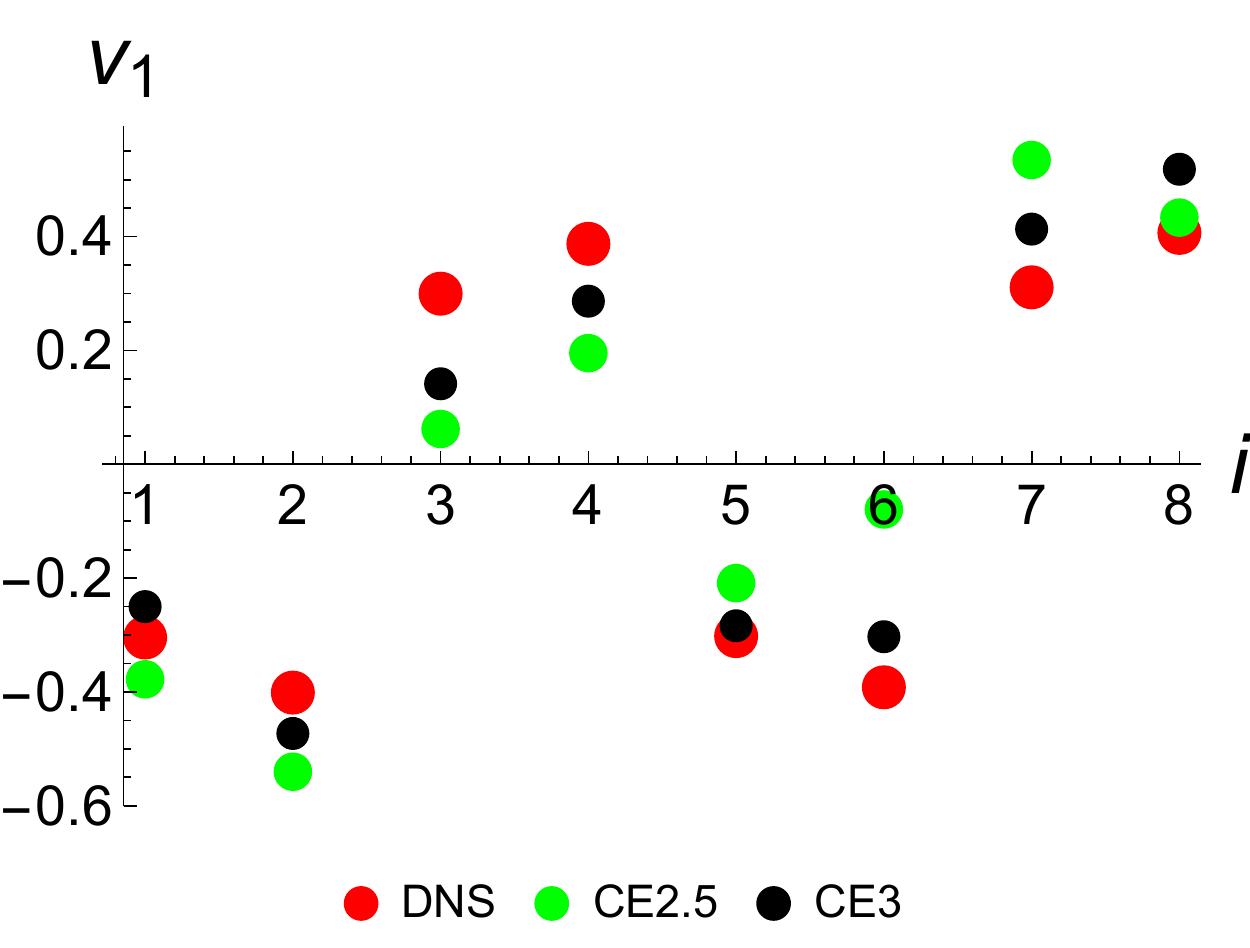}
}
\subfigure[]
{
	\includegraphics[width=0.45\hsize]{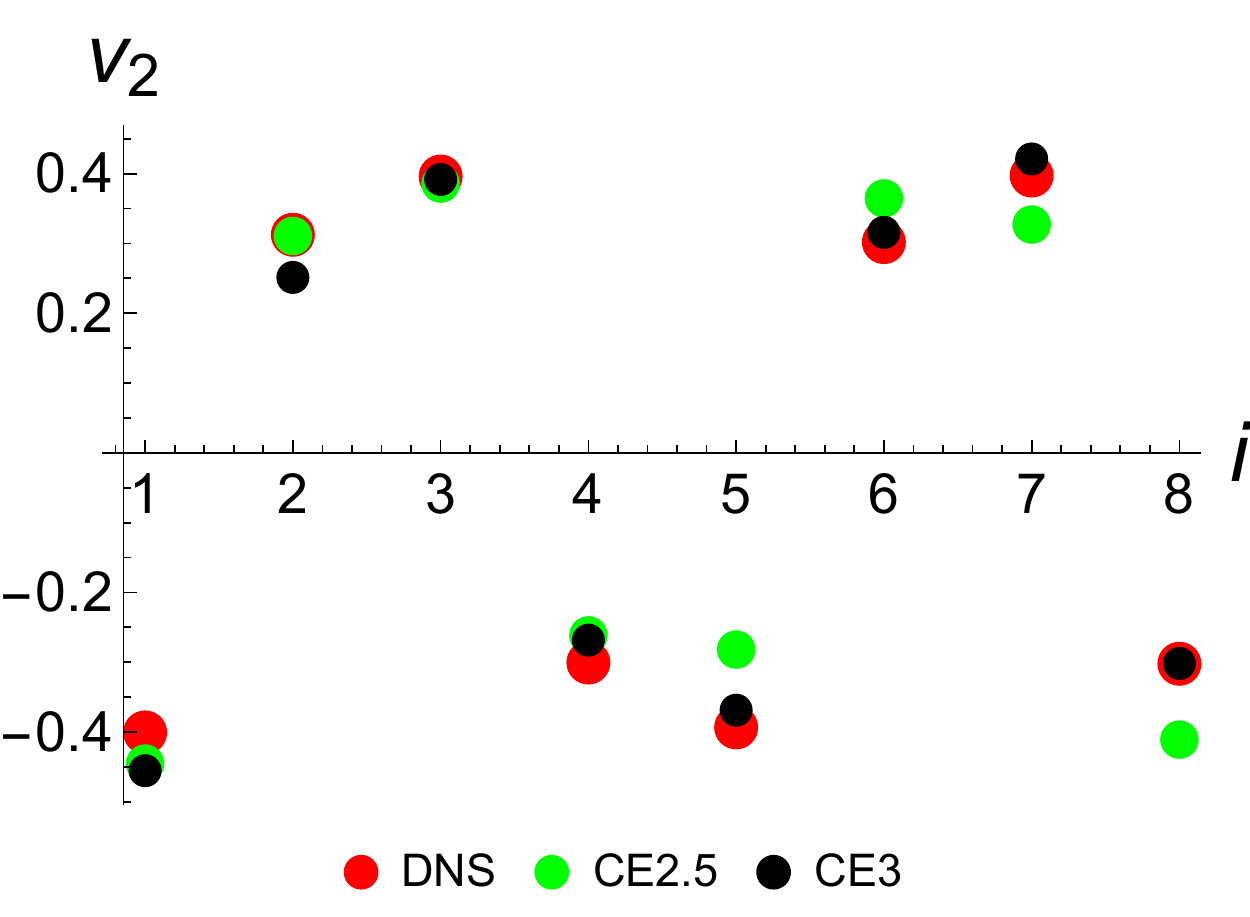}
}
\subfigure[]
{
	\includegraphics[width=0.45\hsize]{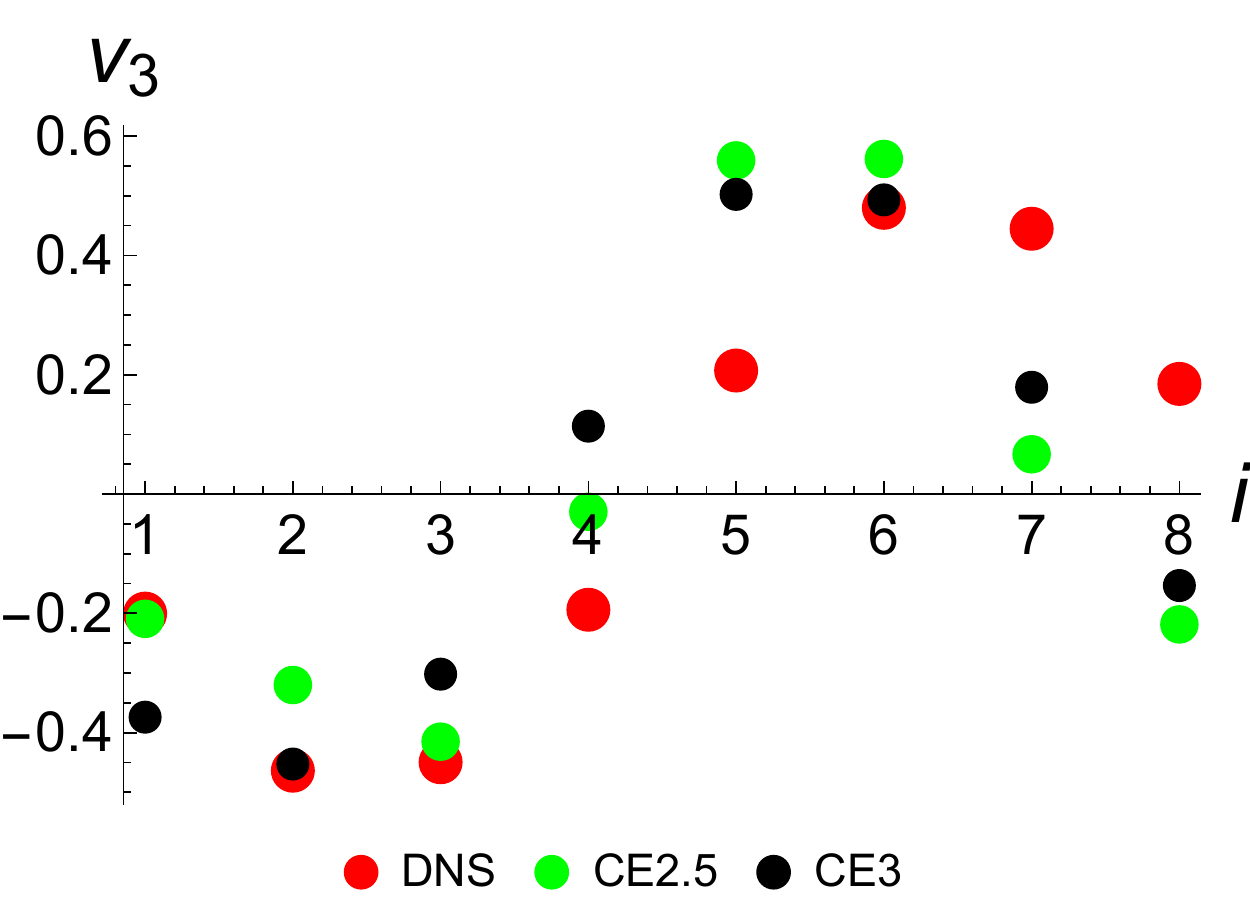}
}
\caption{The illustration of a variety of dynamical and statistical properties obtained in DNS and DSS for the Lorenz96 system in the periodic state for $c =1.05$ and $f_{i}=20$ for $i=2,3,\cdots 8$, where the eddy damping parameter is chosen to be $\tau_d^{-1}=20$ for both CE2.5 and CE3 computation. The time series of the dynamical model in DNS is shown in (a), the first cumulants, $C_{x_i}$ obtained in DNS and DSS are plotted in (b), the eigenvalues of the covariance matrix in DNS and DSS are in (c) and the plots of the first three eigenvectors of the covariance in DNS and DSS are illustrated in (d)--(f).}
\label{inho4}
\end{figure}
\begin{figure}
\centering
\subfigure[]
{
	\includegraphics[width=0.45\hsize]{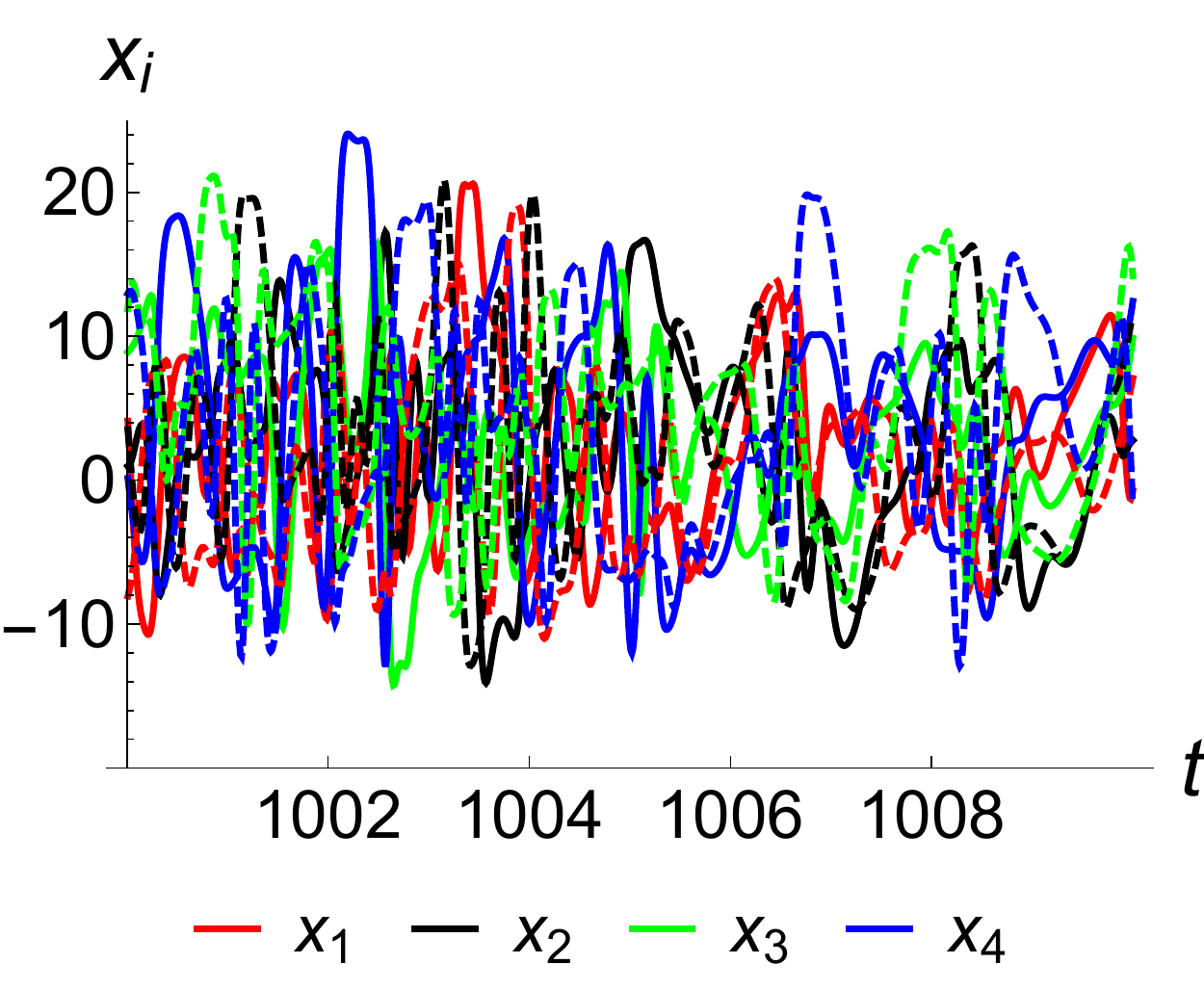}
}
\subfigure[]
{
	\includegraphics[width=0.45\hsize]{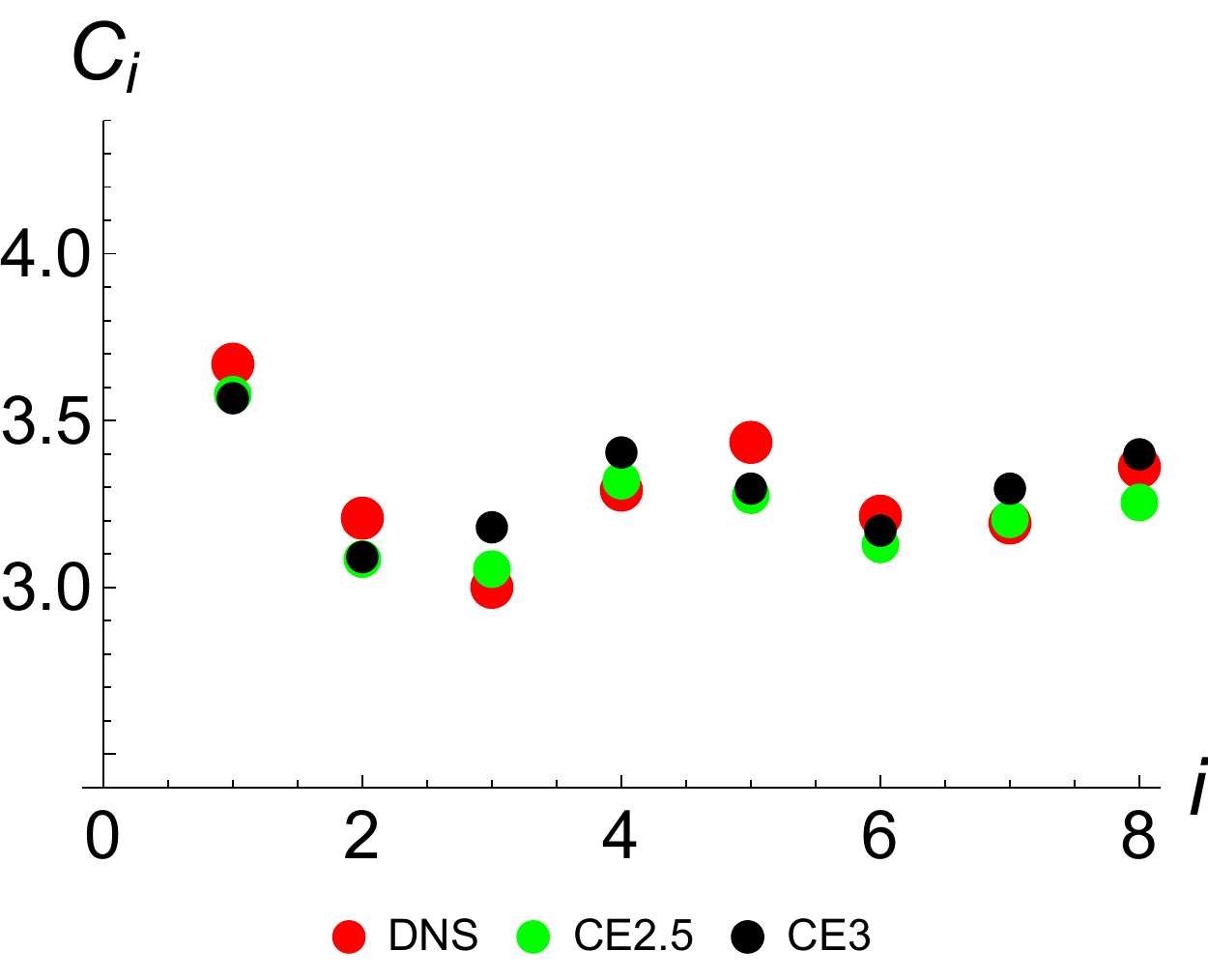}
}
\subfigure[]
{
	\includegraphics[width=0.45\hsize]{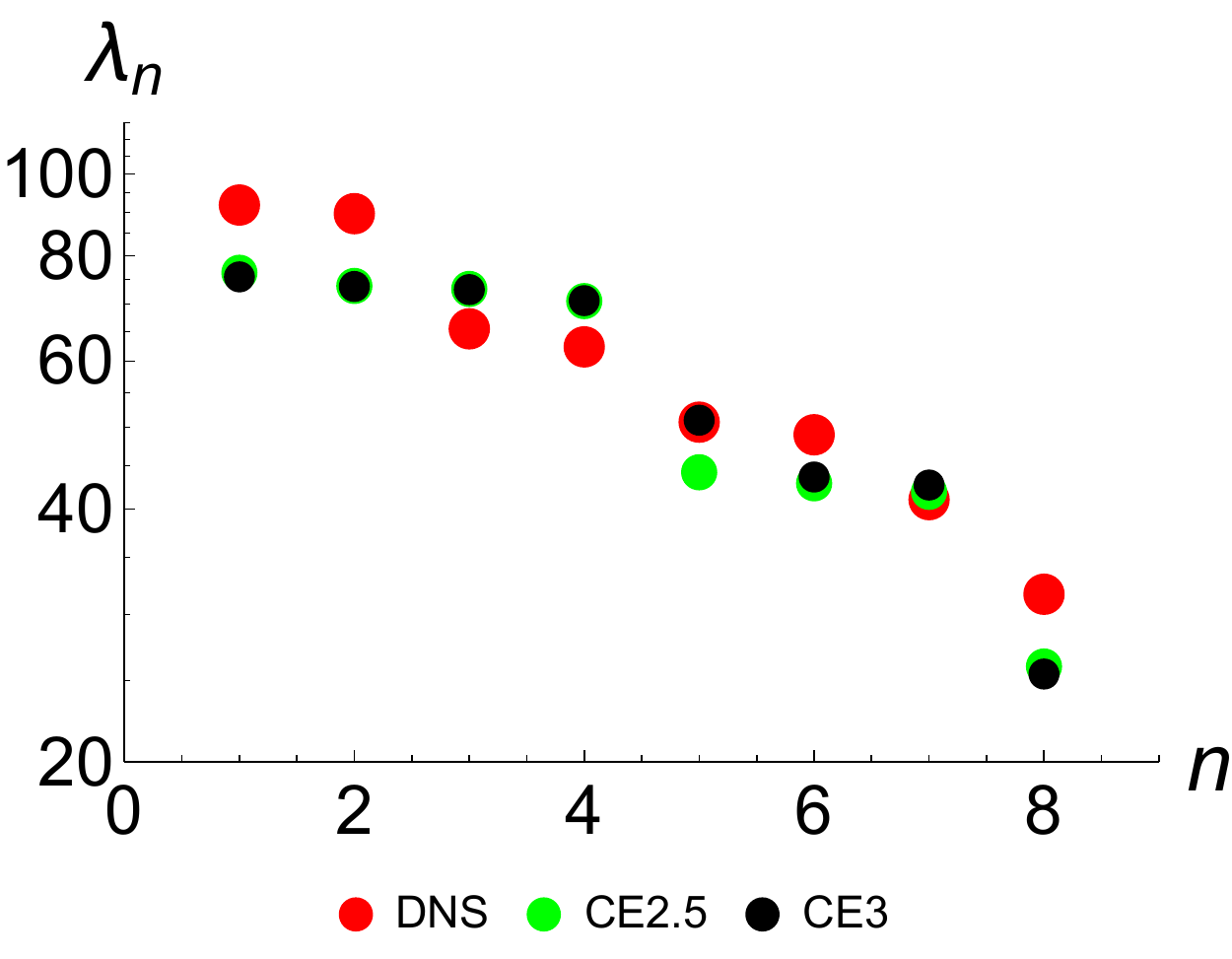}
}
\subfigure[]
{
	\includegraphics[width=0.45\hsize]{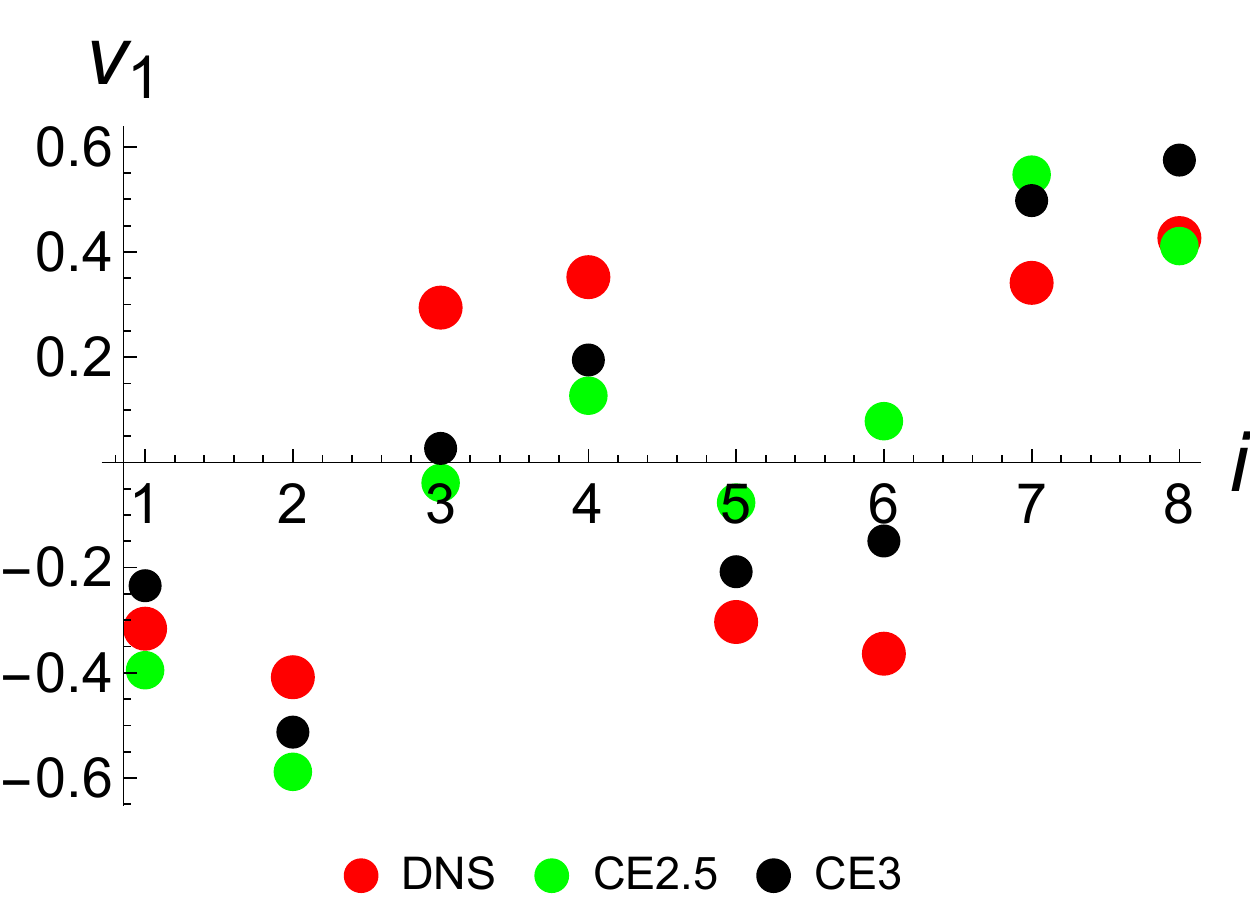}
}
\subfigure[]
{
	\includegraphics[width=0.45\hsize]{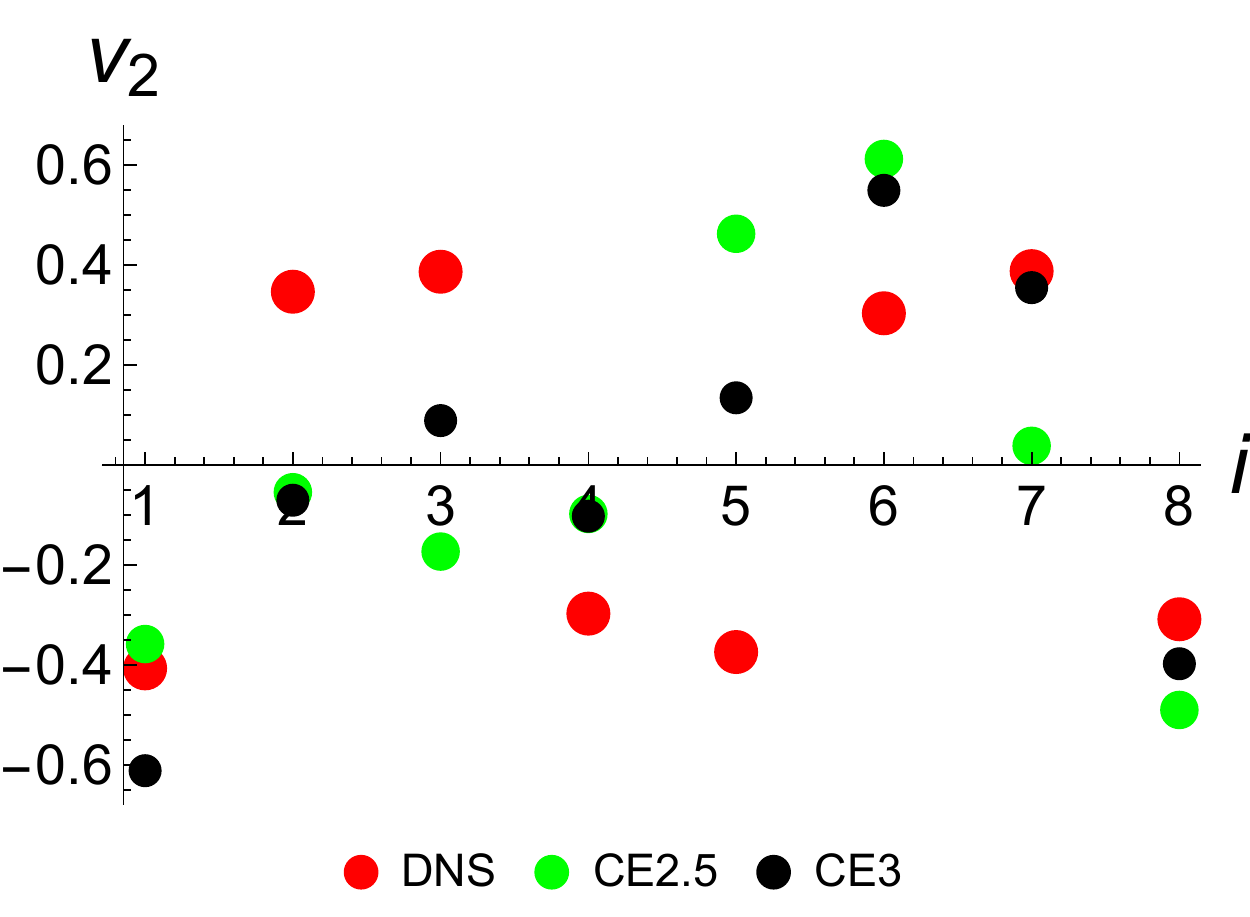}
}
\subfigure[]
{
	\includegraphics[width=0.45\hsize]{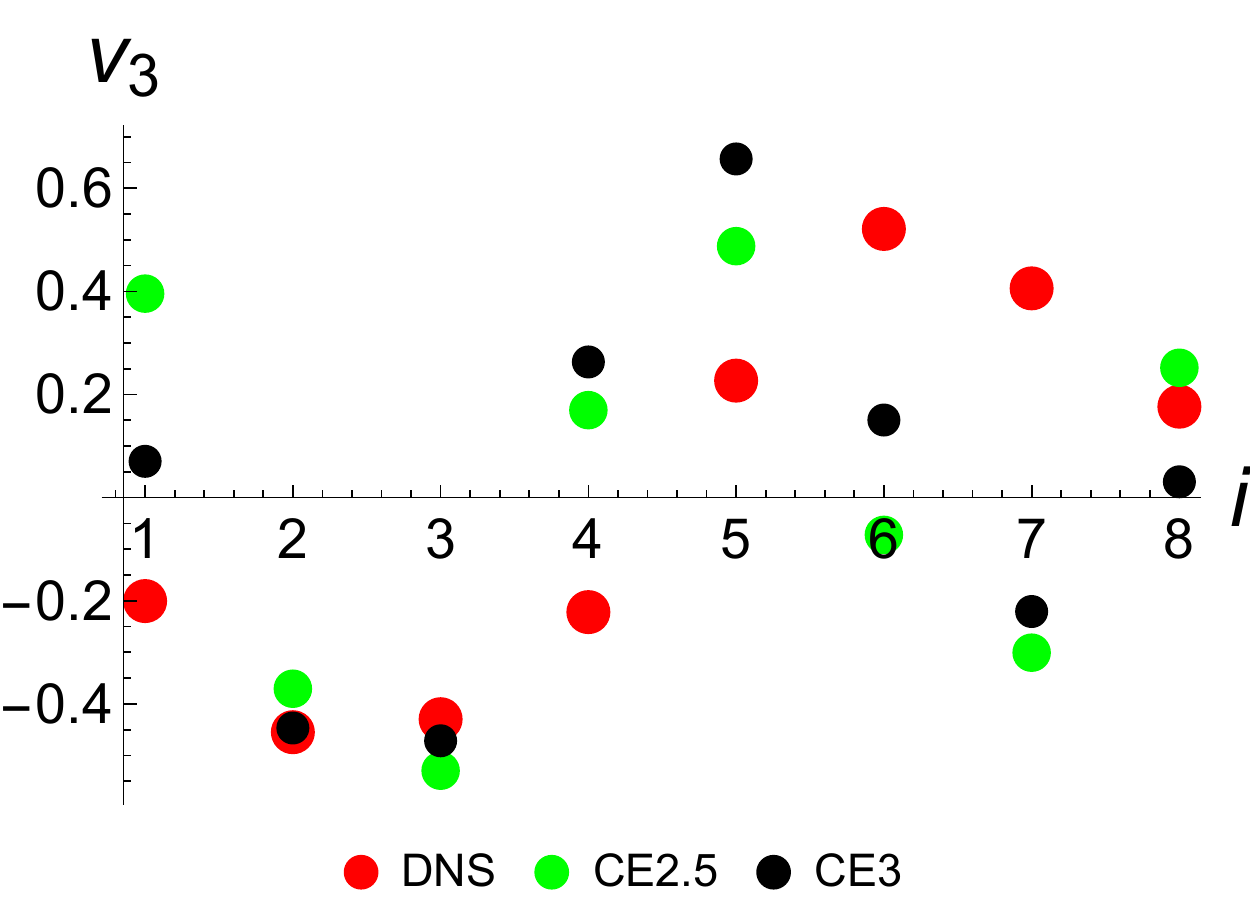}
}
\caption{The illustration of a variety of dynamical and statistical properties obtained in DNS and DSS for the Lorenz96 system in the periodic state for $c =1.2$ and $f_{i}=20$ for $i=2,3,\cdots 8$, where the eddy damping parameter is chosen to be $\tau_d^{-1}=20$ for both CE2.5 and CE3 computation. The time series of the dynamical model in DNS is shown in (a), the first cumulants, $C_{x_i}$ obtained in DNS and DSS are plotted in (b), the eigenvalues of the covariance matrix in DNS and DSS are in (c) and the plots of the first three eigenvectors of the covariance in DNS and DSS are illustrated in (d)--(f).}
\label{inho5}
\end{figure}
\begin{figure}
\centering
\subfigure[]
{
	\includegraphics[width=0.45\hsize]{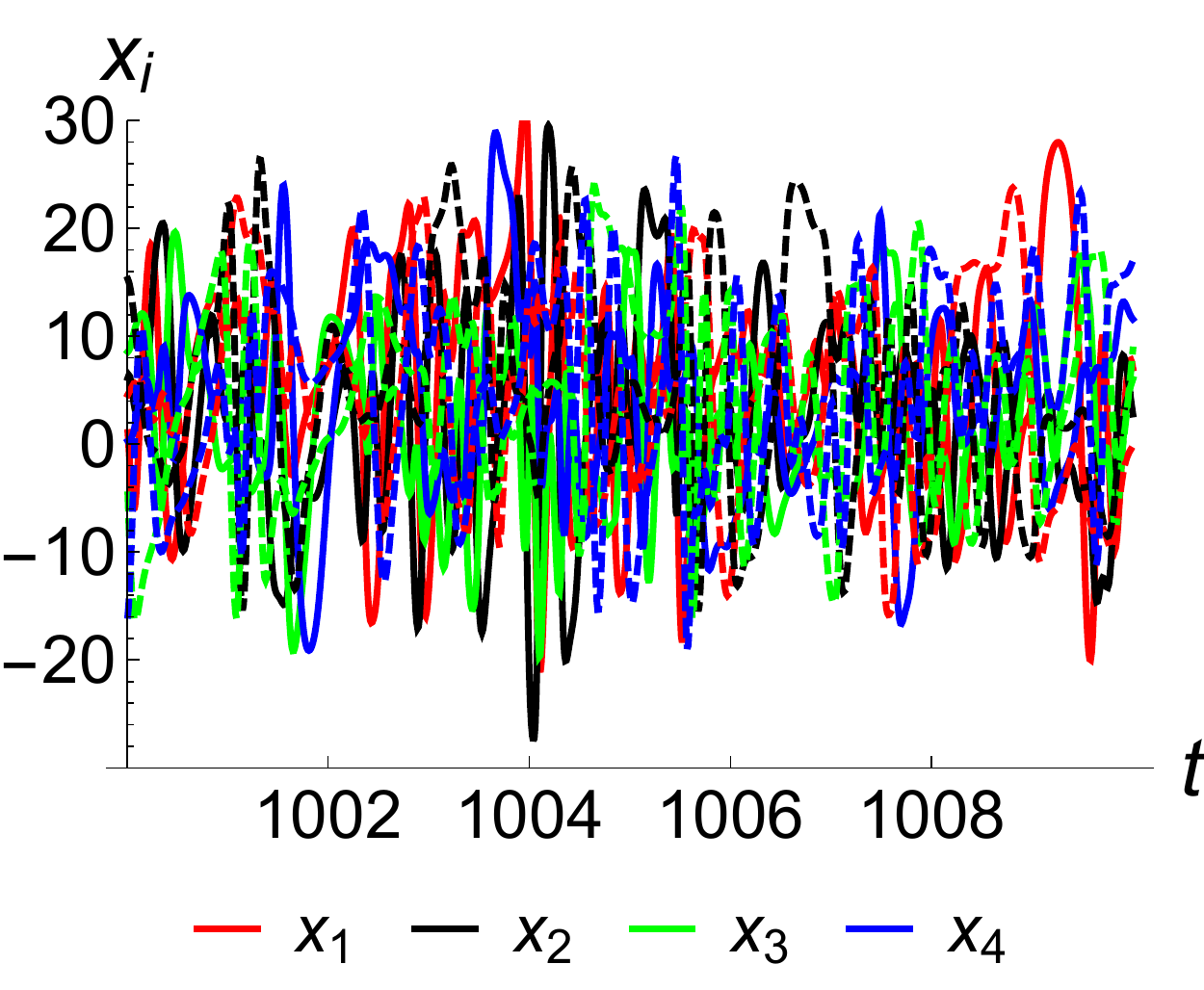}
}
\subfigure[]
{
	\includegraphics[width=0.45\hsize]{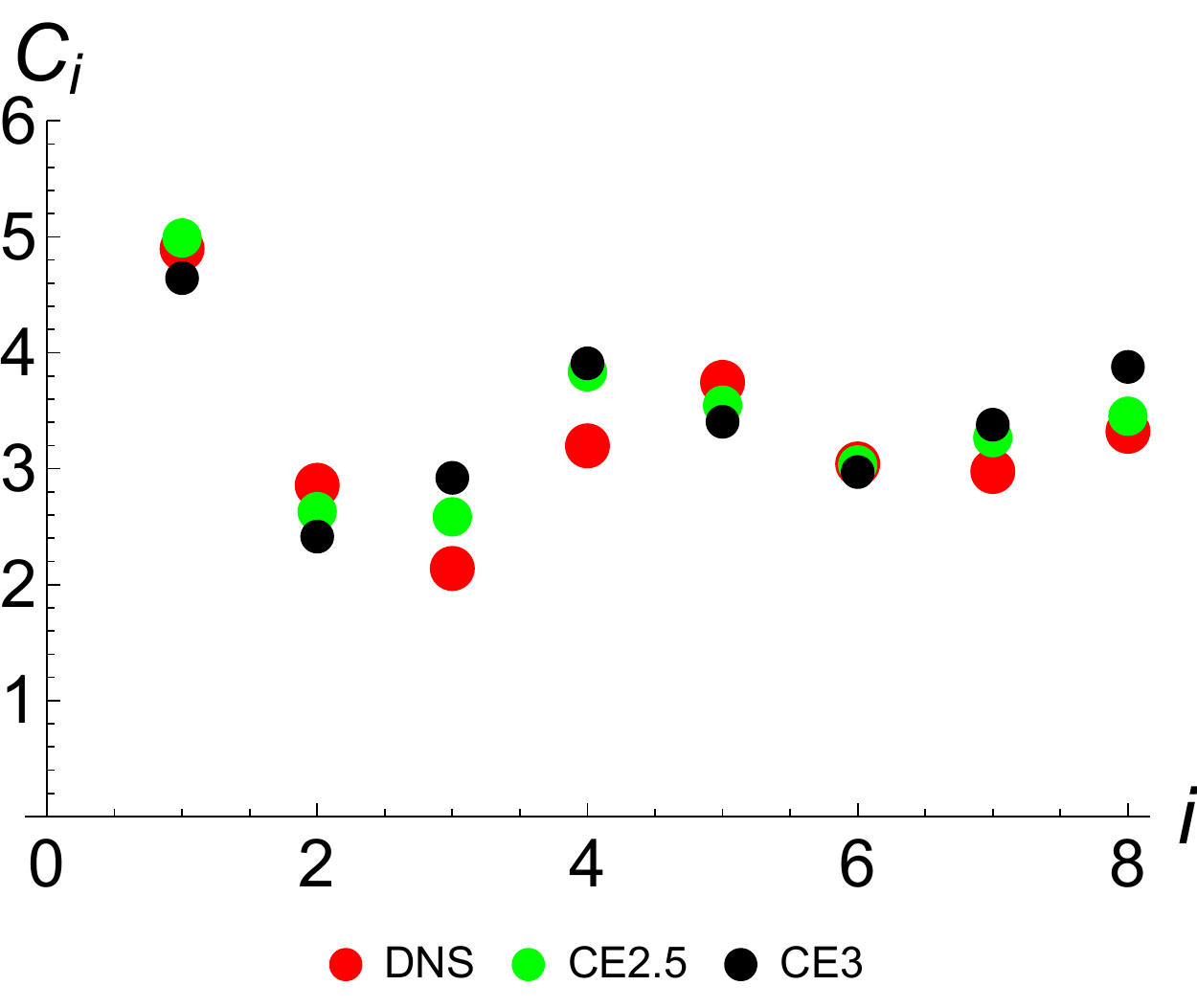}
}
\subfigure[]
{
	\includegraphics[width=0.45\hsize]{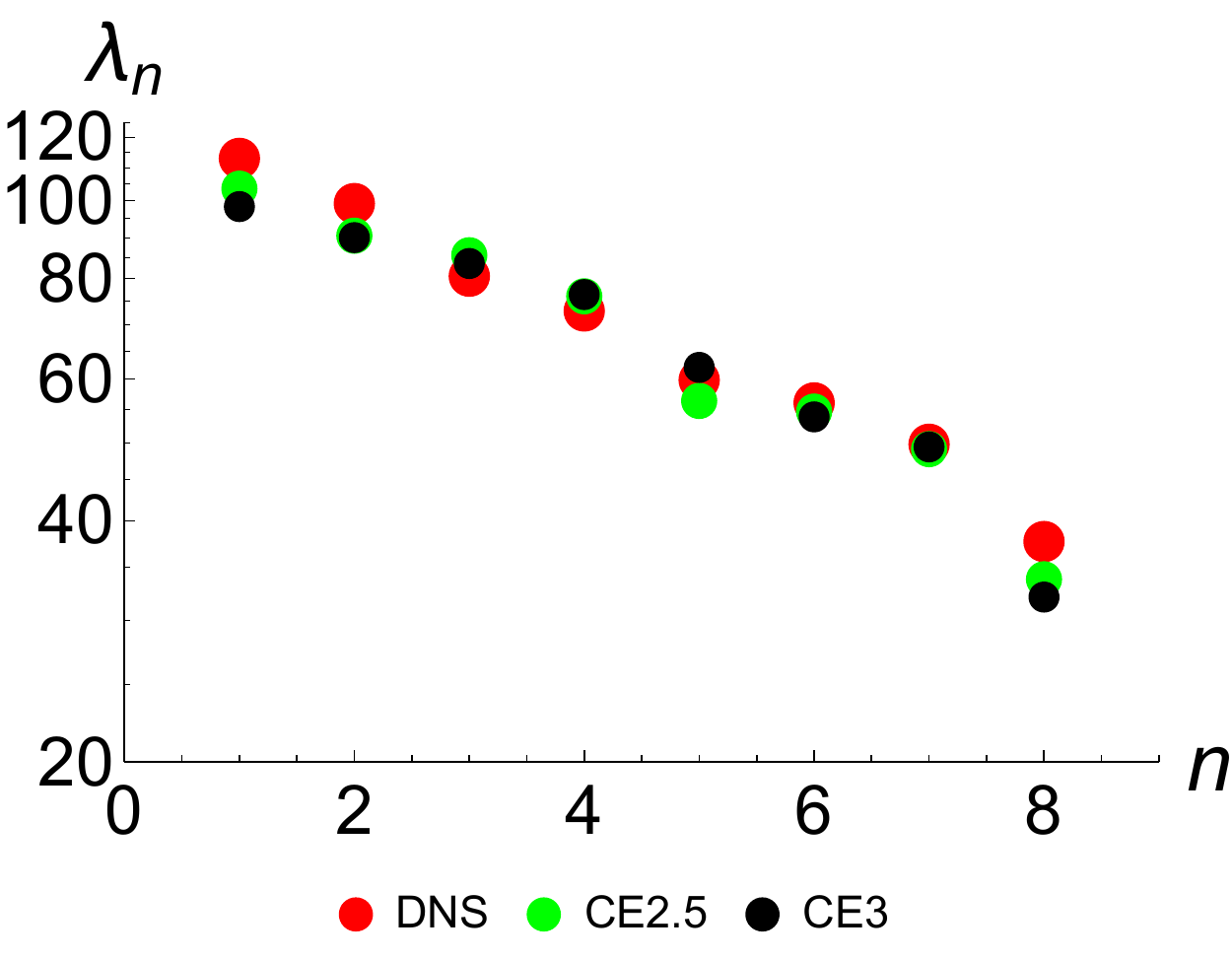}
}
\subfigure[]
{
	\includegraphics[width=0.45\hsize]{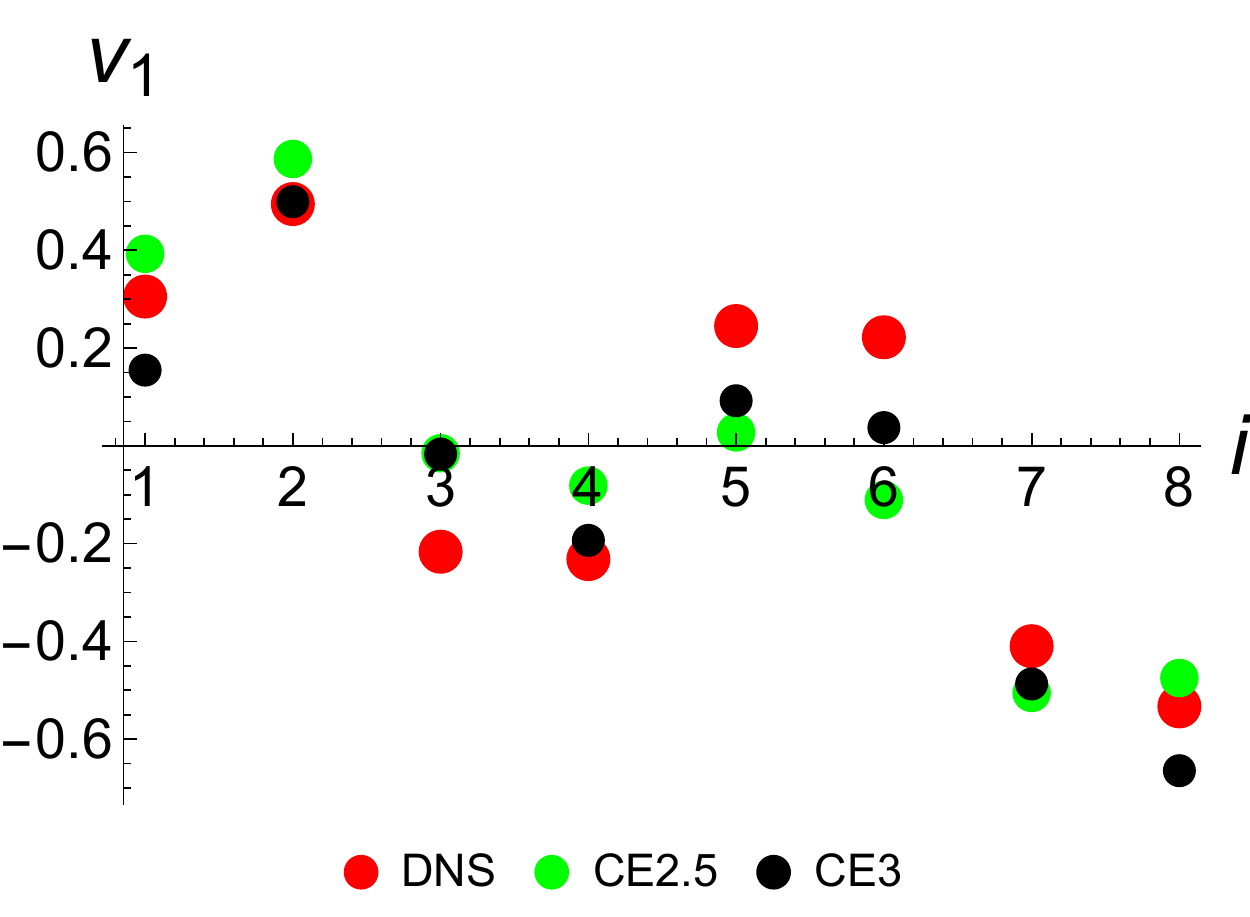}
}
\subfigure[]
{
	\includegraphics[width=0.45\hsize]{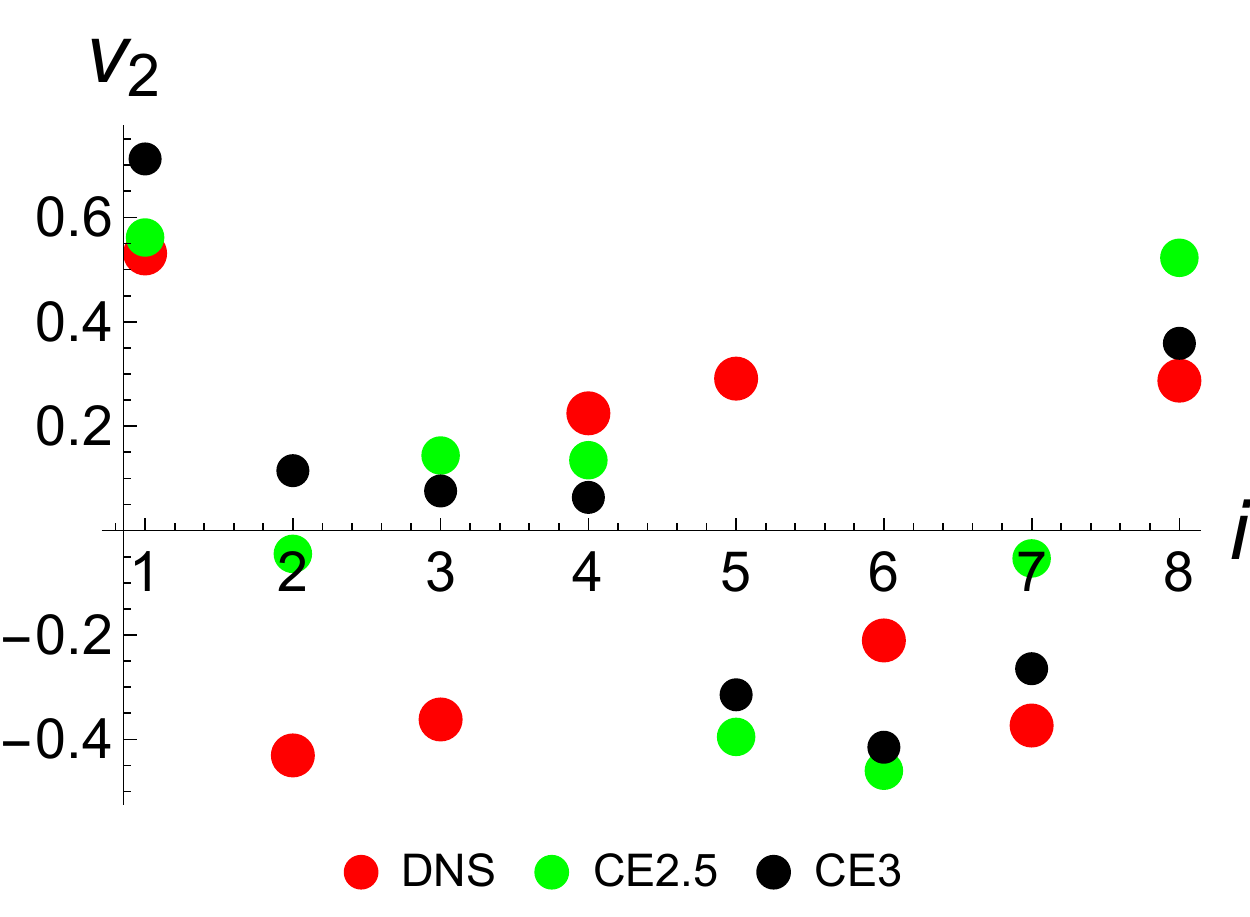}
}
\subfigure[]
{
	\includegraphics[width=0.45\hsize]{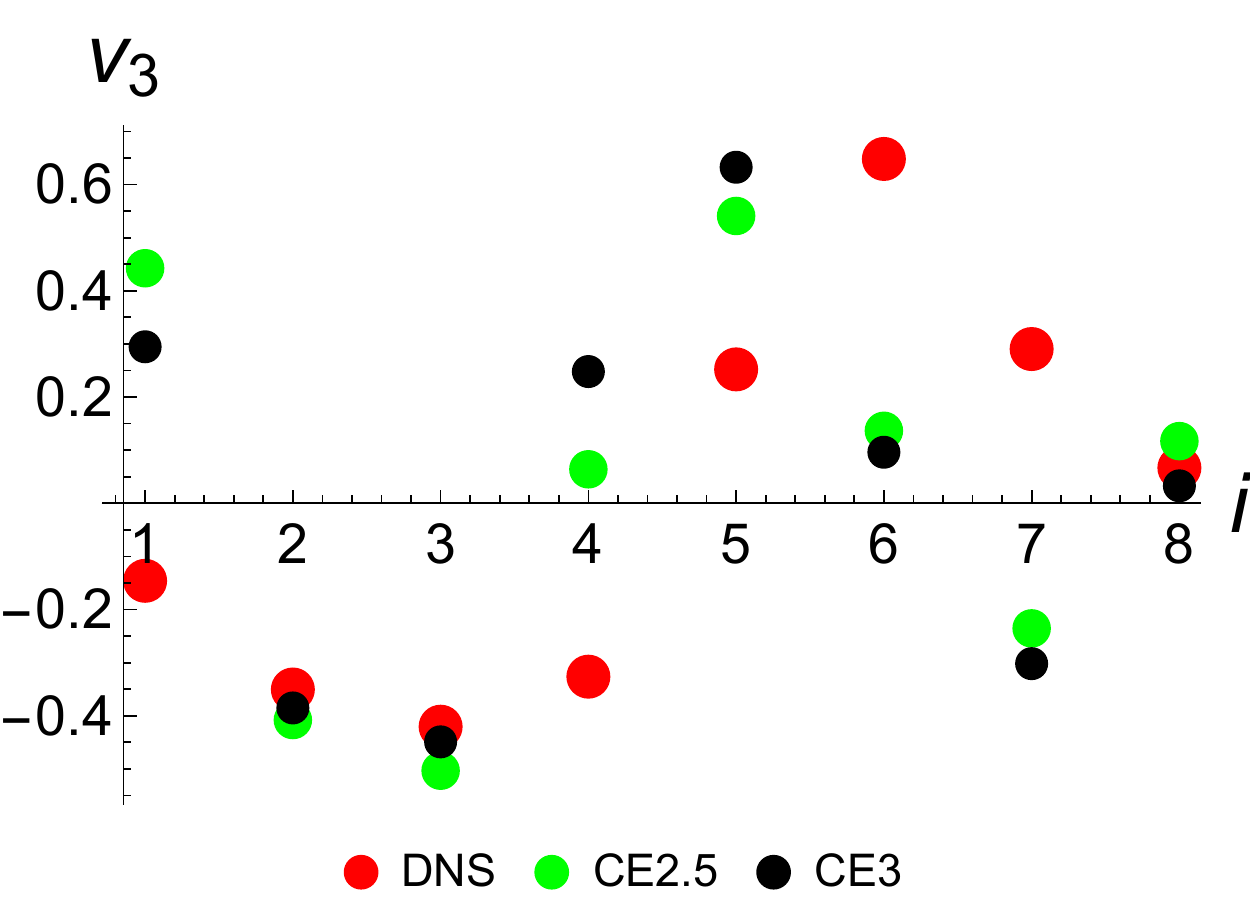}
}
\caption{The illustration of a variety of dynamical and statistical properties obtained in DNS and DSS for the Lorenz96 system in the periodic state for $c =2$ and $f_{i}=20$ for $i=2,3,\cdots 8$, where the eddy damping parameter is chosen to be $\tau_d^{-1}=20$ for both CE2.5 and CE3 computation. The time series of the dynamical model in DNS is shown in (a), the first cumulants, $C_{x_i}$ obtained in DNS and DSS are plotted in (b), the eigenvalues of the covariance matrix in DNS and DSS are in (c) and the plots of the first three eigenvectors of the covariance in DNS and DSS are illustrated in (d)--(f).}
\label{inho6}
\end{figure}

In this section, we study the dynamics of Lorenz96 system driven by the inhomogeneous force, $f_i$, in DNS and DSS. We use a parameter, $c$, to quantify the inhomogeneity of the external force, $f_i$, where for all cases the first component of the force, $f_1$, which acts on the first node, is set to be $c$ times as large as the other components, i.e., $f_1 = c f_i$ for $i > 1$; whilst the other components remain equal, i.e., $f_i=f_j$ for $i, j\neq 1$. We repeat the numerical experiments for the homogeneous cases in the periodic and chaotic states and compare the low-order statistics obtained in DNS and DSS. The numerical results are illustrated and compared in Figs. (\ref{inho1} -- \ref{inho6}), where the time series of the state vector, ${\bf x}$ obtained by solving the dynamical equation~(\ref{govEq}) and the low-order statistics obtained in DNS and DSS are plotted in each figure.


The inhomogeneous force has a significant impact on the dynamics in addition to the low-order statistics of the Lorenz96 system in the periodic and chaotic state, even for the system with small inhomogeneities, e.g., for $c=1.05$. The mean trajectory of the first node of the Lorenz96 system, $C_{x_1}$, which is driven by the largest external force, has a larger value than the other components of the trajectory. Owing to the nonlinear coupling, the first cumulants of the $4$th, $7$th and $8$th nodes are as large as $C_{x_1}$, whilst the mean trajectory of the $2$nd node, which continuously transport energy to the first node, is the smallest. Importantly, the unevenly distributed force  breaks the translational symmetry of the Lorenz96 system, i.e., the eigen-pairs of the covariance matrix is no longer doubly-folded and the eigen-decomposition of the matrix becomes unique in contrast with the cases under the homogeneous forcing. We observe that the unevenly distributed force constantly drives the system away from the homogeneous state and the nonlinear coupling further enhances the inhomogeneity introduced by $f_i$, e.g., see Figs. (\ref{inho3} \& \ref{inho5}) for $c=1.2$ \& $f_{i>1}=5,\ 20$ and Figs. (\ref{inho5} \& \ref{inho6}) for $f_{i>1}=20$ \& $c=1.2,\ 2$ for comparison.

For all test cases in the periodic and chaotic state with different level of inhomogeneity, the solutions of the CE2.5/3 equations are found to be numerically stable and accurate for the eddy damping parameter, $\tau_d$, in the range, $\tau_d \sim {\cal O}(0.01) - {\cal O}(0.1))$. The optimal eddy damping parameter, $\tau_d$, for both periodic and weakly chaotic states, i.e., $f_i \leq 5$, are approximately $\tau_d = 0.1$ for all cases. In the periodic state, the low-order cumulants approximated by the CE2.5/3 system perfectly agree with those obtained in DNS, e.g., see Figs. (\ref{inho1} -- \ref{inho2}). In the weakly chaotic state for $f_{i>1} = 5$ for $c=1.05,\ 1.2$ and $2$, the cumulant equations also very accurately approximates the low-order statistics of the Lorenz96 system as compared with the results obtain in DNS. Importantly, we find that in the weakly turbulent regime, the DSS equations is more accurate for approximating the low-order statistics of the inhomogeneous system than the homogeneous ones, e.g., see Fig. (\ref{inho3}) for $f_{i>1}=5$ and $c=1.2$, where the eigenvalues of the covariance matrix also agree very well with the eigenvalues found by the ensemble averaging of the solutions in DNS. However, the associated eigenvectors are less accurate, due to the sensitivity of the eigen-decomposition of the covariance matrix of this system. In the strongly chaotic regime for $f_{i>1}=20$ for all level of inhomogeneities, the low-order cumulants obtained by solving the CE2.5/3 equations are less accurate but the error remains less than $2\%$ for the first cumulants. In this regime, the maximum and also the optimal eddy damping parameter, $\tau_d$, is found to be $\tau_d = 1/20$, which indicates the greater significance of the fourth cumulants in the cumulant approximations for the strongly chaotic regime than for the periodic and weakly chaotic regimes. We note that setting $\tau_d=1/10$ leads to the solution of the cumulant equations oscillating in time, and so it does not converge to statistical equilibrium. However, if $\tau_d$ is set smaller than $1/20$, the cumulant equations tend to underestimate the inhomogeneity of the system. For example, the difference between the largest two eigenvalues, $\lambda_1$ and $\lambda_2$, of the covariance matrix, which are split by the inhomogeneous forcing, is less estimated by the cumulant equations, e.g., see Fig. (\ref{inho4} \& \ref{inho5}) for comparison.

Furthermore, we solve the cumulant equations using the model reduction strategies, {\it i}) and {\it ii}). In the periodic state, where a smaller number of eigen-pairs of the covariance matrix are excited as compared with the spatial dimension of the matrix, $\xmax{n}=8$, e.g., for $f_{i>1}=1.2$, we obtain $4$ significant eigen-pairs. Here, we can always reduce the complexity of the covariance matrix using the strategy {\it i}) and the numerical solutions of the reduced CE2.5/3 approximations are as accurate as the solutions of the original cumulant equations. In contrast, in the chaotic state with all  levels of inhomogeneity, the covariance matrix cannot be reduced using the strategy {\it i}). But the diagonalised cumulant equations can always be obtained by using the strategy {\it ii}), and the solutions of the transformed CE2.5/3 equations are always found to be identical to the solutions of the original cumulant equations. However, for the inhomogeneous case, the eigenbasis of the covariance matrix is no longer a Fourier basis, instead it varies as a function of $c$, e.g. see Table (\ref{tb5})
\begin{table*}
\begin{ruledtabular}
\begin{tabular}{l|r|r|r|r|r|r|r|r|r}
         &             & ${\cal F}_0$ & ${\cal F}_1$ & ${\cal F}_2$ & ${\cal F}_3$ & ${\cal F}_4$ & ${\cal F}_5$ & ${\cal F}_6$ & ${\cal F}_7$ \\ \hline
$c=0$    & $v_1$ & $0$          & $0$          & $0$        & ${\bf C}_{1,1}$          & ${\bf C}_{1,2}$          & $0$          & $0$          & $0$                           \\\hline
$c=1.05$ &             &     $-1.7E{-1}$ & $-1.9E{-1}$ & $-1.6E{-1}$ & $1.2E{-1}$ & ${\bf 6.3E{-1}}$ & ${\bf -6.1E{-1}}$ & $-1.5E{-1}$ & $-3.2E{-1}$                                     \\\hline
$c=1.2$   &           &       $-1.7E{-1}$ & $ -2.1E{-1} $ & $ -1.6E{-1} $ & $1.3E{-1}$ & ${\bf 6.0E{-1}}$ & ${\bf - 6.2E{-1}}$ & $ - 1.1E{-1}$ & $ - 3.5E{-1}$ \\\hline
$c=2$    &             &       $-1.6E{-1}$ &  $-2.9E{-1}$ &  $-1.3E{-1}$ & $1.4E{-1}$ & ${\bf 5.5E{-1}}$ & $-5.4E{-1}$ & ${\bf 6.1E{-2}}$ & $-4.9E{-1}$          \\\hline\hline
$c=0$    & $v_2$ & $0$          & $0$          & $0$          & ${\bf C}_{2,1}$        & ${\bf C}_{2,2}$          & $0$          & $0$          & $0$                            \\\hline
$c=1.05$ &             &             $-9.1E{-2}$ & $5.3E{-2}$ & $-9.9E{-2}$ & $4.4E{-1}$ & ${\bf-7.1E{-1}}$ & ${\bf -5.2E{-1}}$ & $-9.8E{-2}$ & $-1.1E{-1}$                     \\ \hline
$c=1.2$ &  & $-7.7E{-2}$ & $6.7E{-2}$ & $ -1.0E{-1} $ & $4.1E{-1}$ & ${\bf -7.3E{-1}} $ & ${\bf -5.1E{-1}}$ & $- 1.0E{-1}$ & $ -1.1E{-1}$ \\\hline
$c=2$    &             &  $-8.1E{-2}$ & $1.1E{-1}$ & $-1.8E{-1}$ & $2.4E{-1}$ & ${\bf -7.2E{-1}}$ & ${\bf -5.7E{-1}}$ & $-1.7E{-1}$ & $-1.2E{-1}$       \\\hline             
\end{tabular}
\end{ruledtabular}
\caption{The projection of the eigenvectors with two largest eigenvalues on to the Fourier basis, ${\cal F}_k$, for the Lorenz96 system in the chaotic state for $f_{i>1}=20$ with different inhomogeneity, $c=0, 1.05, 1,2$ \& $2$, where for each eigenbasis, the largest two coefficients in amplitude are marked in bold. For $c=0$ case, the projection is non-unique due to the translational symmetry, where the coefficients satisfies, ${\bf C}_{1,1}^2 + {\bf C}_{1,2}^2 = {\bf C}_{2,1}^2 + {\bf C}_{2,2}^2 = 1$.}
\label{tb5}
\end{table*}

for the coefficients of the eigenvectors with the two largest eigenvalues, which are projected on to the Fourier basis, for the Lorenz96 system in the chaotic state for $f_{i>1}=20$ with different homogeneity, $c=0, 1.05, 1.2$ and $2$. As expected for large inhomogeneity parameter, $c$, the eigenbasis, which comprises more than one Fourier modes, becomes complex in space. In the study, we diagonalise the cumulant equations using the eigenbasis of the covariance matrix computed by solving the original cumulant equations.

\section{Conclusion}
\label{con}

We have implemented two effective approaches to reduce the statistical descriptions of a dynamical system for solving the low-order DSS equations closed by the CE2.5/3 approximations. In this paper, we have demonstrated the effectiveness of the methods by solving a representative Lorenz96 model in a variety of dynamical regimes to synthesise various characteristics of astrophysical and geophysical fluid dynamical systems in nature. By varying a single control parameter of the Lorenz96 system -- the external forcing term, we access the periodic, doubly periodic and chaotic regimes of the Lorenz96 system and further introduce inhomogeneity of the external forcing to break the translational symmetry of the system.

Primarily, we show that DSS is an accurate method for describing the statistical behaviour of the Lorenz96 system in all dynamical states; we obtain the accurate time invariant solutions of the low-order cumulants as compared with those obtained in DNS for the Lorenz96 system. We also show that the CE2.5/3 equations are numerically stable for the eddy damping parameter, $\tau_d$, in the range between $10^{-3}$ and $10^{-1}$, which is approximately $10$ to $100$ times smaller than a characteristic time scale of the travelling wave of the Lorenz96 system. The optimal $\tau_d$ is found to be approximately $0.1$ for all cases. For this system, the covariance matrix is always found to be dominated by the diagonal elements with off-diagonal elements found to be approximately $10$ to $100$ times smaller. Interestingly, we find that, when the spectrum of the covariance becomes dense as the Lorenz96 system settles in the weakly chaotic regime, e.g., for the external force to be $f_i=5$, the eigen-decomposition of the covariance matrix becomes sensitive to the off-diagonal elements of the covariance matrix. However as we increase the chaoticity of the system by increasing $f_i$, the accuracy of the eigen-decomposition increases. Translational symmetry for the Lorenz96 system is preserved when the external force, $f_i$, is an equal constant over all nodes or satisfies an identical probability distribution. In this scenario, the Fourier basis function, which quantifies the correlations of the travelling wave of different wave numbers, is the natural eigenbasis function of the covariance matrix. Furthermore the eigensystem of the matrix has a degeneracy and the eigen-decomposition is non-unique. When the translational symmetry of the Lorenz96 system is broken by the application of an external inhomogeneous force, the degeneracy of the eigensystem of the covariance matrix is lifted, and so a Fourier basis is no longer appropriate; here the eigen-decomposition becomes unique.

We then focus our studies on the model reduction strategies for solving DSS equations. In the periodic state of the Lorenz96 system with and without the translational symmetry, the fluctuations of the non-coherent components of the Lorenz96 system are quantified by a few eigenmodes, which results in a sparse spectrum of the eigen-system of the covariance matrix, i.e., a relatively small number of eigenvalues are significantly greater than zero as compared with the spatial resolution of the system. For this type of problem, we may always apply a model reduction strategy, {\it i}) to reduce the complexity of the covariance matrix by retaining only the leading order eigen-pairs of the covariance matrix in the numerical computation. This strategy is well suited for CE2/2.5 systems, as the third cumulants in CE2.5 approximation are determined by the quadratic interactions of the second cumulants. However, the strategy, {\it i}), becomes less effective, when all eigenvectors are excited in the covariance matrix with the associated eigenvalues significantly greater than zero, e.g., for the Lorenz96 system in a chaotic state. On the other hand, the covariance matrix is symmetric and non-negative definite and so can always be diagonalised. The solution of the diagonalised DSS equations is equivalent to the original DSS equations by the unitary transform. This model reduction strategy, {\it ii}) works well for all cases of the Lorenz96 system, both in the periodic and chaotic regimes with and without the translational symmetry for CE2.5/3 approximations. In numerical computations, we solve for the diagonal elements of the second cumulants and set all off-diagonal ones zero. This simplification further reduces the computational complexity of the third cumulants. However, the application of this method may be limited by the determination of the eigenvector of the covariance matrix prior to the observation. For the Lorenz96 system with the broken translational symmetry, the eigenvectors of the covariance matrix vary as a function of the inhomogeneous external force, which is unknown before we solve the DNS or the original DSS equations.

The fluctuations of the non-coherent components of the PDE systems are expected to be accurately described by a few eigen-pairs of the covariance matrix, which enables us to simply the DSS equations using the model reduction strategy {\it i}) via the eigen-decomposition. When the eigenvectors of the covariance are priorly known, the diagonalised DSS equations can be obtained in a straightforward manner. Then the statistical description of the PDE systems may be further reduced by using the combined strategies of {\it i}) and {\it ii}). We therefore believe that the dynamical systems describing turbulent fluid dynamics and magnetohydrodynamical problems in two and three dimensions may be efficiently and accurately solved using DSS in the near future.

\section{Acknowledgements}

This is supported in part by European Research Council (ERC) under the European Unions Horizon 2020 research and innovation program (grant agreement no. D5S-DLV-786780) and by a grant from the Simons Foundation (Grant number 662962, GF). 

\bibliography{Lorenz}

\end{document}